\documentclass[12pt, english]{article}

\newcommand{\E}{\mathbb{E}}

\usepackage{changes}
\usepackage{verbatim}
\usepackage{url}
\usepackage{amsmath}
\usepackage{amsfonts}
\usepackage{amssymb}
\usepackage{todonotes}
\usepackage{footmisc}
\usepackage{relsize,etoolbox}
\usepackage{bbm}
\usepackage{color, colortbl}
\definecolor{Gray}{gray}{0.95}
\usepackage{booktabs}
\usepackage{pdflscape}
\usepackage{amsfonts}
\usepackage{graphicx}
\usepackage{multirow}
\usepackage{epstopdf}
\usepackage{adjustbox}
\usepackage[titletoc,title]{appendix}
\usepackage{lineno}
\usepackage{rotating}

\definecolor{lavander}{cmyk}{0,0.48,0,0}
\definecolor{violet}{cmyk}{0.79,0.88,0,0}
\definecolor{burntorange}{cmyk}{0,0.52,1,0}

\def\lav{lavander!90}
\def\oran{orange!30}

\tikzstyle{peers}=[draw,circle,violet,bottom color=\lav,
                  top color= white, text=violet,minimum width=10pt]
\tikzstyle{superpeers}=[draw,circle,burntorange, left color=\oran,
                       text=violet,minimum width=20pt]
\tikzstyle{legendsp}=[rectangle, draw, burntorange, rounded corners,
                     thin,bottom color=\oran, top color=white,
                     text=burntorange, minimum width=2.5cm]
\tikzstyle{legendp}=[rectangle, draw, violet, rounded corners, thin,
                     bottom color=\lav, top color= white,
                     text= violet, minimum width= 2.5cm]
\tikzstyle{legend_general}=[rectangle, rounded corners, thin,
                           burntorange, fill= white, draw, text=violet,
                           minimum width=2.5cm, minimum height=0.8cm]

\usepackage[most]{tcolorbox}
 \tcbset{colback=white}

\usepackage[square]{natbib}

\usepackage{setspace}
\usepackage[margin=1in]{geometry}
\usepackage[latin9]{inputenc}
\usepackage{geometry}
\usepackage{array}
\usepackage{float}
\usepackage{booktabs}
\usepackage{enumitem}
\usepackage{bm}
\usepackage{amsmath}
\usepackage{amssymb}
\usepackage{graphicx}
\usepackage{setspace}
\definecolor{sangre}{rgb}{0.6,0.18,0.19}
\definecolor{dullmagenta}{rgb}{0.4,0,0.4}
\definecolor{darkblue}{rgb}{0,0,0.6}
\usepackage[capitalize]{cleveref}

\makeatletter

\usepackage{bm}
\usepackage{caption}
\usepackage{enumitem}
\usepackage{epstopdf}
\usepackage{dsfont}
\usepackage{amsfonts}
\usepackage{amsthm}

\newtheorem{assumption}{Assumption}
\newenvironment{thmbis}[1]
  {\renewcommand{\theassumption}{\ref{#1}$'$}%
   \addtocounter{assumption}{-1}%
   \begin{assumption}}
  {\end{assumption}}
\usepackage{lscape}
\usepackage{lipsum}

\usepackage{babel}

\@ifundefined{showcaptionsetup}{}{%
 \PassOptionsToPackage{caption=false}{subfig}}
\usepackage{subfig}
\makeatother

\begin{document}

\title{Signaling and Employer Learning with Instruments%
\thanks{\protect\linespread{1}\protect\selectfont We thank Thomas Lemieux, three anonymous referees, Peter Arcidiacono, Leora Friedberg, John Bodian Klopfer, Edwin Leuven, Emily Nix, John Pepper, and participants at the 2018 Cowles Conference in Honor of Joseph Altonji, 2019 Essen Health Conference, 2021 WEAI conference, and seminar participants at University of Alberta, University of Bergen, University of British Columbia, University of Calgary, Deakin University, University of Delaware, Georgetown University, Goethe University Frankfurt,   Harris School of Public Policy, London School of Economics and Political Science, Norges Bank, Ohio State University, Ragnar Frisch Centre for Economic Research, University of Rochester, University of Southern California,  University of Sussex, Tinbergen Institute, and University of Waterloo for helpful comments and suggestions. This project received generous financial support from the Research Council of Norway through grants 194339, 250516, 267428 and 275123, the Social Science and Humanities Research Council of Canada and the Canada Research Chair program.}}

\author{Gaurab Aryal\thanks{ Department of Economics, University of Virginia, { aryalg@virginia.edu}} \qquad
 Manudeep Bhuller\thanks{Department of Economics, University of Oslo; and Statistics Norway,
 {manudeep.bhuller@econ.uio.no}}  \qquad 
 Fabian Lange\thanks{ Department of Economics, McGill University; and NBER, {fabolange@gmail.com}}
}

\date{\today}

\maketitle

\begin{abstract}
This paper considers the use of instruments to identify and estimate private and social returns to education within a model of employer learning. What an instrument identifies depends on whether it is hidden from, or transparent (i.e., observed) to, the employers. A hidden instrument identifies private returns to education, and a transparent instrument identifies social returns to education. We use variation in compulsory schooling laws across non-central and central municipalities in Norway to, respectively, construct hidden and transparent instruments. We estimate a private return of 7.9\%, of which 70\% is due to increased productivity and the remaining 30\% is due to signaling.

\end{abstract}

\newpage
\noindent 

\section{Introduction\label{sec:Intro}}
Two competing models rationalize the positive relationship between earnings and education that is universally found in data. Since \cite{Becker1962}, proponents of the human capital model argue that education increases skills that employers value. By contrast, the job-market signaling model of \cite{Spence1973} posits that education signals differences in innate abilities among workers.\footnote{For more on signaling and the human capital model, see \cite{Wolpin1977, TylerMurnaneWillett2000, Bedard2001, Fang2006, Hopkins2012, ClarkMartorell2014, FengGraetz2017} and \cite{Arteaga2018}.} Signaling, however, tends to be inefficient because workers expend valuable resources only to signal their abilities. Thus, signaling creates a wedge between private returns and social returns to education.\footnote{We define private returns as the effect of education on individual wages and social returns as the effect of education on individual productivity, and the difference between them is the signaling value. This definition abstracts from productive externalities beyond the employer-employee relationship \citep{acemoglu2000large, moretti2004estimating}, non-production and non-pecuniary benefits \citep{Lochner2011, OreopoulosSalvanes2011}, and fiscal externalities and distributional impacts \citep{Stiglitz1975}.}

Education policy requires empirical guidance on how large this wedge is. However, the literature has long recognized the difficulties in separating signaling effects from human capital effects. See \cite{LangeTopel2006} and the references therein. One way forward is to recognize that if workers use education to signal their abilities at the start of their careers, then over time, employers may also learn about workers' abilities.

Two influential papers, \citet{FarberGibbons1996} and \citet{AltonjiPierret2001} (FG and AP, respectively) use changes across experience in how earnings correlate with schooling and a proxy of unobserved ability (e.g., IQ score) to test employer learning. Following FG and AP, \cite{Lange2007} shows that employers learn fast, and using the first-order conditions that characterize schooling decisions, he identifies an upper bound on how much signaling contributes to the private returns to education over the career. In general, researchers following this line of literature need a valid proxy of unobserved ability. Even then, they can only identify bounds on social returns after imposing strong and unverifiable assumptions.

We add to the set of empirical methods available for distinguishing human capital returns from signaling returns to schooling by asking: what do instrumental variable (IV) estimates of the returns to education identify within the employer learning framework of FG, AP, and \cite{Lange2007}? We show that, under suitable assumptions about the instrument's observability, IV estimates of the causal effects of education on wages point-identify the private and the social returns to education.

{ Particularly, to make progress in this setting with asymmetric information, we introduce two new concepts related to IVs: 
 a \emph{hidden} IV that employers do not observe and a \emph{transparent} IV that they do. Then we show that, in a general setting where returns to skills can vary with experience, a hidden IV identifies \emph{experience-varying} private returns to education, whereas a transparent IV identifies \emph{experience-varying} social returns to education. Thus, the central message of the paper and its main contribution is to show that what employers know about the IV determines what we can estimate with it. More broadly, this approach can guide the identification strategies in similar contexts where  asymmetric information is important.

To identify the signaling returns to education, we need to identify both the private \emph{and} social returns to education. We show that we need to lean more heavily on the model when only a hidden IV is available. In particular, if we assume that the returns to skill are constant across work experience, then we can identify the private and social returns to education. Intuitively, when the productivity returns to education are constant across experience, the private returns, identified using the hidden IV, change only because employers' information changes. Furthermore, with work experience these hidden IV estimates converge to the experience-invariant social return, which is thus also identified. We can then use the gap between these social and private returns to identify the signaling value of education.\footnote{In addition, if we are willing to impose parametric assumptions, such as normality of idiosyncratic shocks in productivity, as in \cite{Lange2007}, then we can also identify the speed of employer learning.}

Unsurprisingly, when the returns to skill vary with work experience, a hidden IV alone can not identify the private and social returns to education. However, access to both hidden \textit{and} transparent IVs is sufficient for identification because then we can disentangle the effects of experience-varying productivity on wages from employer learning.\footnote{In practice, two IVs are seldom available for the same population of workers. To combine estimates from two samples (exposed to hidden IV and transparent IV), we have to assume that the effect of education on productivity varies in the same way by work experience across these two samples. This homogeneity assumption still allows for differences in the overall productivity effect of education across these samples.\label{footnote:homogeneity1}} 
Intuitively, we can first rely on how the transparent IV estimates vary with experience to identify the effect of education on productivity --the social returns-- across the work experience. Then we can combine these estimates with the hidden IV estimates to separate the social and the private returns to education.}

We implement these ideas on a unique dataset of earnings histories between 1967 and 2014 for Norwegian males born between 1950 and 1980. From 1960 to 1975, Norway extended compulsory schooling from 7 years to 9 years, but it did not implement this change on the entire country all at once. Instead, compulsory schooling requirements increased at different times for different municipalities. The differential implementation of this reform provides us with a plausibly exogenous source of variation in schooling attainment.

We use this historical episode to construct two instruments, one of which we interpret as plausibly hidden and the other as plausibly transparent. In particular, to define these two IVs, we rely on an important feature of how the compulsory schooling reform was implemented across municipalities and the local labor markets structure in Norway. In 1960, the Norwegian territory consisted of 732 municipalities, and the timing of reform implementation varied substantially across these municipalities, even within a local labor market.\footnote{Norway is classified into 160 local labor markets. In each market, we define municipalities with the largest population as central municipalities and the rest as non-central. The average population in non-central municipalities was 3,500, and around 40\% lived in central municipalities.} Employment was primarily concentrated in the central municipalities, with employers hiring workers from both central and non-central locations. For workers born and raised in non-central municipalities, it is reasonable to assume that employers were uninformed about the reform's timing, given that there were a large number of such municipalities, each with a potentially different reform timing. 

In contrast, employers were more likely informed about the timing of the reform in the dominant population center within each labor market. Consequently, they observe the reform exposure status for workers who grew up in central locations. Building on this distinction, we interpret our instruments as hidden for those growing up in the non-central municipalities and transparent for those in the central municipalities.

Using these data, we first examine how the IV returns to schooling vary across work experience and interpret the patterns through the lens of our model. The hidden IV estimates from the non-central municipalities suggest that the private returns to education are initially high during the first few years in the labor market, but they decline rapidly and stabilize around 5.5\% after 15 years of work experience. Thus, our results are in line with those by \cite{Lange2007} who found that employers learn fast. In contrast, the transparent IV estimates from the central municipalities suggest that social returns to education do not vary systematically with experience. In particular, in the transparent IV sample, the social returns to education average at 8.8\% across experience levels, close to 10\% in the initial year of work experience and 8\% after 30 years. Due to this constancy in the social returns, we cannot reject the standard employer learning model with constant returns to skill.

Second, we quantify how much signaling and human capital acquisition contributes to the lifetime returns to education. The signaling value is determined by how quickly employers learn and how strong the signal is, i.e., how large a difference in ability can be inferred based on education. Combining estimates from the transparent and hidden IVs, we determine the internal rate of return to an additional year of education to be 7.9\%. Our estimates suggest that 70\% of this private return represents the productivity-enhancing effect of education, and 30\% represents the signaling effect of education. These estimates are from the model with experience-varying returns to skill. Our conclusions do not change even if we estimate the model, assuming that the returns to skills are constant over experience.\footnote{In a companion work \cite{AryalBhullerFabian2021b}, we consider an extension to allow a hidden correlate of ability, as traditionally assumed in the employer learning literature, such as AP, FG, \cite{Lange2007} and \cite{Arcidiacono2010}. In that paper, we report results relying on an IQ test score from the Norwegian Armed Forces to estimate the social and private returns to education. These results are broadly consistent with those in the literature and the IV estimates reported here.}

Finally, we consider two extensions of our model. First, we study the identification when it is not clear whether the IV is hidden or transparent. In particular, we consider the case of a \emph{partially-transparent} IV, where the IV is transparent with some probability and hidden with the complementary probability. We show that with experience-invariant returns to skill, a partially transparent IV identifies the lower bounds on the private returns and the signaling value while point-identifying the speed of learning. With experience-varying returns to skill, access to a transparent IV \emph{and} a partially-transparent IV is sufficient to identify the lower bounds on the private returns and the signaling value. 
Access to a hidden IV \emph{and} a partially-transparent IV, however, is sufficient to point-identify all the parameters under a stronger homogeneity condition on the social returns across samples.

For the second extension, we revisit the assumption about homogeneous returns. Recent studies have provided evidence that the returns can vary across individuals (\cite{Card1999, heckman2006earnings, CarneiroHeckmanVytlacil2011}). To accommodate heterogeneous returns, we reframe our analysis within the potential outcomes framework, e.g., \cite{ImbensAngrist1994}, with binary (low or high) schooling. We show that our identification results extend to this setting, with a proviso that now a hidden IV identifies the \emph{average} private returns among compliers, and a transparent IV identifies the \emph{average} social returns among compliers.

The rest of our paper proceeds as follows. Section \ref{sec:Model} describes our model and defines the private and the social returns. Section \ref{sec:Identification} establishes the identification of the private and the social returns using instruments. Section \ref{sec:DataSetting} presents our data and Section \ref{sec:Results} contains our empirical findings and robustness exercises. Section \ref{sec:Extension} considers extensions and Section \ref{sec:Discussion} presents a discussion. Section \ref{sec:Conclusion} concludes. All additional results are in the Online Appendix.

\section{Model\label{sec:Model}}

We present a model of employer learning in a perfectly competitive labor market. Our model extends FG's original formulation, which has become the standard model in the empirical employer learning literature, by allowing the returns to skill to vary with experience.

Let worker $i$'s log-productivity with work experience $t\in\mathbb{T}:=\{0,1,2,\ldots, \infty\}$ be  
\begin{eqnarray}
\psi_{it} := \ln \chi_{it} =\lambda_{t}\times\left(\beta_{ws}S_{i}+A_{i}+\varepsilon_{it}\right)+H(t),\label{eq:chi}
\end{eqnarray}
where $S$ is the years of schooling, $A$ is the unobserved ability (to both employers and researchers), and $\varepsilon$ is transient variation in productivity independent of all other variables.\footnote{In FG, there are also variables capturing information available to employers but not the researchers. An example of such a variable could be knowledge of a foreign language, typically mentioned in job applicants' r\'esum\'es and verified. FG also allows for a correlate of ability available to researchers but not employers. These are necessary to derive how the partial correlation between earnings, ability proxy, and schooling changes with experience. For notational convenience, we suppress these variables here. Our results go through unchanged if we allow for them; see the Online Appendix \ref{section:IQ_new} for more.}

Thus, in our setting, work experience affects productivity through $H(t)$ and $\lambda_t\in[0,\infty)$. 
Here, the function $H(t)$ captures how log-productivity commonly varies
with work experience across individuals.
While we allow $H(t)$ to be a nonparametric function of $t$, we assume that it does not depend on either $S$ or $A$.\footnote{In our empirical application, we specify $H(t, X)$ so that the experience profile can vary flexibly with individual characteristics $X$ that include dummy variables for birth cohort and childhood municipality.} 
The parameters $\{\lambda_t: t\in \mathbb{T}\}$ capture the experience-specific effect of composite individual skill $(\beta_{ws}S_{i}+A_{i}+\varepsilon_{it})$.\footnote{The composite skill $(\beta_{ws}S_{i}+A_{i}+\varepsilon_{it})$ and skill returns ($\lambda_{t}$) are assumed to be multiplicatively separable in (\ref{eq:chi}). Relaxing this assumption will in most cases require other functional form assumptions about the productivity effect of ability for the identification. We, however, can allow the productivity effect of schooling to be experience-varying, while the ability effect to be experience-invariant as a special case of the model.} 

Throughout the paper, we normalize the experience-specific effect at the start of a career to be one, i.e., at time $t=0, \lambda_0=1$. So, $\lambda_t$ is the returns to skill in period $t>0$ relative to the start of one's career.  
When $\lambda_t$ is constant across $t$, i.e., $\lambda_t=1$, for all $t\in\mathbb{T}$, we say that the returns to skill are experience-invariant. When $\lambda_t$ varies with $t$, then the effect of the individual skill component $(\beta_{ws}S_{i}+A_{i}+\varepsilon_{it})$ on productivity varies with experience and we say that the returns to skill are experience-variant.

Under perfect competition, workers are
paid the conditional expected output, given the information available to the employers.
At time $t$, besides knowing $S_i$, employers also observe the total output ($\chi_{it}$), which is equivalent to observing a signal $\xi_{it}:= A_{i}+\varepsilon_{it}$ about $i$'s productivity. 
Let $\mathcal{E}_{it}=(S_{i},\xi_{i}^{t})$
denote employers' information about $i$ in period $t$ where $\xi_{i}^{t}=\left\{ \xi_{i\tau}\right\} _{\tau<t}$ is the history of all past signals.\footnote{Throughout the paper, we assume that employers have symmetric information about workers' abilities and past outputs. See Section \ref{section:asymmetric learning} for more on this assumption.} 
So, the wage in period $t$ is equal to the expected productivity given ${\mathcal E}_{it}$, i.e., $
W_{it}=\E\left[\chi_{it}|{\mathcal E}_{it}\right] = \E\left[\exp(\psi_{it})|{\mathcal E}_{it}\right]$.

Next, we make distributional assumptions to simplify the wage equation and keep the employer learning process tractable. 
In particular, we follow \cite{Lange2007} and assume that $\varepsilon_{it}\stackrel{i.i.d}{\sim}\mathcal{N}(0,\sigma_{\varepsilon}^{2})$ and $(S_i, A_i)\stackrel{i.i.d}{\sim}\mathcal{N}(\bm{\mu},\Sigma)$, across workers and across experience. 
Let $\sigma_0^2=Var(A|S)$ be the conditional variance of $A$ given $S$.
Thus $\exp(\psi_{it})$ in (\ref{eq:chi}) is a lognormal random variable with parameters $\E\left[\psi_{it}|{\mathcal E}_{it}\right]$ and $v_{t}:= Var\left(\lambda_t (A_{i}+\varepsilon_{it})|\mathcal{E}_{it}\right)$, so its mean is given as $\E\left[\exp(\psi_{it})|{\mathcal E}_{it}\right]=\exp\left(\E\left[\psi_{it}|{\mathcal E}_{it}\right]+\frac{1}{2} v_t \right)$. 

Substituting this expression for the mean of $\exp(\psi_{it})$ in the wage equation gives a simpler form $W_{it}=\exp\left(\E\left[\psi_{it}|{\mathcal E}_{it}\right]+\frac{1}{2} v_t \right).
$
Then, taking the log on both sides of this equation, and using (\ref{eq:chi}), we get the log-earnings equation
\begin{eqnarray}
\ln W_{it} = \lambda_t\times (\beta_{ws}S_{i}+\E\left[A_{i}|\mathcal{E}_{it}\right])+H\left(t\right)+\frac{1}{2}v_{t}=\lambda_t\times (\beta_{ws}S_{i}+\E\left[A_{i}|\mathcal{E}_{it}\right])+\tilde{H}\left(t\right),\label{eq:Wage}
\end{eqnarray}
where $\tilde{H}(t)\equiv H\left(t\right)+\frac{1}{2}v_{t}$ collects the terms that vary only with $t$ but not across the realizations
of $\xi_{i}^{t}$. 
For notational simplicity, we suppress $\tilde{H}\left(t\right)$ until the empirical analysis. 
Next, we define the \emph{social returns} to education and the \emph{private returns} to education.  

\paragraph{\bf Social Returns to Education.}
Above, we defined $A_i$ as the component of skill that correlates with but is not caused by schooling. Together with assuming that $\varepsilon_{it}$ is orthogonal to $S_i$, it follows that the coefficient $\delta_t^{\psi|S}:=\lambda_t\times\beta_{ws}$ of $S$ in (\ref{eq:chi}) is the causal effect of education on productivity. 
For notational simplicity, letting $u_{it}=\lambda_t\times\left({A}_{i}+\varepsilon_{it}\right)$ and suppressing $H(t)$ in (\ref{eq:chi}) we get our key equation: 
\begin{eqnarray}
\psi_{it} =\delta_{t}^{\psi|S}\times S_{i}+u_{it}.\label{eq: causal schooling 2}
\end{eqnarray}
Thus, the social returns to education, $\delta_{t}^{\psi|S}$, can vary with work experience. 

\paragraph{\bf Private Returns to Education.} In contrast to the social returns, the \emph{private return} to education is the causal effect of schooling on wages, which captures the fact that besides productivity, schooling also affects (log) wages through employers' expectations about the worker's ability. 
Thus, the private return to education is the partial derivative of log-wages in (\ref{eq:Wage}) with respect to $S$, which can vary with work experience. The private returns depend on how $S$ affects $\mathbb{E}[A_i|{\mathcal E}_it]$, which in turn depends on the employers' learning process. So, next, we model employers' learning process. 

To this end, first, note that the joint normality of $(S,A)$ implies that the conditional expectation of $A_i$ given information at $t=0$, $\E\left[A_{i}|{\mathcal E}_{i0}\right]=\E\left[A_{i}|S_i\right]$, is linear in $S$, i.e., 
\begin{equation}
A_{i}=\phi_{A|S}\times S_{i}+\varepsilon_{A_i|S_i},\label{eq:corr A on S,Q}
\end{equation}
where $\varepsilon_{A_i|S_i}:= A_{i}-\E\left[A_{i}|S_i\right]$.\footnote{For much of what follows, $S$ need not be Gaussian, but it simplifies the exposition of the argument.  
Without it, equation (\ref{eq:corr A on S,Q}) would represent a linear projection of $A_{i}$ on $S_{i}$ instead of a conditional expectation.}
Furthermore, the normality assumptions allow us to write the conditional expectation of ability $\E\left[A_{i}|\mathcal{E}_{it}\right]$ in linear form as 
\begin{eqnarray}
\E\left[A_{i}|\mathcal{E}_{it}\right]=\theta_{t}\E\left[A_{i}|S_i\right]+\left(1-\theta_{t}\right)\bar{\xi_{i}^{t}},\label{eq:Kalman}
\end{eqnarray}
where $\bar{\xi_{i}^{t}}=\frac{1}{t}\sum_{\tau<t}\xi_{i\tau}$ is the
average of signals up to period $t$ and $\theta_{t}=\frac{1-\kappa}{1+(t-1)\kappa}\in[0,1]$ is the weight on the initial signal $(S_i)$ with $\kappa=\frac{\sigma_{0}^{2}}{\sigma_{0}^{2}+\sigma_{\varepsilon}^{2}}\in[0,1]$. 
In particular, (\ref{eq:Kalman})
shows that the conditional expectation of ability 
at time $t$ is a weighted average of the conditional expectation of ability at $t=0$, before any 
additional information about productivity has been received, and the average of all additional signals received up to period $t$, i.e., $\bar{\xi_{i}^{t}}$.

The weight $\theta_t$ declines with $t$ because, as time passes, the accumulated observed output measures become a better predictor of individual productivity differences than $S$, which was the information available at $t=0$. The rate at which $\theta_t$ declines with $t$ depends on the parameter $\kappa$ that \cite{Lange2007} refers to as the ``speed of learning." The speed of learning governs how quickly information about productivity accumulates in the market. If the signal-to-noise ratio $\sigma_0^2/\sigma_{\varepsilon}^2$ is high, then $\kappa$ will be close to 1, and the market learns quickly about the worker's ability $A$. However, irrespective of $\kappa$, when workers have spent enough time in the market and enough signals have been accumulated, the employers will put all weight only on the information accumulated after joining the labor market, i.e., $\lim_{t\rightarrow\infty}\theta_t=0$.

Thus, schooling can affect expected log-earnings at $t$ in two different ways: (i) directly because employers
use schooling to form expectations about productivity; and (ii) through learning, because schooling affects productivity, which employers learn over time by observing workers'
outputs. Substituting (\ref{eq:corr A on S,Q}) and (\ref{eq:Kalman})
in (\ref{eq:Wage}), and using $\bar{\xi_{i}^{t}}=A_{i}+\overline{\varepsilon}_{i}^{t}$ and the fact that $\E(A_i|S_i)$ is linear in $S_i$, and for notational simplicity making a change of variable $\tilde{u}_{it}\equiv \lambda_t\left(1-\theta_{t}\right)\left({A}_{i}+\overline{\varepsilon}_{i}^{t}\right)$ we get  
{\begin{eqnarray}
\ln W_{it}
=\lambda_t\left(\beta_{ws}+ \theta_{t}\phi_{A|S}\right)\times S_{i}+\lambda_t\left(1-\theta_{t}\right)\left({A}_{i}+\overline{\varepsilon}_{i}^{t}\right)=\delta_{t}^{W|S}\times S_{i}+\tilde{u}_{it}.\qquad \label{eq:final_wage_eq} 
\end{eqnarray}}
Therefore, the coefficient of schooling ($\delta_{t}^{W|S}$) in (\ref{eq:final_wage_eq}) is the private return to education. 

Comparing the private returns in (\ref{eq:final_wage_eq}) with the social returns in (\ref{eq: causal schooling 2}) gives us the following crucial relationship between the two: 
\begin{equation}
\underbrace{\delta_{t}^{W|S}}_{\texttt{private returns}}=\underbrace{\delta_{t}^{\psi|S}}_{\texttt{social returns}}+\underbrace{\theta_{t}}_{\texttt{weight}}\times\underbrace{\lambda_t \times\phi_{A|S}}_{\texttt{adjustment term}}.\label{eq:private return}
\end{equation}
Thus, the private returns $\delta_{t}^{W|S}$ differs from the social return $\delta_{t}^{\psi|S}$ if the marginal effect of schooling on expected $A$ based on the information available to firms at $t$, $\lambda_t \times \phi_{A|S}$, is non-zero. 
In fact, the signaling literature \citep{Spence1973} assumes that the ``adjustment term'' in (\ref{eq:private return}) is non-negative, so that education has signaling value, and hence $\delta_{t}^{W|S}\geq \delta_{t}^{\psi|S}$. 
However, employers eventually learn everything, as $\lim_{t\rightarrow\infty}\theta_t=0$, so, the wedge between the private and the social returns disappears with work experience, i.e., $\lim_{t\rightarrow\infty}\delta_{t}^{W|S}=\lim_{t\rightarrow\infty}\delta_{t}^{\psi|S}$.

\section{Identification \label{sec:Identification}}
In this section, we study how to identify the social and private returns using IVs.
The primary concern in the empirical literature on the returns to education is the ability bias, i.e., education correlates with determinants of productivity that are unobserved to the researcher. IV estimates of the returns to education require that unobserved ability is orthogonal to the IV. Our paper's core methodological message is as follows: to interpret the causal effects of education on wages based on IVs as either private or social returns, we must also take a stand on whether the IV is hidden from, or known to, the employers.

Before we present formal identification arguments, we highlight the intuition behind our approach. To this end, consider an IV that is uncorrelated with individual ability $A$ except through schooling. The question then is how earnings are related to this instrument and how that relation depends on whether employers observe the instrument and how informed they are about the ability $A$. We contrast two polar cases: first, the start of workers' careers when employers only observe schooling and possibly the instrument, and second, after the worker has spent sufficient time in the market so that $A$ has been revealed to the employers. 

Consider the second case first. Once employers know $A$, earnings equal productivity, and any differences in earnings associated with variation in the instrument reflect only the causal effect on productivity through education. It is so because, by assumption, the instrument is unrelated to ability $A.$ Thus, if we consider workers with sufficiently long work experience, whether the IV is hidden or transparent is immaterial--it identifies the social returns to education.  

In the first case, it matters whether the IV is hidden or transparent for workers at the start of their careers. 
When the instrument is hidden, employers observe only schooling, and any variation in schooling induced by the instrument will therefore lead employers to deduce higher ability. Therefore, hidden IVs identify the causal effect of education on wages but not on productivity, i.e., they identify the private but not the social returns to education. 

However, when the IV is transparent, average compensation, conditional on the instrument, will equal expected productivity conditional on the instrument. Any variation in compensation with the instrument status will therefore capture the causal effect of the instrument on productivity. Moreover, the IV is uncorrelated with $A$, so the causal effect only runs through education. Transparent instruments thus identify the social returns to education, not the private returns. 

The remainder of this section formalizes these intuitions and shows what happens during the learning period between these two polar cases and how having hidden \emph{and} transparent IV is sufficient to identify the signaling value of education.

\subsection{Instrumental Variables}
Throughout the paper, we maintain the assumption that the IV is orthogonal to unobserved productivity components. For simplicity, we develop the analysis using a binary IV, $D_{i}\in\{0,1\}$, and we assume that this IV satisfies the following assumption:

\begin{assumption} 
(Instrumental Variables)
\begin{enumerate}
\item (Exogeneity): $u_{it}\perp D_{i}$, where $u_{it}$ is
defined in (\ref{eq: causal schooling 2});
\item (Relevance): $\E[S_{i}|D_{i}=0]\neq\E[S_{i}|D_{i}=1]$. 
\end{enumerate}
\label{ass:1} \end{assumption} 
Under Assumption \ref{ass:1}, the Wald estimate in period $t$ gives 
\begin{eqnarray}
\texttt{plim}\!\!\quad\!\!\hat{b}^{IV}_{t} & := & \frac{\E\left[\ln W_{it}|D_{i}=1,t\right]-\E\left[\ln W_{it}|D_{i}=0,t\right]}{\E\left[S_{i}|D_{i}=1,t\right]-\E\left[S_{i}|D_{i}=0,t\right]}\label{eq:Binary Instrument},
\end{eqnarray}
which is one of the main identifying equations that we use throughout the paper.\footnote{We are also implicitly assuming stable unit treatment value assumption (SUTVA) \cite[see, e.g.,][assumption 1]{AngristImbensRubin1996} is valid. This assumption holds that a treated unit's potential outcomes are not affected by other units' exposure to the instrument. Signaling models are inherently subject to general equilibrium effects since workers' schooling decisions and employers' inference problems, central to these models, depend on the population distributions of skills and schooling. Also see related discussion on general equilibrium effects in Section \ref{sec:Discussion}.}

For workers with sufficiently long work experience, an IV identifies the social returns to education. 
In particular, for workers with sufficiently long work experience, $\lim_{t\rightarrow\infty}\theta_t=0$ so, with a slight abuse of notation, we get $\lim_{t\rightarrow\infty}\ln W_{it}=\psi_{i\infty}$, and because $S$ does not vary with $t$, from (\ref{eq:Binary Instrument}) we get 
\begin{eqnarray}
\texttt{plim}\!\!\quad\!\!\left(\lim_{t\rightarrow\infty}\hat{b}^{IV}_{t}\right)=\frac{\E\left[\psi_{i\infty}|D_{i}=1\right]-\E\left[\psi_{i\infty}|D_{i}=0\right]}{\E\left[S_{i}|D_{i}=1\right]-\E\left[S_{i}|D_{i}=0\right]}=\lim_{t\rightarrow\infty}\delta_t^{\psi|S},\label{eq:iv-separable}
\end{eqnarray}
where the second equality follows from Assumption \ref{ass:1}-(1). Thus the part of the productivity, $\psi_{it}$, not caused by schooling, $S$, is orthogonal to the instrument, $D$. 
So, with sufficient work experience, i.e., $t\rightarrow \infty$, the IV identifies the causal effect of schooling on productivity. Intuitively, in the long run, everything about a worker's ability is revealed to the
employers, and thus, knowledge of the IV itself becomes irrelevant for wage setting. It is only through its effect on schooling, which affects productivity that the IV affects wages in the long run. In the case of intermediate work experience, i.e., $t<\infty$, what $D$ identifies depends on whether it is hidden from, or known to, the employers.

\paragraph{Hidden and Transparent Instruments.}
To focus on the role of observability, we distinguish between two types of IVs depending on whether they are observed and priced-in by the employers. A \emph{hidden} instrument is unobserved by the employers and therefore does not enter into wage-setting beyond its effect on schooling. We use the superscript ${{\mathfrak h}}$ in $D^{\mathfrak h}$ to refer to a hidden IV. In particular, a hidden IV, $D^{\mathfrak h}$, satisfies the following assumption: 

\begin{assumption}\label{ass:2} (Hidden Instrument) For all $i, D^{{\mathfrak h}}_{i}\not\in{\mathcal E}_{it}$
which implies $\ln W_{it}\text{\ensuremath{\perp}}D^{{\mathfrak h}}_{i}|(S_{i},\xi^{t}_{i})$. \end{assumption}

Assumption \ref{ass:2} is conceptually different from Assumption \ref{ass:1}-(1). Assumption \ref{ass:1}-(1) asserts that the IV is conditionally independent of the determinants of productivity not caused by schooling, whereas Assumption \ref{ass:2} captures the idea that wages do not depend on the instrument $D^{{\mathfrak h}}$, given the information ${\mathcal E}_t$. So, $\ln W_{it}=\E\left[\psi_{i}|\mathcal{E}_{it},D_{i}^{\mathfrak h}\right]=\E\left[\psi_{i}|\mathcal{E}_{it}\right]$.

In many settings, Assumption \ref{ass:2} is a natural assumption. The clearest examples relate to field experiments of interventions that incentivize some students to take more schooling (treatment) and not others (control). In these cases, whether a student is in the treatment or the control group is typically not known to employers. Other examples of plausibly hidden instruments from quasi-experimental settings include (i) the interaction of draft lottery number and year of birth in \cite{AngristKrueger1992}; (ii) the interaction of policy intervention, family background, and season of birth in \cite{PonsGonzalo2002}; (iii) parents' education and the number of siblings in \cite{Taber2001}; and (iv) the elimination of student aid programs interacted with an indicator for a deceased father in \cite{Dynarski2003}. Besides these, many studies also exploit interactions of birth year and location of birth with locally implemented policy reforms, e.g., \cite{Duflo2001} and \cite{MeghirPalme2005}, which are similar to our IV.

In contrast to the \emph{hidden} instrument, we say that an instrument is a \emph{transparent} instrument if it satisfies Assumption \ref{ass:1}, but is known to the employers and ``priced-in" the wages. We use the superscript ${{\mathfrak t}}$ in $D^{{\mathfrak t}}$ to refer to a transparent IV.

\begin{assumption}\label{ass:3} (Transparent Instrument) Employers observe $D^{{\mathfrak t}}_i$ and $\ln W_{it}= \E[\psi_{it}|{\mathcal E}_{it},D^{{\mathfrak t}}_i]$. \end{assumption}

Assumption \ref{ass:3} implies that $D^{{\mathfrak t}}$ is not orthogonal to wages, conditional on schooling and other controls. From Assumption \ref{ass:1}, however, we know that $D^{{\mathfrak t}}$ satisfies the exclusion restriction with respect to productivity $\psi$. So, while $D^{{\mathfrak t}}$ violates the exclusion restriction for wages and thus does not identify the causal effect of schooling on wages (i.e., the private return), it identifies the causal effect of schooling on productivity (i.e., the social return). We formalize these intuitions in the following two subsections. 

Examples of instruments used in the literature that are plausibly transparent are (i) tuitions at two- and four-year state colleges in \cite{KaneRouse1995}; (ii) a dummy for being a male aged 19-22 from Ontario in \cite{LemieuxCard2001}; (iii) local labor market conditions in \cite{CameronHeckman1998, CameronTaber2004} and \cite{CarneiroHeckmanVytlacil2011}; (iv) change in minimum school-leaving age in the U.K. from 14 to 15 in \cite{Oreopoulos2006}; and (v) the distance to the college in \cite{Card1993}, \cite{KaneRouse1995}, \cite{Kling2001} and \cite{CameronTaber2004}. 

In practice, the distinction between hidden and transparent instruments can and often will be blurry. Thus, researchers need to be explicit about their assumptions about employers' information regarding the IV when interpreting the IV estimates.

\paragraph{Private and Social Returns.} Next, we show that a hidden IV identifies the private returns to education. 
As discussed earlier, the intuition behind this is as follows. When employers observe a worker with more years of schooling, they do not know that some have higher schooling because of the IV and not because of higher abilities. Thus, workers who are affected by the hidden IV are, on average, paid more relative to the others because Assumption \ref{ass:1}-(2) implies that the former group has, on average, more years of schooling than the latter group. Then Assumption \ref{ass:1}-(1) implies that the effect of the IV on the wages is only through schooling, thereby identifying the causal effect of schooling on earnings.

In particular, with a hidden IV, $D^{{\mathfrak h}}_{i}$, that satisfies Assumptions \ref{ass:1} and \ref{ass:2}, the numerator in (\ref{eq:Binary Instrument}) for $D^{{\mathfrak h}}_{i}=d, d\in\{0,1\}$ becomes 
$
\E\left[\ln W_{it}|D^{{\mathfrak h}}_{i}=d,t\right] = \E\left[\beta_{ws}S_{i}+\E\left[A_{i}|S_{i},\xi^{t}_{i}\right]|D^{{\mathfrak h}}_{i}=d,t\right].
$
Here, $\ln W_{it}$ does not directly depend on $D^{{\mathfrak h}}_{i}$ because the employers do not use $D^{{\mathfrak h}}_{i}$ when setting the wages.  
Instead, $D^{{\mathfrak h}}_{i}$ affects $\ln W_{it}$ indirectly via $\left(S_{i},\xi^{t}_{i}\right)$ that makes up the information
$\mathcal{E}_{it}$ about productivity available to the employers in period $t$. 
Then (\ref{eq:corr A on S,Q}) allows us to express the conditional log-wages given above as 
 \begin{eqnarray*}
\E\left[\ln W_{it}|D^{{\mathfrak h}}_{i}=d,t\right] = \left(\beta_{ws}+\theta_{t}\phi_{A|S}\right)\E\left[S_{i}|D^{{\mathfrak h}}_{i}=d\right]= \lambda_t\left(\delta^{\psi|S}+\theta_{t}\phi_{A|S}\right)\E\left[S_{i}|D^{{\mathfrak h}}_{i}=d\right].
\end{eqnarray*}
Subtracting the conditional log-wages at $D^{{\mathfrak h}}_i=0$ from the conditional log-wages at $D^{{\mathfrak h}}_i=1$ and taking the probability limit of (\ref{eq:Binary Instrument}) gives  
\begin{eqnarray}
\texttt{plim}\!\!\quad\!\! \hat{b}^{IV^{\mathfrak h}}_{t} = \frac{\E\left[\ln W_{it}|D^{{\mathfrak h}}_{i}=1,t\right]-\E\left[\ln W_{it}|D^{{\mathfrak h}}_{i}=0,t\right]}{\E\left[S_{i}|D^{{\mathfrak h}}_{i}=1\right]-\E\left[S_{i}|D^{{\mathfrak h}}_{i}=0\right]}=\lambda_t\times\left(\delta^{\psi|S}+\theta_{t}\phi_{A|S}\right)= \delta_{t}^{W|S}.\quad\label{eq:IV_hidden_time}
\end{eqnarray}

Next, we show that transparent IV identifies the social returns to education. 
Intuitively, employers know the economy's structure and know that the instrument is orthogonal to productivity except for its effect on schooling. Because employers pay the expected product conditional on their information, any average wage difference between treated and untreated populations reflects average productivity differences between the two groups. Since, by Assumption \ref{ass:1}, all productivity differences between groups are causal, the transparent IV using wages also identifies the social return to education.

Therefore, if the employers are informed about the instrument, the IV estimate of returns to education is a consistent estimate of the productivity effect of education on earnings, i.e.,
\begin{eqnarray*}
\E\left[\ln W_{it}|D^{{\mathfrak t}}_{i}=d, t\right] =\E\left[\delta^{\psi|S}S_{i}+u_{it}|D^{{\mathfrak t}}_{i}=d, t\right]=\lambda_t\times\delta^{\psi|S}\times\E\left[S_{i}|D^{{\mathfrak t}}_{i}=d\right],
\end{eqnarray*}
where the first equality follows from (\ref{eq: causal schooling 2}) and the second equality from Assumption \ref{ass:1}. 
Using this expression for the conditional log-wages in (\ref{eq:Binary Instrument}) and taking the probability limit gives 
\begin{eqnarray}
\texttt{plim}\!\!\quad\!\! \hat{b}^{IV^{\mathfrak t}}_{t} &=& \frac{\E\left[\ln W_{it}|D^{{\mathfrak t}}_{i}=1,t\right]-\E\left[\ln W_{it}|D^{{\mathfrak t}}_{i}=0,t\right]}{\E\left[S_{i}|D^{{\mathfrak t}}_{i}=1\right]-\E\left[S_{i}|D^{{\mathfrak t}}_{i}=0\right]}=\lambda_t\times\delta^{\psi|S}= \delta_{t}^{\psi|S}.\quad\label{eq:IV_transparent_time}
\end{eqnarray}
Hence the Wald estimator for a transparent IV, $D^{{\mathfrak t}}$, identifies the social returns to education at all $t$, i.e., $\texttt{plim}\!\!\quad\!\!\hat{b}^{IV^{\mathfrak t}}_{t}=\delta_t^{\psi|S}$. 

In summary, access to a hidden IV is sufficient to identify the experience-varying private returns to education, and access to a hidden IV is sufficient to identify the experience-varying social returns to education. The two of them together identify the signaling value of schooling and the speed of employer learning, even when the returns to skills vary with experience. 
Next, we show that if we have access to only a hidden IV, we can still identify the signaling value of education, but it requires the additional assumption that the returns to skills are experience-invariant.

\subsection{Experience-Invariant Returns to Skill\label{sec:identification_timeinvariant}}
With a hidden IV, setting $\lambda_t=1$ in (\ref{eq:IV_hidden}) gives  
\begin{eqnarray}
\texttt{plim}\!\!\quad\!\! \hat{b}^{IV^{\mathfrak h}}_{t} =\delta^{\psi|S}+\theta_{t}\phi_{A|S}.\quad\label{eq:IV_hidden}
\end{eqnarray}
Comparing (\ref{eq:IV_hidden}) with the private returns defined in (\ref{eq:private return}), we conclude that with experience-invariant returns to skill, at every work experience level $t$, the hidden IV identifies the private returns to education, i.e., $\texttt{plim}\!\!\quad\!\!\hat{b}^{IV^{\mathfrak h}}_{t}=\delta^{W|S}$. 
Furthermore, from (\ref{eq:IV_hidden}) we can see that as the worker gets more experience, i.e., $t\rightarrow\infty$, $\lim_{t\rightarrow \infty}\theta_t=0$ and so $\texttt{plim}\!\!\quad\!\!\hat{b}^{IV^{\mathfrak h}}_{t}$ converges to the social return to education, $\delta^{\psi|S}$. Thus, having access to a hidden IV and a repeated cross-section of wages across workers' careers is sufficient to identify the productivity effect of education, i.e., the social returns to education.\footnote{Although three periods suffice to identify the model, data from additional periods improve the precision of the estimates.} Using private and social returns, we can identify the signaling value of education.

Besides the private and the social returns, a hidden IV also identifies the speed of learning $\kappa$. 
Let $b^{IV^{\mathfrak h}}_{0}= \delta^{\psi|S}+\phi_{A|S}$ denote the private return to education at $t=0$ and $b^{IV^{\mathfrak h}}_{\infty}=\delta^{\psi|S}$ denote the social return to education. We can then re-write the IV estimate in (\ref{eq:IV_hidden}) at experience $t$ as a weighted average of private and social returns, as 
\begin{eqnarray}
\texttt{plim}\!\!\quad\!\! \hat{b}^{IV^{\mathfrak h}}_{t}=\theta_{t}\times b^{IV^{\mathfrak h}}_{0}+\left(1-\theta_{t}\right)\times b^{IV^{\mathfrak h}}_{\infty}.\label{eq:nlls}
\end{eqnarray}
Then, using the IV estimates for $t=0,\ldots, T$, for a large $T$ we can estimate the RHS parameters using the non-linear least squares (NLLS). This means that we can ``solve" for $\{b_{0}^{IV}, b_{\infty}^{IV}, \kappa \}$ from $\hat{b}^{IV^{\mathfrak h}}_{t}$ for $t=0,\ldots, T$, and once we know $\kappa$ we can determine $\theta_{t}$ for each $t$.

Similarly with a transparent IV, setting $\lambda_t=1$ in (\ref{eq:IV_transparent}) gives 
\begin{eqnarray}
\texttt{plim}\!\!\quad\!\! \hat{b}^{IV^{\mathfrak t}}_{t} =\delta^{\psi|S}.\quad\label{eq:IV_transparent}
\end{eqnarray}
Hence the Wald estimator for a transparent IV, $D^{{\mathfrak t}}$, identifies the social returns to education at all $t$, i.e., $\texttt{plim}\!\!\quad\!\!\hat{b}^{IV^{\mathfrak t}}_{t}=\delta^{\psi|S}$. Unlike the hidden IV, however, access to a transparent IV is not sufficient to identify the signaling value of education or the speed of learning.

\subsection{Experience-Varying Returns to Skill\label{sec:identification_timevarying}}
Next, we consider the case when the returns to skill vary with experience. Like the experience-invariant case, a hidden IV identifies the private returns to education, and a transparent IV identifies the experience-varying social returns to education. When the returns to skill vary with experience, then the social returns also vary with experience. Thus, to identify the speed of learning, we have to separately identify the effect of employers' learning from the variation in the returns to the skill with experience. 

In this case, access to only one IV, hidden or transparent, is insufficient to identify the speed of learning. 
If we have access to both hidden and transparent IVs and if $\lambda_t$ satisfies homogeneity (defined below), we can identify the speed of learning.  
Our approach to identifying the model with experience-varying returns to skill relies on combining estimates from (\ref{eq:IV_hidden_time}) and (\ref{eq:IV_transparent_time}). 
However, in practice, two IVs are seldom available for the same sample. So, to combine estimates from different samples, we need homogeneity.

\begin{assumption} (Homogeneous Experience-Varying Returns to Skill) Let $\lambda_t$ be homogenous across the hidden IV and transparent IV samples, for each $t\in\mathbb{T}$.\label{ass:4}
\end{assumption} 

Assumption \ref{ass:4} requires that the two samples exposed to hidden and transparent IVs have the same experience-varying returns to skills. This assumption, however, does not impose any restriction on the experience-invariant components of social returns, i.e., the parameter $\delta^{\psi|S}$ in (\ref{eq:IV_hidden_time}) and (\ref{eq:IV_transparent_time}), so that the overall social returns can be different across the samples. We will discuss the validity of this assumption in our setting later.

The identification then proceeds in the following steps. First, from (\ref{eq:IV_transparent_time}) at $t=0$ we can identify $\texttt{plim}\!\!\quad\!\!\hat{b}_{0}^{IV^{\mathfrak t}} = \delta^{\psi|S}$. Then, substituting $\delta^{\psi|S}$ in (\ref{eq:IV_transparent_time}) identifies $\lambda_t=\frac{\texttt{plim}\hat{b}_{t}^{IV^{\mathfrak t}}}{\texttt{plim}\hat{b}_{0}^{IV^{\mathfrak t}}}$ for $t>0$. 
Once we have identified the parameters $\{\lambda_t: t>0\}$, then under Assumption \ref{ass:4} we can use these parameters with $D^{\mathfrak h}$ in (\ref{eq:IV_hidden_time}) to identify $\{b^{IV^{\mathfrak h}}_{0}, b^{IV^{\mathfrak h}}_{\infty}, \kappa\}$ in the same way as we identified the private returns using $D^{\mathfrak h}$ in (\ref{eq:nlls}). We can either estimate the parameters sequentially, by following the identification steps, or jointly-estimate the parameters from (\ref{eq:IV_hidden_time}) and (\ref{eq:IV_transparent_time}).

\section{Data and Empirical Setting\label{sec:DataSetting}}

In this section, we first describe our data sources, how we construct our sample, and the key variables. Then we describe the Norwegian compulsory schooling reform that we use to construct hidden and transparent IVs. Finally, we discuss the empirical model specifications.

\subsection{Data Sources and Sample Construction\label{sec:Data}}

We use several registry databases maintained by Statistics Norway to construct a rich longitudinal dataset containing all Norwegian males from 1967 to 2014. We observe each of these individuals' demographic information (e.g., birth cohort and childhood municipality of residence) and socio-economic information (e.g., years of education and annual earnings). Notably, the dataset also includes a unique personal identifier, which allows us to follow individuals' earnings over time and merge our dataset with the information on IQ test scores from the Norwegian Armed Forces.

The Norwegian earnings data have several advantages over those available in most other countries. 
First, there is no attrition from the original sample other than the natural attrition due either to death or out-migration. Second, the data pertain to all residents and are not limited to some sectors or occupations. 
Third, we can construct long earnings histories to estimate the returns to education at each year of labor market experience.

We restrict our sample to Norwegian males born between 1950 and 1980, and our sample includes several cohorts for whom we observe earnings for most of their adult life.\footnote{In our annual income panel data from 1967 to 2014, we observe the oldest cohort (1950) between the ages 17 and 64 and the youngest cohort (1980) up to age 34.} 
We restrict the sample to males because females in these cohorts had comparatively lower labor force participation, more variation in hours worked over the life-cycle, and the military IQ test scores are available only for males born after 1950. We further exclude immigrants and Norwegian males with missing information on schooling, childhood municipality of residence, IQ test score, or ``exposure" to the compulsory schooling reform. Applying these restrictions, we retain 732,163 Norwegian males.

Our primary outcome variable is the natural logarithm of pre-tax annual labor earnings.\footnote{These earnings exclude income from capital and self-employment. They also exclude unconditional cash transfers from the government, e.g., social and economic assistance, housing assistance, child allowance.} 
Although hourly wages would have been an ideal outcome to measure, they are unavailable for most of our sample period. We thus restrict our sample to workers with earnings above the SGA threshold (which was U.S. \$10,650 in 2015) to minimize variation in earnings induced by variation in working hours.\footnote{In Section \ref{sec:Results_robustness}, we show that our results are robust to varying the earnings threshold.} We are left with $718,237$ males forming a panel comprising $14,746,755$ person-year observations. On average, an individual works full-time for approximately $20.5$ years. We use this sample in our empirical analysis, but the sample is unbalanced across work experience. We observe earnings for $579,984$ individuals at the start, i.e., experience $t=0$, and $190,900$ individuals after thirty years, i.e., experience $t=30$.\footnote{Death or migration contribute little to the decline in the number of observations between $t=0$ and $t=30$. The main reason the number of observations declines is because the sample ends in 2014. Thus, we can follow the first cohort (1950) across all ages from 17 to 64, while we can only follow the last cohort (1980) until they are 34 years old.
In Section \ref{sec:Results_robustness}, we show results for alternative samples, e.g., by limiting the analysis to cohorts 1950--1965 and focusing on the early part of the workers' careers, for which the panel is balanced.}

\subsection{Measures of Schooling and IQ\label{sec:Schooling}}

Our primary variable of interest is years of schooling, corresponding to the highest level of completed education. This variable comes from Statistics Norway's Education Register. Statistics Norway uses reports on educational attainment submitted directly by the relevant educational establishments to determine each individual's schooling years. Thus the chance of misreporting is slight. Using the years of schooling variable and the age at the start of each year, we measure potential experience as age minus years of schooling minus school starting age.\footnote{We measure age of individuals at the start of each year. We then follow their earnings from the first calendar year after they graduate. We do not use earnings from the year of their graduation because most individuals would be in school for some time during the said calendar year. As discussed above, we drop observations for the years when workers have either low earnings or are unemployed because we restrict our sample to workers with earnings above the SGA threshold.} 

We also have information on conscripts' IQ test scores from the Norwegian Armed Forces. In Norway, military service was compulsory for all able males in the birth cohorts in our sample. Before each male entered the service, the military assessed his medical and psychological suitability. Most eligible Norwegian males in our sample took this test around their $18^{th}$ birthday. We use the IQ test score that is a composite unweighted mean from three speeded tests--arithmetics, word similarities, and figures.\footnote{Arithmetic and word tests mirror respective tests in the Wechsler Adult Intelligence Scale, and the figures test is comparable to the Raven Progressive Matrix test, see \cite{SundetBarlaugTorjussen2004}, and \cite{Thrane1977}.}

\begin{figure}[t!]
\begin{centering}
\includegraphics[width=0.6\columnwidth]{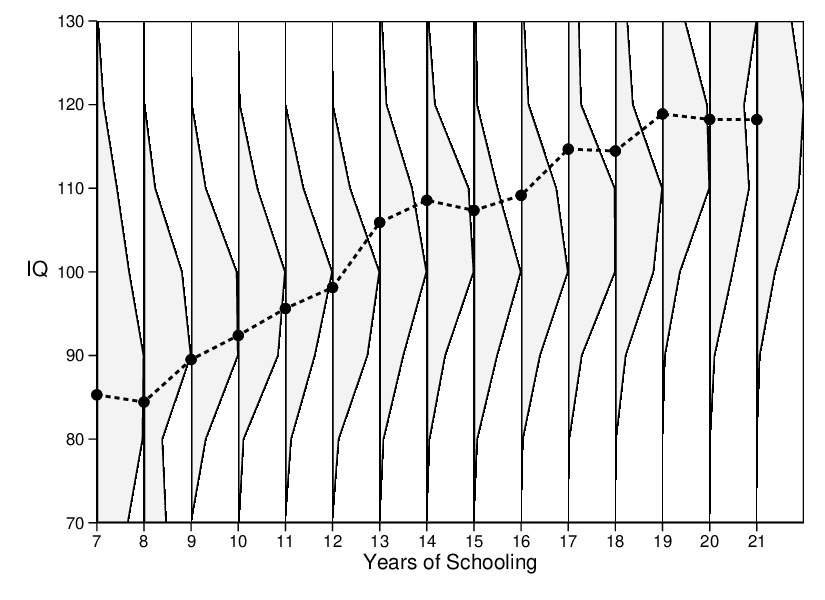}
\par\end{centering}
\caption{Conditional Probability Density of IQ Test Scores on Years
of Schooling.
 \label{fig:IQ_Schooling}}
\emph{\scriptsize{}Note:}{\scriptsize{} The sample consists of Norwegian males born 1950-1980 observed in earnings data over the years 1967-2014 with years of potential experience between 0 and 30 years with annual earnings above 1 SGA threshold (N=14,746,755). The IQ test score along the y-axis is the residual from a regression of the raw IQ test score on birth cohort dummies, removing secular changes over time, and further standardized to have a mean of 100 and a standard deviation of 15. The black dotted line plots the average IQ test score by individuals' years of schooling, while the shaded areas plot the conditional probability density of IQ.}{\scriptsize\par}
\end{figure}

 In Figure \ref{fig:IQ_Schooling}, we show the average and conditional density of IQ for each year of schooling between 7 and 21 years. This figure illustrates two striking patterns in our data worth noting. First, the measures of IQ and schooling are strongly correlated, with a correlation of almost 0.5. Second, sharp increases in the average IQ score occur around the entry years of college (13/14 years) and master's degree (16/17), with more gradual increases at other schooling stages. This pattern could be due to substantial ability-related (psychic) costs or particular requirements for enrolling in higher education in Norway.\footnote{Public education in Norway is meritocratic \citep{KirkeboenLeuvenMogstad2016}. Students with higher GPAs in high school are more likely to select into fields with high demand and have higher IQ scores.} This strong correlation in our data suggests that schooling predicts ability, satisfying a necessary condition for schooling to have a signaling value in our setting.

\subsection{The Compulsory Schooling Reform\label{sec:Setting}}
Between 1960 and 1975, Norway enacted a school reform that increased the minimum required schooling from 7 to 9 years. Different municipalities, which are the most decentralized local administration, implemented this reform in different years. Thus, for more than a decade, the length of compulsory schooling required of Norwegian children depended on their birth year and municipality of residence at age 14, henceforth, the childhood municipality. We use the timing differences across municipalities induced by the reform's staggered implementation as an IV for school years. For more on this reform, see \cite{BlackDevereuxSalvanes2005}.\footnote{This compulsory schooling reform in Norway has been used previously, albeit in different contexts, by \cite{MonstadPropperSalvanes2008, AakvikSalvanesVaage2010, MachinSalvanesPelkonen2012}, and \cite{BhullerMogstadSalvanes2017}. However, besides increasing the minimum required schooling, the law also standardized curricula to improve schooling quality, as noted by \cite{BlackDevereuxSalvanes2005}. This change in quality may raise some concern about whether an IV based on this reform satisfies the exclusion restriction (Assumption \ref{ass:1}-(1)). Although we do not show these results for brevity, our model can incorporate this effect and provide a sharp prediction about the bias. In particular, suppose we assume that the causal effect of $D$ on $A$ is linear and additively separable, i.e., $A_{i} =\delta^{A|D}D_{i}+\tilde{A}_{i}$, such that $\delta^{A|D}>0$ is the effect of the reform on school quality (in terms of log-productivity). Then we can show that the estimates of the private and social returns have the same bias, $\lambda_{t}(\delta^{A|D}/\varkappa)$, where $\varkappa>0$ is the effect of the reform on years of schooling. As the difference between the private and the social return equals the signaling value, the bias term would cancel out, and the signaling value estimate would remain unchanged despite a violation of the exclusion restriction.}

\begin{figure}[t!]
\begin{centering}
\subfloat[Exposure across Birth Cohorts]{
\centering{}\includegraphics[width=0.5\columnwidth]{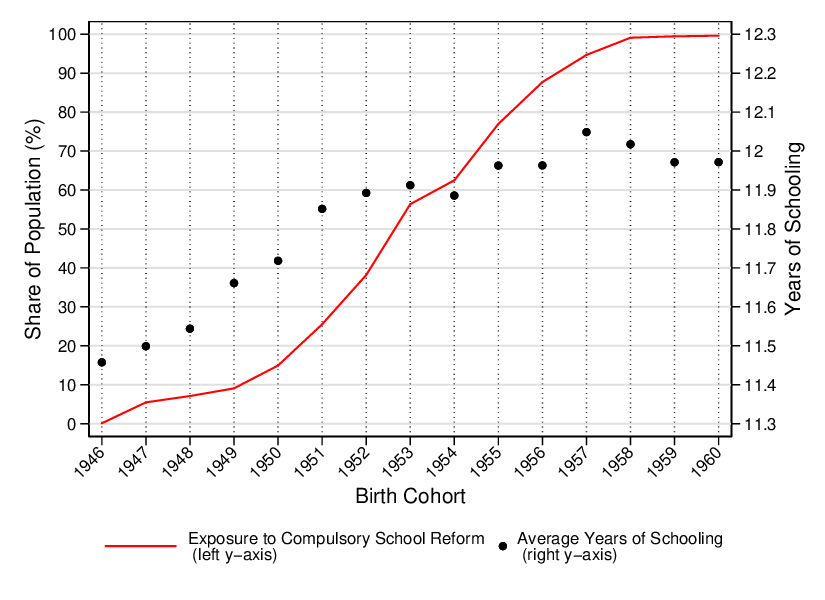}}
\subfloat[Timing within Local Labor Markets]{
\centering{}\includegraphics[width=0.5\columnwidth]{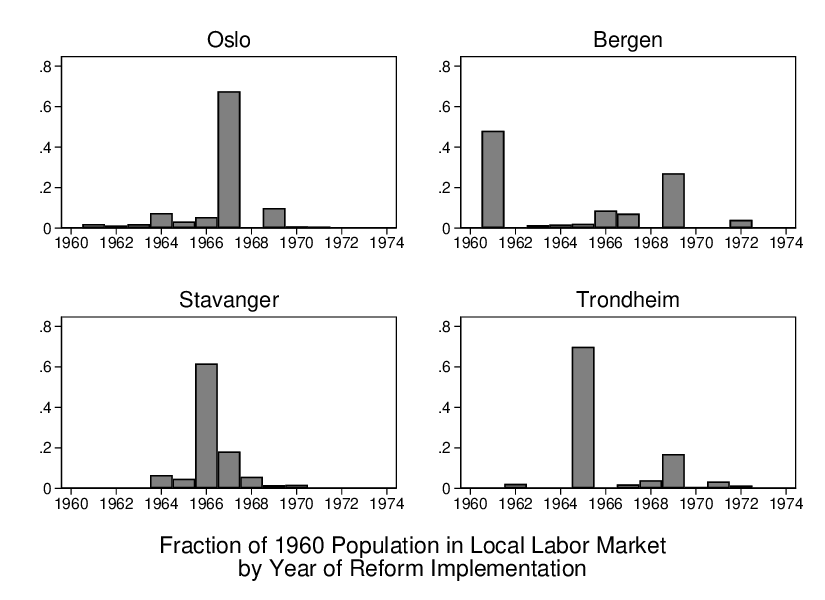}}
\par\end{centering}
\caption{Compulsory School Reform Across Birth Cohorts and Local Labor Markets.\label{fig:Reform_Schooling_Cohorts}}
\emph{\scriptsize{}Note:}{\scriptsize{} The red line in plot (a) shows the cohort-specific share of population exposed to the compulsory school reform, while the black dots indicate the average years of schooling for Norwegian male cohorts born 1946-1960. Plot (b) shows the fraction of 1960 population in the four biggest local labor markets (concentrated around the four major cities) by the year of reform implementation. Using the 1960 classification of municipalities, there were 40 municipalities in the Oslo region, 27 municipalities in the Trondheim region, and 25 municipalities each in the Bergen and Stavanger regions. The variation in the timing of reform within local labor markets (LLMs) is due to variation in the timing of reform across municipalities within LLMs.}{\scriptsize\par}
\end{figure}

For $672$ out of the $732$ municipalities in 1960, we can date the reform's implementation using historical records. This information is missing for the remaining $60$ municipalities \citep{MonstadPropperSalvanes2008}. As shown in Figure \ref{fig:Reform_Schooling_Cohorts}, there is considerable variation in the fraction of birth cohort exposed to the reform (Figure \ref{fig:Reform_Schooling_Cohorts}-(a)) and in the timing of reform even within local labor markets (Figure \ref{fig:Reform_Schooling_Cohorts}-(b)). In particular, panel (a) shows that nobody born before 1946 and everybody born after 1960 were required to attend school for nine rather than seven years. 

Figure \ref{fig:Reform_Schooling_Cohorts}-(b) shows that there is considerable variation even within the four largest local labor markets (the four most significant metropolitan areas in Norway). For instance, Oslo city's municipality, which in 1960 accounted for two-thirds of the Oslo labor market region population, implemented the reform in 1967. In contrast, the reform's timing varied between 1961 and 1971 across the remaining population living in one of the other 39 municipalities.\footnote{We use the classification of Norway into 160 local labor markets based on geographic commuting patterns constructed by \cite{GundersenJuvkam2013}. On average, each market has five municipalities.}

As shown in Section \ref{sec:Identification}, to separately identify the private and the social returns to education in a model with experience-invariant returns to skill, the instrument should satisfy the standard IV assumption (Assumption \ref{ass:1}) and also be a hidden instrument (Assumption \ref{ass:2}). 
For an instrument to be called a hidden instrument in our setting, it must mean that employers are not informed about the interaction between a worker's birth cohort and the timing of compulsory school reform in the worker's childhood municipality. On the contrary, for an instrument to be transparent (Assumption \ref{ass:3}), the employer must be informed about the worker's birth cohort and the timing of reform in the worker's childhood municipality. 

While the information on a worker's birth cohort and residence is readily available for the employer (e.g., from worker's r\'esum\'e), the employer may not know the year when the law was implemented in the worker's childhood municipality. In contrast to compulsory schooling laws legislated centrally in many countries or by the states in the U.S., the Norwegian compulsory school reform was decentralized and was implemented at the local municipal level. As indicated by Figure \ref{fig:Reform_Schooling_Cohorts}-(b), this decentralized implementation leads to substantial variation in the reform's timing across municipalities.
These features related to the implementation of the reform, combined with the structure of local labor markets in Norway, provide exogenous variation in schooling that can be utilized either as a hidden or a transparent instrument. 

On the one hand, the substantial variation in reform timing illustrated in Figure \ref{fig:Reform_Schooling_Cohorts} implies that retrieving information on reform exposure for each applicant would be onerous and costly for the employers. Information on the timing of compulsory schooling reform for all municipalities was, until recently, not readily available in any public databases. 
 Therefore, for the 1946-1960 cohorts, graduating in an era long before the Internet, this information would not have been easily traceable for employers. These institutional features suggest that it is reasonable to assume that exposure to the reform can serve as a hidden IV.

On the other hand, employers have incentives to remain informed about significant changes to their local labor market's education system. When major cities or population centers implemented such schooling reforms, newspapers and other media were more likely to report, and employers more likely to learn. We thus believe it is more plausible that compulsory schooling regulations pertaining to the residents of the central municipalities were understood and priced into wages by employers. Based on this reasoning, we construct a plausibly hidden instrument and a plausibly transparent instrument by partitioning our analytical sample into two, based on the relative size of workers' childhood municipality in their local labor markets. 

To construct a transparent instrument, we include workers who grew up in population centers, defined as municipalities with the largest population in each local labor market, and assume that employers are informed about their reform exposure status. We refer to this sample as the \emph{transparent IV sample}. It includes 295,488 individuals and 6,048,776 person-year observations, i.e., 41\% of the entire sample. 

Similarly, to construct a hidden instrument, we include workers who grew up in non-central municipalities, defined as the complement of central municipalities in each local labor market. We assume that employers are uninformed about these workers' reform exposure status. We refer to this sample as the \emph{hidden IV sample}, and it includes 422,749 individuals and 8,697,979 person-year observations, i.e., 59\% of the entire sample.

By relying on central and non-central municipalities to construct transparent and hidden IVs, respectively, we assume that (1) firms everywhere can verify a worker's education level, (2) firms know when the most populous municipality in the local labor market implemented a compulsory schooling reform, and (3) firms do not know when compulsory schooling reforms were implemented in the remaining municipalities in the local labor market. 
As illustrated in Figure \ref{fig:Reform_Schooling_Cohorts} the non-central municipalities are typically very small, and the timing of the reform varies widely across them. Furthermore, the timing of compulsory schooling reform across municipalities was not readily available in online databases until recently. These features support our assumptions. It would have been onerous and costly for the employers to retrieve the information on the timing of the reforms for employees stemming from small municipalities.

Finally, as discussed in Section \ref{sec:identification_timevarying}, to identify the model with experience-varying returns to skill, we also need to maintain homogeneity of experience-varying skill returns across samples (Assumption \ref{ass:4}). In our context, this implies that the dynamic path of skill returns is identical across individuals who grew up in central and non-central municipalities. One justification for this assumption is that there existed substantial ``leaving the nest'' migration in early adulthood such that while the childhood municipality in which an individual grew does determine whether an individual was exposed to the compulsory schooling reform, this does not predict the labor market conditions that the individual faced as an adult. Indeed, we find that almost 47\% of people had migrated out of their childhood municipality by age 30. Furthermore, the vast majority of these migrated from non-central to central municipalities. These features may imply that the assumption of homogeneity of experience-varying skill returns across those who grew up in central and non-central municipalities could be reasonable in our context. Naturally, the plausibility of this assumption should be assessed in the particular context considered.

\subsection{Empirical Specifications\label{sec:Specifications}}
Our empirical approach estimates the IV specification using 2SLS. In particular, at each experience $t$, the second stage projects log-earnings ($\ln W_{it}$) on control variables ($X_i$) as well as the projected schooling ($\hat{S}_i$) from the first stage, i.e., 
 \begin{equation}
\ln W_{it}=a_{t}^{IV^{{\mathfrak r}}}+b^{IV^{{\mathfrak r}}}_{t}\hat{S}_{i}+e_{t}^{IV^{{\mathfrak r}}}X_{i}+u_{it}^{IV^{{\mathfrak r}}}.\label{eq:Specification 1}
\end{equation}
By estimating (\ref{eq:Specification 1}) separately for each $t$ and $\mathfrak r$, we obtain parameters that vary freely across experience and samples.
The control variables include a full set of dummies for birth cohort and childhood municipality, and ${\mathfrak r}\in\{{\mathfrak h},{\mathfrak t}\}$ indicates whether we use the hidden IV or the transparent IV sample. The first-stage equation is 
\begin{eqnarray}
S_{i}={o^{{\mathfrak r}}}+{\varkappa^{{\mathfrak r}}}D^{{\mathfrak r}}_{i}+{n^{{\mathfrak r}}}X_{i}+v^{{\mathfrak r}}_{i}, \label{eq:First Stage}
\end{eqnarray}
where the binary instrument $D^{{\mathfrak r}}_{i}\in\{0,1\}$ is equal to 1 if $i$ was ``exposed" to compulsory schooling reform and 0 otherwise. We record individual $i$ as exposed if the reform was implemented in $i$'s childhood municipality when $i$ was 14 years old or younger.\footnote{Unlike (\ref{eq:Specification 1}), the coefficient $\varkappa^{{\mathfrak r}}$ of the instrument $D^{{\mathfrak r}}$ in the first-stage equation do not change with $t$ because $D^{{\mathfrak r}}$ and $S$ are experience-invariant. However, with an unbalanced panel and separate estimations by experience, the first-stage estimates of $\varkappa^{{\mathfrak r}}$ can vary with $t$. In practice, estimates of $\varkappa^{{\mathfrak r}}$ are stable across the experience range that we consider despite differences in the sample composition by experience.} 

We use the childhood municipality indicators to control for unobservable determinants of earnings or schooling fixed at the municipality level. By adding these indicators, we can compare individuals who grew up in the same municipality but were exposed to different compulsory schooling requirements. Thus, we exploit variation in their schooling stemming from differential exposure to the compulsory schooling law. By including birth cohort indicators, we control for aggregate changes in schooling and earnings across cohorts.

Our parameters of interest are $\{b^{IV^{{\mathfrak h}}}_{t},b^{IV^{{\mathfrak t}}}_{t}\}$, the coefficients on schooling at experience $t=0,1,\ldots$, resulting from the hidden IV and the transparent IV, respectively. We maintain the assumptions that (i) conditional on $X$, both instruments $\{D^{{\mathfrak h}},D^{{\mathfrak t}}\}$ satisfy Assumption \ref{ass:1}, and as discussed in Section \ref{sec:Setting}, that in our setting (ii) $D^{{\mathfrak h}}$ satisfies the hidden IV Assumption \ref{ass:2} and $D^{{\mathfrak t}}$ satisfies the transparent IV Assumption \ref{ass:3}.\footnote{The reform timing is also uncorrelated with baseline municipality characteristics \citep{BhullerMogstadSalvanes2017}.} 

As discussed in Section \ref{sec:identification_timeinvariant}, under experience-invariant returns to skill and maintaining Assumption \ref{ass:2}, $\hat{b}^{IV^{\mathfrak h}}_{t}$ converges to the experience-invariant social return to education, $\delta^{\psi|S}$, as $t\rightarrow\infty$, and moreover, $\hat{b}^{IV^{\mathfrak h}}_{t}$ provides a consistent estimate of the private return at $t$, $\delta^{W|S}_{t}$, for any $t<\infty$. Thus, we can estimate the social returns to education as $\hat{b}^{IV^{\mathfrak h}}_{t\rightarrow\infty}$, and use the rate at which $\hat{b}^{IV^{\mathfrak h}}_{t}$ converges to $\hat{b}^{IV^{\mathfrak h}}_{\infty}$ to estimate the speed of learning, $\kappa$. 

For the model with experience-varying returns to skill and maintaining Assumption \ref{ass:2} on $D^{{\mathfrak h}}$ and Assumption \ref{ass:3} on $D^{{\mathfrak t}}$, we can recover the private returns $\delta^{W|S}_{t}$ for the hidden IV sample from $\hat{b}^{IV^{\mathfrak h}}_{t}$ and the experience-varying social returns $\delta^{\psi|S}_{t}$ for the transparent IV sample from $\hat{b}^{IV^{\mathfrak t}}_{t}$. As discussed in Section \ref{sec:identification_timevarying}, using $\hat{b}^{IV^{\mathfrak t}}_{t}$, we can also recover the experience-varying skill component $\hat{\lambda}_{t}$ under location normalization $\lambda_0=1$. And, finally, maintaining Assumption \ref{ass:4} of homogeneity of $\lambda_t$ across the two samples, we can infer the speed of employer learning, $\kappa$, and the experience-invariant part of the social returns $\delta^{\psi|S}$ for the hidden IV sample.

A challenge to identifying returns to schooling based on (\ref{eq:Specification 1}) is that individuals who grew up in different municipalities could have had different growth rates in schooling and earnings even in the absence of compulsory schooling reform. Following \cite{BhullerMogstadSalvanes2017}, we also test the stability of our first-stage and IV estimates to the inclusion of extrapolated linear and quadratic municipality-specific trends in educational attainment and lifetime earnings estimated using data on pre-reform cohorts as additional controls. We refer to estimates based on (\ref{eq:Specification 1})--(\ref{eq:First Stage}) as obtained from the baseline specification and estimates that we get after controlling for municipality-specific trends.

Finally, note that by estimating (\ref{eq:Specification 1}) separately for each $t$, we also allow the work experience to interact with observable characteristics $X_{i}$, further weakening the functional form assumption embedded in (\ref{eq:chi}). In particular, in (\ref{eq:Specification 1}) we have specified $H(t,X_{i})=a_t^{IV^{\mathfrak r}} + e_t^{IV^{\mathfrak r}} X_{i}$, where coefficients $a_{t}^{IV^{\mathfrak r}}$ and $e_t^{IV^{\mathfrak r}}$ can vary flexibly by experience, and thus flexibly capture both a common experience profile and its interactions with $X_{i}$.

\section{Empirical Results\label{sec:Results}}
This section contains our main empirical results, beginning with IV estimates by work experience for both the hidden and the transparent IV samples. These allow us to identify both the private and the social returns to education. We use these to quantify the contribution of signaling to the returns to education before we provide results from sensitivity analyses.

\subsection{IV Estimates of the Returns to Education\label{sec:ResultsIV}}

In Table \ref{tab:FS} column (1), we display the estimates of (\ref{eq:First Stage})--the first-stage effect of our compulsory schooling reform instrument on years of schooling--for the entire sample. This estimate indicates that exposure to compulsory schooling reform increased completed schooling by 0.237 years. Columns (2) and (3) show the same first-stage estimates for the hidden IV and the transparent IV samples, respectively. All three first-stage coefficients are of similar magnitude and precisely estimated, confirming that the IV relevance condition is satisfied in all samples and that weak IV bias is not a concern for our analysis.

\begin{table}[h]
\begin{centering}
\caption{First-Stage Estimates on Years of Schooling.\label{tab:FS}}
\par\end{centering}
\begin{centering}
{\scriptsize{}{}}%
\begin{tabular*}{1\textwidth}{@{\extracolsep{\fill}}@{\extracolsep{\fill}}lccccc}
\toprule 
 & \multicolumn{1}{c}{{\footnotesize{}Full Sample}} & & \multicolumn{1}{c}{{\footnotesize{}Hidden IV}} & & \multicolumn{1}{c}{{\footnotesize{}Transparent IV}}\tabularnewline
 & \multicolumn{1}{c}{{\footnotesize{}}} & & \multicolumn{1}{c}{{\footnotesize{} Sample}} & & \multicolumn{1}{c}{{\footnotesize{}Sample}}\tabularnewline 
\cmidrule{2-2} \cmidrule{4-4} \cmidrule{6-6} 
 & {\footnotesize{}(1)} & & {\footnotesize{}(2)} & & {\footnotesize{}(3)} \tabularnewline
\midrule
{\footnotesize{}Instrument:} & & & & & \tabularnewline
\emph{\footnotesize{}{}Exposure to Compulsory Schooling Reform} & {\footnotesize{}{}0.237}\textsuperscript{{*}{*}{*}} & & {\footnotesize{}{}0.228}\textsuperscript{{*}{*}{*}} & & {\footnotesize{}{}0.240}\textsuperscript{{*}{*}{*}} \tabularnewline
 & {\footnotesize{}{}(0.025)} & & {\footnotesize{}{}(0.034)} & & {\footnotesize{}{}(0.038)}\tabularnewline
\midrule
{\footnotesize{}Controls:} & & & & & \tabularnewline
{\footnotesize{}{}Municipality Fixed Effects} & $\checkmark$ & & $\checkmark$ & & $\checkmark$ \tabularnewline
{\footnotesize{}{}Cohort Fixed Effects} & $\checkmark$ & & $\checkmark$ & & $\checkmark$ \tabularnewline
\midrule
{\footnotesize{}{}F-statistic (instrument)} & {\footnotesize{}{}87.7} & & {\footnotesize{}{}45.7} & & {\footnotesize{}{}55.5} \tabularnewline
{\footnotesize{}{}Sample Mean Years of Schooling} & {\footnotesize{}{}12.36} & & {\footnotesize{}{}12.36} & & {\footnotesize{}{}12.27} \tabularnewline
{\footnotesize{}{}Standard Deviation Years of Schooling} & {\footnotesize{}{}2.50} & & {\footnotesize{}{}2.50} & & {\footnotesize{}{}2.46} \tabularnewline
{\footnotesize{}{}Number of Observations} & {\footnotesize{}{}14,746,755} & & {\footnotesize{}{}8,697,979} & & {\footnotesize{}{}6,048,776} \tabularnewline
\bottomrule
\end{tabular*}
\par\end{centering}
\smallskip{}

\emph{\scriptsize{}{}Note:}{\scriptsize{}{} 
The full sample (column 1) consists of Norwegian males born in 1950-1980 observed any time
in earnings data over 1967-2014 with years of potential experience
between 0 and 30 years and annual earnings above 1 SGA threshold (N=14,746,755).
The hidden IV sample (column 2) further drops individuals who grew
up in the municipality with the largest population size in each of
the 160 labor market regions in Norway (N=8,697,979), while the transparent IV sample (column 3) retains only individuals who grew up in the municipality with the largest population size in each labor market (N=6,048,776). All estimations include fixed effects for birth cohort and childhood municipality. We cluster the standard errors at the local labor market region.}{\scriptsize\par}

{\scriptsize{}{}{*} p \textless{} 0.10, {*}{*} \textless{} 0.05,
{*}{*}{*} p \textless{} 0.01.}{\scriptsize\par}
\end{table}

\begin{figure}[h!]
\begin{centering}
\subfloat[Hidden IV Estimates]{\centering{}\includegraphics[width=0.5\columnwidth]{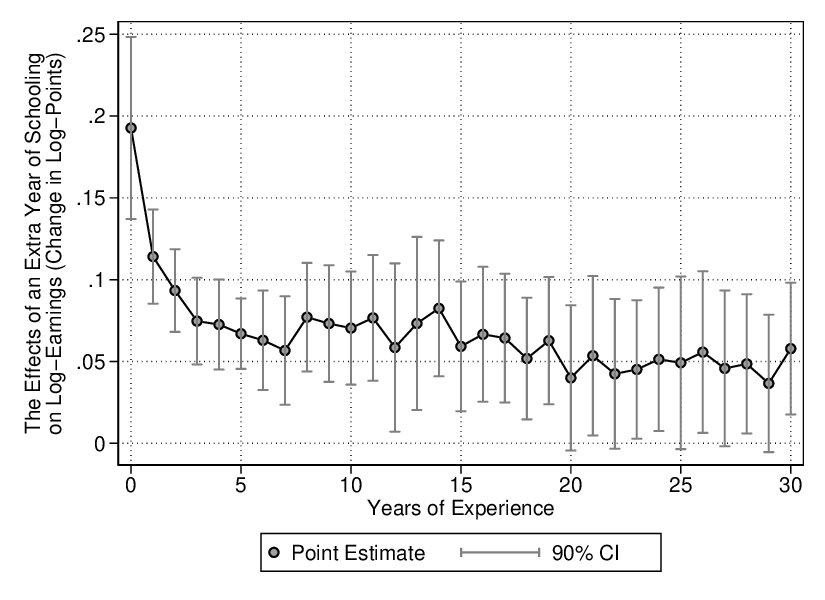}}
\subfloat[Transparent IV Estimates]{\centering{}\includegraphics[width=0.5\columnwidth]{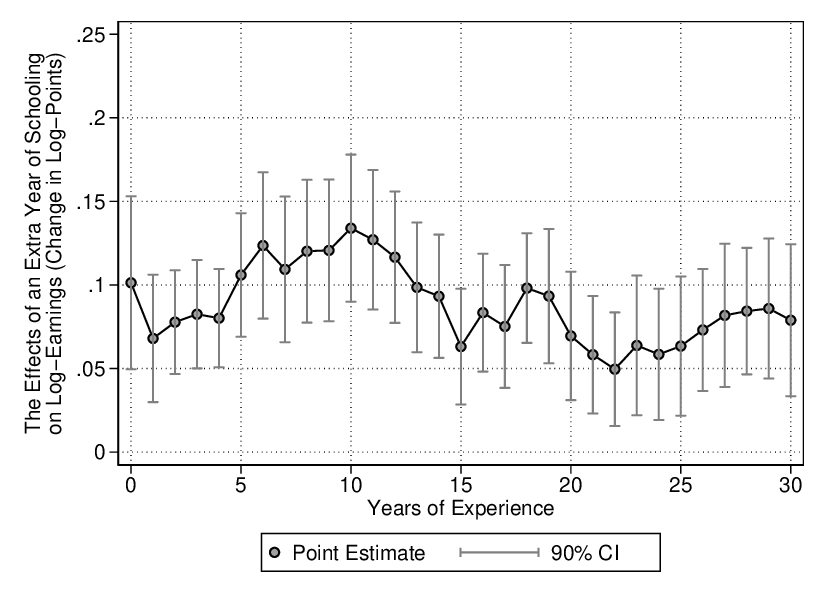}}
\par\end{centering}
\caption{Hidden and Transparent IV Estimates of the Returns to Schooling.\label{fig:IVestimates}}
\emph{\scriptsize{}Note:}{\scriptsize{} The estimation sample
consists of Norwegian males born 1950-1980 observed in earnings data
over years 1967-2014 with years of potential experience between 0
and 30 years and annual earnings above 1 SGA threshold (N=14,746,755).
The hidden IV sample (plot (a)) further drops individuals who grew
up in the municipality with the largest population size in each of
the 160 labor market regions in Norway (N=8,697,979), while the transparent IV sample (plot (b)) retains only individuals who grew up in the municipality with the largest population size in each labor market (N=6,048,776). Plots (a) and (b) display IV estimates from separate estimations of (\ref{eq:Specification 1}) for each year of experience using the hidden and the transparent IV samples. All estimations include fixed effects for birth cohort and childhood municipality. Standard errors are clustered at the local labor market region. Vertical bars denote the 90\% confidence intervals. }{\scriptsize\par}
\end{figure}

We now turn our attention to the second-stage IV estimates of (\ref{eq:Specification 1}) for the hidden IV sample. Figure \ref{fig:IVestimates}-(a) displays the hidden IV estimates at different work experiences. Under the assumption that this IV is hidden from the employers, this provides consistent estimates of private returns to schooling at each year of work experience. The estimates suggest initially high private returns to schooling, followed by a steep decline during the first five years of work. Then, the returns gradually stabilize and approach 5.5\% for those with 15 or more years of work experience.

Next, we turn to the transparent IV estimates of (\ref{eq:Specification 1}). These estimates are displayed in Figure \ref{fig:IVestimates}-(b), and they correspond to the social returns to schooling across the experience. In contrast to the hidden IV estimates in Figure \ref{fig:IVestimates}-(a), the transparent IV estimates do not decline with experience and are generally flatter across work experience. These estimates are indeed consistent with constant social returns, and, in turn, we fail to reject the standard model of employer learning with experience-invariant returns to skill.\footnote{Formally, we use $\hat{\lambda}_{t}=({\hat{b}_{t}^{IV^{\mathfrak t}}} \slash {\hat{b}_{0}^{IV^{\mathfrak t}}})$, and test the hypothesis that for every $t\in\mathbb{T}_{>0}$ the ratio $\hat{\lambda}_{t}=1$. The F-statistic from this joint test is 0.9, which means we cannot statistically reject the hypothesis of constant social returns. The $\hat{\lambda}_{t}$ estimates and the associated 90\% confidence intervals are presented in the Online Appendix, Figure \ref{fig:IV_lambda}. By comparison, the F-statistic for the joint test of constant hidden IV estimates is 3.3, and we can easily reject the hypothesis that the private returns are constant over experience.} 

The transparent IV estimates in Figure \ref{fig:IVestimates}-(b) imply that the average social return (across experience) is 8.8\%. This social return is higher than the long-term value that the hidden IV estimates, as shown in Figure \ref{fig:IVestimates}-(a) converge towards, indicating that the experience-invariant part of the social returns is not homogenous across the transparent IV and the hidden IV samples. While we do not explore the sources of such heterogeneity in social returns, our identification approach allows differences in the experience-invariant part of the social returns across workers from central and non-central locations.\footnote{Notably, Assumption {\ref{ass:4}} requires that the \emph{experience-varying} component of returns to skill is identical across samples. However, this assumption does not restrict the \emph{experience-invariant} component of social returns. Such heterogeneity in social returns could, for instance, exist due to differences in inputs factor in the production of human capital during childhood across workers who grew up in central and non-central locations (e.g., child-care access, parental resources, see discussions in \cite{CunhaHeckman2007}).}

\subsection{The Speed of Learning and the Signaling Value of Education}\label{sec:Results_Signaling}

We now use the hidden IV estimates and the transparent IV estimates, shown in Figure \ref{fig:IVestimates} to determine the private and social returns to schooling, the speed of employer learning, and the signaling value of education. 

We first present estimates that allow for experience-varying returns to skill, which are constructed by combining estimates from the hidden and the transparent IVs, as discussed in Section \ref{sec:identification_timevarying}. We report the NLLS estimates in Table \ref{tab:MainResults_IV_Speed_InitalLimit} columns (1)-(2), based on sequential (i.e., estimate (\ref{eq:IV_transparent_time}) before (\ref{eq:IV_hidden_time})) and joint estimations (i.e., estimate (\ref{eq:IV_transparent_time}) and (\ref{eq:IV_hidden_time}) jointly), respectively. Consistent with Figure \ref{fig:IVestimates}-(a), in column (1), we estimate an initial private return to education ($b^{IV^{\mathfrak h}}_{0}$) of 19.8\%, a limit return ($b^{IV^{\mathfrak h}}_{\infty}$) of 5.5\% and a speed of learning ($\kappa$) estimate of $0.505$, which implies rapid learning on the part of employers. In particular, this estimate of $\kappa$ implies that after the first five years of employment, employers put 16.4\% weight on the initial signal they received from workers. As we can see from panel B, column (1), after 15 years, this weight declines further to 6.1\%. Comparing across columns (1)-(2), we find that our estimates are robust to whether we sequentially or jointly estimate (\ref{eq:IV_hidden_time})-(\ref{eq:IV_transparent_time}).

In addition to estimates of our key model parameters ($b^{IV^{\mathfrak h}}_{0}, b^{IV^{\mathfrak h}}_{\infty}, \kappa$), we also report the average return ($b^{IV^{\mathfrak t}}$) implied by the transparent IV estimates, and similarly, the average return ($b^{IV^{\mathfrak h}}$) implied by the hidden IV estimates across years of experience 0 to 30. The average return for the transparent IV sample is estimated to be 8.8\%, compared to an average return of 6.4\% and a limit return of 5.5\% for the hidden IV sample.

\begin{table}[t!]
\begin{centering}\caption{IV Estimates of the Speed of Employer Learning, and Initial and Limit Returns. \label{tab:MainResults_IV_Speed_InitalLimit}}\par\end{centering}
\begin{centering}{\scriptsize{}}
\scalebox{0.97}{\begin{tabular*}{1\textwidth}{@{\extracolsep{\fill}}lcccc}\toprule 
{\footnotesize{}Model Specifications:} & \multicolumn{2}{c}{{\footnotesize{}Experience-Varying Skill Returns}} & & \multicolumn{1}{c}{{\footnotesize{}Experience-Invariant Skill Returns}}\tabularnewline
\cmidrule{2-3} \cmidrule{5-5} & \emph{\footnotesize{}Sequential} & \emph{\footnotesize{}Joint}\tabularnewline & \emph{\footnotesize{} Estimation} & \emph{\footnotesize{} Estimation}\tabularnewline & {\footnotesize{}(1)} & {\footnotesize{}(2)} & & {\footnotesize{}(3)}\tabularnewline\midrule{\footnotesize{}A. Parameters of Interest:} & & & & \tabularnewline\cmidrule{1-1} {\footnotesize{} \quad Speed of Learning $\kappa$} & {\footnotesize{}0.505}\textsuperscript{{*}{*}{*}} & {\footnotesize{}0.550}\textsuperscript{{*}{*}{*}} & & {\footnotesize{}0.532}\textsuperscript{{*}{*}{*}}\tabularnewline & {\footnotesize{}(0.107)} & {\footnotesize{}(0.126)} & & {\footnotesize{}(0.058)}\tabularnewline{\footnotesize{} \quad Initial Return $b_{0}^{IV^{\mathfrak{h}}}$} & {\footnotesize{}0.198}\textsuperscript{{*}{*}{*}} & {\footnotesize{}0.195}\textsuperscript{{*}{*}{*}} & & {\footnotesize{}0.192}\textsuperscript{{*}{*}{*}}\tabularnewline & {\footnotesize{}(0.015)} & {\footnotesize{}(0.012)} & & {\footnotesize{}(0.010)}\tabularnewline{\footnotesize{} \quad Limit Return $b_{\infty}^{IV^{\mathfrak{h}}}$} & {\footnotesize{}0.055}\textsuperscript{{*}{*}{*}} & {\footnotesize{}0.055}\textsuperscript{{*}{*}{*}} & & {\footnotesize{}0.050}\textsuperscript{{*}{*}{*}}\tabularnewline & {\footnotesize{}(0.006)} & {\footnotesize{}(0.008)} & & {\footnotesize{}(0.003)}\tabularnewline{\footnotesize{} \quad Average Return $b^{IV^{\mathfrak{h}}}$} & {\footnotesize{}0.064}\textsuperscript{{*}{*}{*}} & {\footnotesize{}0.066}\textsuperscript{{*}{*}{*}} & & {\footnotesize{}0.067}\textsuperscript{{*}{*}{*}}\tabularnewline & {\footnotesize{}(0.003)} & {\footnotesize{}(0.002)} & & {\footnotesize{}(0.007)}\tabularnewline {\footnotesize{} \quad Average Return $b^{IV^{\mathfrak{t}}}$} & {\footnotesize{}0.088}\textsuperscript{{*}{*}{*}} & {\footnotesize{}0.089}\textsuperscript{{*}{*}{*}} & & -- \tabularnewline & {\footnotesize{}(0.024)} & {\footnotesize{}(0.019)} & & \tabularnewline 
{\footnotesize{}B. Weight on Initial Signal:} & & & & \tabularnewline\cmidrule{1-1} {\footnotesize{} \quad $\theta_{t}$ at $t=5$} & {\footnotesize{}16.4\%} & {\footnotesize{}14.1\%} & & {\footnotesize{}15.0\%} \tabularnewline{\footnotesize{} \quad $\theta_{t}$ at $t=10$} & {\footnotesize{}8.9\%} & {\footnotesize{}7.6\%} & & {\footnotesize{}8.1\%} \tabularnewline{\footnotesize{} \quad $\theta_{t}$ at $t=15$} & {\footnotesize{}6.1\%} & {\footnotesize{}5.2\%} & & {\footnotesize{}5.5\%} \tabularnewline 
{\footnotesize{}C. Internal Rate of Return:} & & & & \tabularnewline\cmidrule{1-1} {\footnotesize{}\quad Private IRR} & {\footnotesize{}0.079} & {\footnotesize{}0.076} & & {\footnotesize{}0.072}\tabularnewline{\footnotesize{}\quad Social IRR} & {\footnotesize{}0.055} & {\footnotesize{}0.055} & & {\footnotesize{}0.050}\tabularnewline
{\footnotesize{}\quad Signaling Value} & {\footnotesize{}30.4\%} & {\footnotesize{}28.0\%} & & {\footnotesize{}30.6\%}\tabularnewline
{\footnotesize{}D. Present Discounted Value:} & & & & \tabularnewline\cmidrule{1-1} {\footnotesize{}\quad Private PDV} & {\footnotesize{}442,769} & {\footnotesize{}453,501} & & {\footnotesize{}448,906}\tabularnewline{\footnotesize{}\quad Social PDV} & {\footnotesize{}294,222} & {\footnotesize{}317,260} & & {\footnotesize{}298,379} \tabularnewline
{\footnotesize{}\quad Signaling Value} & {\footnotesize{}33.5\%} & {\footnotesize{}30.0\%} & & {\footnotesize{}33.5\%}\tabularnewline
\bottomrule
\end{tabular*}}{\scriptsize\par}\par\end{centering}
\begin{singlespace}
\smallskip{}
 \emph{\scriptsize{}Note:}{\scriptsize{} The estimation sample consists of Norwegian males born 1950-1980 observed in earnings data over years 1967-2014 with years of potential experience between 0 and 30 years and annual earnings above 1 SGA threshold (N=14,746,755), partitioned in a hidden IV sample (N=8,697,979) and a transparent IV sample (N=6,048,776) as discussed in Section \ref{sec:Setting}. Panel A provides estimates of key model parameters for different model specifications. Columns (1)-(2) provide parameter estimates for the specification with experience-varying returns to skill, which combine the hidden, and the transparent IV estimates plotted in Figure \ref{fig:IVestimates}(a)-(b). In the sequential estimation approach (column (1)), we first estimate the $\hat{\lambda}_{t}$ profile based on the transparent IV estimates using (\ref{eq:IV_transparent_time}) under location normalization $\lambda_{0}=1$, and then insert $\hat{\lambda}_{t}$ in (\ref{eq:IV_hidden_time}) and solve for other model parameters using the non-linear least-squares estimation. In the joint estimation approach (column (3)), we jointly solve for all model parameters and $\lambda_{t}$ using (\ref{eq:IV_hidden_time})-(\ref{eq:IV_transparent_time}) by non-linear least-squares estimation. Column (3) provides parameter estimates for the specification with experience-invariant returns to skill, which are based solely on the hidden IV estimates plotted in Figure \ref{fig:IVestimates}(a) and constructed using non-linear least-squares estimation of (\ref{eq:IV_hidden_time}) with $\lambda_t=1$. Panel B provides the implied weight $\theta_{t}$ on the initial signal corresponding to each estimate of the speed of employer learning $\kappa$. Using the estimates reported in Panel A for each model specification, Panel C provides the implied internal rates of private and social returns (IRR), while Panel D provides the implied present discounted value (PDV) of the lifetime earnings return from an additional year of schooling. The PDV estimates are reported in 2015 Norwegian Kroner (1USD $\approx$ 8NOK) and are calculated using an interest rate of 2.3\%, which corresponds to the average real interest rate on deposits and loans in Norway over the period 1967-2010 (\cite{BhullerMogstadSalvanes2017}).}{\scriptsize\par}
{\scriptsize{}{*} p \textless{} 0.10, {*}{*} \textless{} 0.05, {*}{*}{*}
p \textless{} 0.01.}{\scriptsize\par}
\end{singlespace}
\end{table}

Next, we present estimates under the assumption of experience-invariant returns to skill following the results in Section \ref{sec:identification_timeinvariant}. In Table \ref{tab:MainResults_IV_Speed_InitalLimit}, panel A, column (3), we display the estimates of initial private returns ($b^{IV^{\mathfrak h}}_{0}$), limit returns ($b^{IV^{\mathfrak h}}_{\infty}$) and the speed of learning ($\kappa$) obtained solely from the hidden IV estimates shown in Figure \ref{fig:IVestimates}-(a) based on NLLS estimation. In this case, we estimate an initial return of 19.2\%, a limit return of 5\%, and a speed of learning estimate of $0.532$, which again implies rapid learning on the part of employers. Comparing columns (1)-(3), we find that imposing experience-invariant returns to skill does not materially change our conclusions regarding the speed of learning and the evolution of returns. This finding is consistent with the estimates in Figure \ref{fig:IVestimates}-(b), based on which we could not statistically reject the null hypothesis of constant social returns across the experience.

\begin{figure}[t!]
\begin{centering}
\subfloat[Experience-Varying Returns to Skill (Sequential)]{\centering{}\includegraphics[width=0.5\columnwidth]{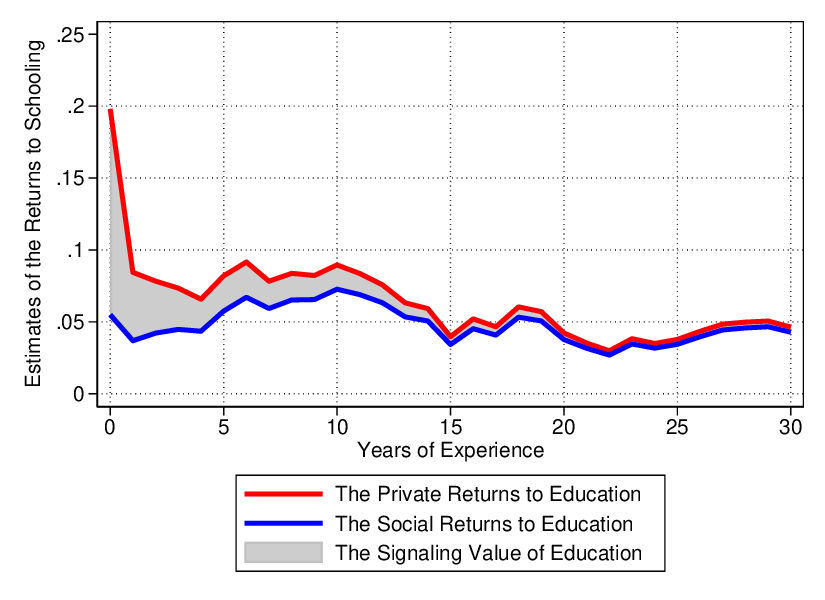}}
\subfloat[Experience-Varying Returns to Skill (Joint)]{\centering{}\includegraphics[width=0.5\columnwidth]{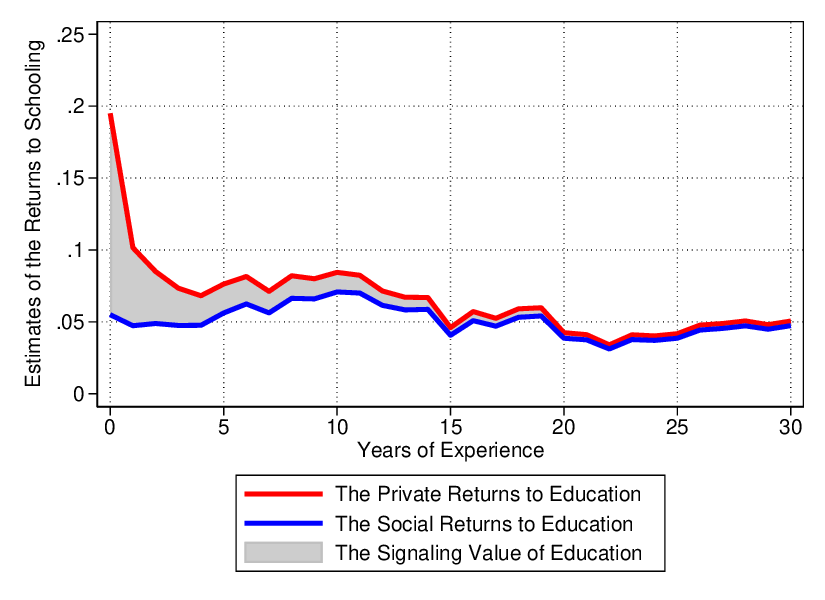}}
\par\end{centering}
\begin{centering}
\subfloat[Experience-Invariant Returns to Skill]{\centering{}\includegraphics[width=0.5\columnwidth]{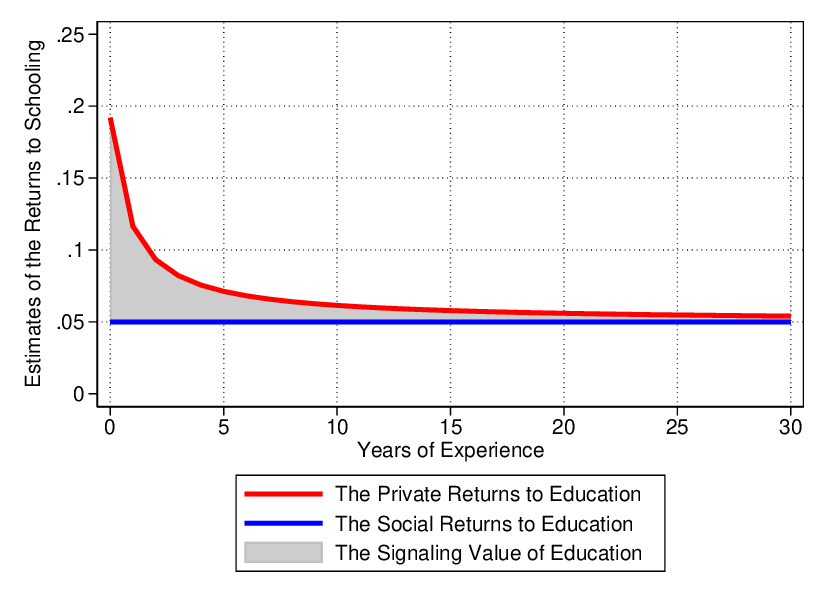}}
\par\end{centering}
\caption{The Private Returns, the Social Returns and the Signaling Value of Education.\label{fig:PrivateSocial}}
\emph{\scriptsize{}Note:}{\scriptsize{} The private and social returns plotted in the red and the blue lines, respectively, are constructed using the estimates in Table \ref{tab:MainResults_IV_Speed_InitalLimit}, columns (1)--(3). The difference between the private and social returns provides the signaling value of education.}{\scriptsize\par}
\end{figure}

Using the estimates in Table \ref{tab:MainResults_IV_Speed_InitalLimit}, we now determine the signaling value of education under different model specifications. We proceed by first graphically illustrating the private and the social returns at each $t$ that are implied by each set of model parameters presented in Table \ref{tab:MainResults_IV_Speed_InitalLimit}. In Figure \ref{fig:PrivateSocial}-(a) we display the private returns (red line) and social returns (blue line) based on the estimates in Table \ref{tab:MainResults_IV_Speed_InitalLimit}, column (1), from a sequential estimation approach. In Figure \ref{fig:PrivateSocial}-(b), we similarly display the returns estimates in Table \ref{tab:MainResults_IV_Speed_InitalLimit}, column (2), from a joint estimation. Both of these estimates allow experience-varying returns to skill ($\hat{\lambda}_t$) based on the point estimates presented in the Online Appendix, Figure \ref{fig:IV_lambda}. By comparison, in Figure \ref{fig:PrivateSocial}-(c), we display the returns estimates from Table \ref{tab:MainResults_IV_Speed_InitalLimit}, column (3), where we impose constant returns to skill. In each sub-figure, we also provide the signaling value of education at each $t$, illustrated by the gap between the implied private and social returns to education.

Figure \ref{fig:PrivateSocial} highlights that a sizable fraction of the private returns that accrue during the early stages of workers' careers are attributable to signaling. To provide a summary measure of the contribution of signaling for the private returns to education over the workers' careers, we calculate the \textit{internal rate of return} (IRR) for an additional year of schooling implied by our estimates. Following \cite{heckman2006earnings}, we define IRR as the discount rate that equalizes the present discounted value of earnings over the career for different choices of schooling. The IRR captures that schooling benefits that accrue early in workers' careers are more valuable. Even with fast employer learning, one could therefore expect a sizable signaling value as the signaling returns occur primarily early in workers' careers.\footnote{Formally, the private IRR $r$ solves $
\sum_{t=0}^{T}\frac{\hat{W}(S+1,t)}{(1+r)^{t+1}}-\sum_{t=0}^{T}\frac{\hat{W}(S,t)}{(1+r)^{t}}=0, 
$
where, for simplicity, we set $\hat{W}(S+1,t)=\hat{W}(S,t)(1+{b^{IV^{\mathfrak h}}_{t}})=\bar{W}(t)(1+{b^{IV^{\mathfrak h}}_{t}})$, $\bar{W}(t)$ denotes average earnings for male workers at experience $t$, and we use a fixed length of work career $T=40$. In this calculation, we rely solely on the opportunity cost of schooling to compute the IRR, and abstract from tuition and psychic costs of schooling and any earnings while in school. For $t>31$, we assume that $\theta_t=0$ and $\lambda_t=1$.} 

Using the experience-specific returns estimates in Figure \ref{fig:PrivateSocial}-(a), we estimate the private IRR to be $7.9\%$, see Table \ref{tab:MainResults_IV_Speed_InitalLimit} panel C, column (1). Thus the private IRR is 2.4 percentage points greater than the social returns at 5.5\%. Based on these estimates, we attribute 70\% of the private return to education, raising workers' productivity and 30\% to signaling. In columns (2)-(3), we perform the same calculation based on the estimates of the returns in Figure \ref{fig:PrivateSocial}, (b)-(c), and again find a signaling component of a similar magnitude. 

As an alternative way to measure the private and social returns, we show in Table \ref{tab:MainResults_IV_Speed_InitalLimit} panel D the present discounted value (PDV) of lifetime returns to an additional year of schooling. As for using the IRR, a comparison of these PDVs estimates also leads us to conclude that about a third of the private returns are attributable to signaling.\footnote{To determine the PDV of private lifetime returns, we calculate $
\sum_{t=0}^{T}\frac{\hat{W}(S+1,t)}{(1+r^{m})^{t+1}}-\sum_{t=0}^{T}\frac{\hat{W}(S,t)}{(1+r^{m})^{t}} 
$, where the discount rate $r^{m}$ is set equal to the average real interest rate of 2.3\% on deposits and loans in Norway from 1960 to 2010 \citep{BhullerMogstadSalvanes2017}.}

\subsection{Robustness and Additional Results\label{sec:Results_robustness}}
In this subsection, we assess whether our findings are robust to allowing for differential trends and using different sample selection criteria. Additionally, we also briefly discuss estimation results following \cite{Lange2007} using the workers' IQ as a hidden correlate of the ability.

\paragraph{Differential Trends.} As discussed in Section \ref{sec:Specifications}, differential trends across individuals who grew up in different municipalities, even in the absence of the compulsory schooling reform, pose a challenge to identifying returns to schooling based on (\ref{eq:Specification 1}). Following \cite{BhullerMogstadSalvanes2017}, we verify that our estimates do not change when we include extrapolated linear and quadratic municipality-specific trends. The results are presented in the Online Appendix Tables \ref{tab:FS_trends} and \ref{tab:IV_Speed_InitalLimit_trends}. We can see that both the first and the second-stage IV estimates of key model parameters are robust to the inclusion of differential trends.\footnote{Indeed, based on the estimates in Table \ref{tab:IV_Speed_InitalLimit_trends} that allow for differential trends we cannot reject that the experience-invariant part of the social returns is equal across the transparent IV and the hidden IV samples.} Similarly, from Figure \ref{fig:IVestimates_trends}, we can see that the experience-specific IV estimates are also robust to differential trends.

\paragraph{Sample Selection.} Next, we consider the sensitivity of our results to using alternative samples. As described in Section \ref{sec:Data}, our estimation sample consists of male workers born 1950--1980, and we retain yearly observations for each of the years when a worker has annual earnings above the SGA threshold. By construction, this sample is unbalanced because our earnings data are limited to calendar years 1967--2014 (data availability) and because some individuals could have been unemployed or worked part-time in some years (partial employment). Due to data availability, the unbalancedness is more severe for younger cohorts as earnings towards the latter part of these workers' careers are not measured by 2014.

To address concerns related to having an unbalanced sample in our baseline, we proceed in two ways. In the first case, we retain our baseline sample of male workers born 1950--1980 but use yearly observations of only up to 10 years of work experience in our estimation. In the second case, we restrict our sample to male workers born 1950--1965 and use observations up to 30 years of work experience, as in our primary analysis. In both of these alternative estimations, the sample is substantially more balanced over time.\footnote{In our baseline hidden IV sample of male workers born 1950--1980, we include $422,749$ individuals and $8,697,979$ person-year observations, among whom we observe earnings for $344,547$ in the initial year of work experience, $355,602$ in the $10^{th}$ year, and $116,385$ in the $30^{th}$ year. In the sample of males born 1950--1965, we include $213,206$ individuals and $5,326,336$ person-year observations, among whom we observe earnings for $173,465$ in the initial year of work experience, $179,027$ in the $10^{th}$ year, and $116,385$ in the $30^{th}$ year.}

The results from these estimations are presented in the Online Appendix Table \ref{tab:IV_Speed_InitalLimit_robust}, columns (2)-(3) and in Figure \ref{fig:IVestimates_robust}-(b). These results show that our returns estimates are robust to using more balanced samples with fewer observations. Furthermore, in Table \ref{tab:IV_Speed_InitalLimit_robust}, column (4), and Figure \ref{fig:IVestimates_robust}-(c), we report estimation results when we extend our sample to male workers born 1940-1980.\footnote{The reason we restricted our baseline sample to birth cohorts born 1950--1980 was that the military IQ test scores are not available for those born before 1950. In Table \ref{tab:IV_Speed_InitalLimit_robust}, column (4), we drop this requirement.} Our findings are also robust to using this extended sample.
 
\paragraph{Earnings Threshold.} In the absence of data on hourly wage rates for most of our sample period, in our baseline estimation, we limited the analysis to individuals with annual labor earnings above the substantial gainful activity (SGA) threshold. This restriction is likely to reduce variability in earnings due to differences in working hours across workers. In the Online Appendix Table \ref{tab:IV_Speed_InitalLimit_robust}, columns (5)-(7) and Figure \ref{fig:IVestimates_robust}-(d)-(f), we provide estimates where we vary the annual earnings threshold to be either 50\%, 75\% or 150\% of the SGA threshold. These estimates again indicate that our findings are robust to using alternative earnings thresholds to select a sample of workers with little variation in working hours.

\paragraph{Definition of the Hidden and Transparent IV Samples.} As discussed in Section \ref{sec:Setting}, to construct the hidden and transparent IVs, we rely on partitioning our sample into two subsamples, based on the relative size of workers' childhood municipality in their local labor markets. We assign workers who grew up in municipalities with the largest population in their respective local labor market to the transparent IV sample, while the remaining workers, those who grew up in non-central municipalities, are part of the hidden IV sample.

We now assess how sensitive our results are to these choices by changing the definition of the hidden IV sample (non-central municipalities) to be further limited by the number of residents living in each municipality at the onset of the compulsory schooling reform in 1960. In the Online Appendix Table \ref{tab:IV_Speed_InitalLimit_robust}, columns (8)-(10) and Figure \ref{fig:IVestimates_robust}-(g-(i), we provide estimates where we vary the population thresholds for being included as non-central municipalities to be either less than 5,000, 10,000 or 25,000 residents. These estimates again indicate that our findings are robust to using alternative sample definitions.

\paragraph{Causal Effect of Education on IQ.} In other related studies \citep[e.g.,][]{BrinchGalloway2012, CarlssonEtAl2015}, researchers have provided causal estimates of education on measures of workers' ability. In a similar spirit, we regress the standardized IQ test score as a function of years of schooling, using the compulsory schooling reform as an instrument for schooling and present the results in Online Appendix Table \ref{tab:IQ-IV}. As mentioned in Section \ref{sec:Schooling}, we use an IQ test score from the Norwegian Armed Forces available for conscripts born 1950--1980. These IV estimates show a strong effect of schooling on IQ: an additional year of schooling at age 18 \emph{causes} approximately 1/4 of a standard deviation increase in IQ.\footnote{The IQ test score is standardized to have a mean of zero and a standard deviation equal to one. Using the same compulsory schooling reform for Norway, \citet{BrinchGalloway2012} documented that an additional year of schooling increased IQ scores measured at age 19 by 3.7 points. Their IQ test score is scaled to have a mean of 100 and a standard deviation of 15. In terms of magnitude, their estimate also corresponds to 1/4 of a standard deviation increase in IQ. \citet{CarlssonEtAl2015} also document similar results for Sweden.} 
Thus, schooling increases worker ability, which we expect to positively correlate with workers' productive skills. A back-of-the-envelope calculation suggests that the schooling-induced increase in IQ results in 3.2\% higher earnings, which is almost 60\% of the 5.5\% social return estimate presented in Table \ref{tab:MainResults_IV_Speed_InitalLimit}, column (1).\footnote{For this calculation, we use the estimated 1/4 of a standard deviation increase in IQ from an additional year of schooling in the Online Appendix Table \ref{tab:IQ-IV}, and an OLS regression of log-wages on IQ, controlling flexibly for experience, which shows that a standard deviation increase in IQ is associated with 12.8\% higher earnings. For the purpose of this exercise, we interpret each coefficient estimates as causal, and conclude that a schooling-induced increase in IQ leads to 3.2\% ($\approx$ 12.8\% $\times$ 1/4) higher earnings.} While we are cautious in interpreting these estimates as ``direct'' evidence of a productivity return to education, they suggest that part of social returns to education in (\ref{eq: causal schooling 2}) could be attributed to the causal effect of education on workers' abilities.

\section{Extensions\label{sec:Extension}}

\subsection{Partially-Transparent Instrument\label{sec:Extension_PartialIV}}

So far, we assumed that researchers could determine whether each individual worker was treated by a hidden or a transparent IV. We now consider an extension that accommodates the situation when researchers are unable to determine the type of IV each worker was exposed to. This is relevant, as in many settings, the distinction between hidden and transparent IVs can be blurry. To operationalize this idea, assume that an IV was transparent for a fraction $\rho$ of workers, and hidden for the remaining fraction $(1-\rho)$. Researchers do not know this fraction $\rho$, nor do they know the type of IV each worker was exposed to. In this situation, we say that we have access to a ``partially-transparent'' instrument. We summarize the main identification results in this setting and leave details to the Online Appendix \ref{sec:Extension_PartialIVappendix}.
 
In this setting, the conditional expectation of log wages becomes a weighted average of the conditional expectation of log wages given hidden and transparent IVs, respectively. When the returns to skill are constant, a partially-transparent IV identifies a lower bound on the private returns to education. Nevertheless, as with the hidden IV, after long enough work experience, this lower bound converges to the social returns. We thus identify the speed of learning and can provide a lower bound for the signaling value. When the returns to skill vary with work experience, we can identify lower bounds on the private returns and the signaling value if we have access to a transparent \emph{and} a partially-transparent IV. To point-identify all parameters, it suffices to have access to a hidden IV in addition to the partially transparent IV if we also strengthen Assumption \ref{ass:4} and instead assume that the social returns are the same across the two IV samples.

\subsection{Heterogeneous Returns\label{sec:Extension_Heterogeneity}}
There is increasing evidence that returns to education are heterogeneous across workers, see \cite{Card1999, heckman2006earnings} and \cite{CarneiroHeckmanVytlacil2011}, among others. 
So, it is important to understand how the heterogeneity affects our identification results. To this end, we consider heterogeneous returns in the Neyman-Rubin potential outcomes framework and study identification with a binary (low or high) schooling choice. We present all the details in the Online Appendix \ref{sec:heterogeneous}, and for brevity, present a summary of our main findings.

In this environment, both the social returns to education and the private returns to education become worker-specific. So, we have to work with the averages across an appropriate subset of workers, which we refer to as the \emph{average social returns} and the \emph{average private returns}, respectively. 
We know that an IV identifies the local average treatment effect for those induced by the IV to switch their schooling status \citep{ImbensAngrist1994}. In particular, if the IV satisfies an appropriate monotonicity condition, it identifies the treatment effect on the compliers, and hence, the causal effect of schooling on workers who obtain ``low" schooling in the absence of the IV but obtain ``high" schooling when ``exposed" to the IV. 

We contribute to this literature by showing that even with heterogeneous returns, a hidden IV identifies the average private returns to education among the compliers, and a transparent IV identifies the average social returns among the compliers. Under our maintained assumption of conditional normality of productivity given schooling, where the mean (but not the variance) of productivity depends on schooling, we can use the hidden IV to identify the speed of learning and the signaling value of education for compliers if the returns to skill are experience-invariant. Moreover, if the returns to skill vary with experience, but if they are homogeneous across compliers of the hidden IV and the transparent IV (Assumption \ref{ass:4}), then having access to the two IVs is again sufficient to identify the speed of learning.

\section{Discussion \label{sec:Discussion}}
We have shown that interpreting the returns to education from IV estimates requires assumptions on what employers know (or do not know) about the instruments. We have done so within a tractable framework, setting aside important issues to convey our core message. In this section, we briefly touch upon a few of these issues, but our discussion here is brief and at best may serve as a marker of topics for future research.

\subsection{Asymmetric Employer Learning\label{section:asymmetric learning}}

Throughout our paper, we have assumed symmetric learning across employers. Symmetry in learning implies that all current and future employers have the same information about an employee. While there has been some empirical support in favor of symmetric learning, at least for employees with low education \citep{Schoenberg2007}, there is also empirical evidence obtained from wage changes across jobs that incumbent employers have an information advantage \citep{GibbonsKatz1991, Kahn2013}, which in turn can lead to the different speed of learning and different wage profiles; see \cite{Pinkston2009}. The learning process becomes significantly more complicated when the information is asymmetric across the employers and when workers and employers act strategically based on this asymmetry. Thus, it is challenging to determine implications for social and private returns in such settings.

\subsection{Market Power}

We also assume that employers perfectly compete for workers, so that workers are paid their expected productivity based on the information available to employers. While convenient, this assumption is at odds with new research suggesting that employers have market power, see \cite{manning2021monopsony} for a review. However, such monopsony power in itself does not imply a systematic bias in our estimates of the social and private returns to education. When employers have market power, they will mark-down wages so that for the marginal worker, expected productivity exceeds the wage. Estimates of the returns to education are, however, based on relative wages across schooling levels. 
For instance, if the mark-downs are proportional to the expected productivity then they ``cancel-out" when we estimate log-linear earnings equations.

This same argument does not apply when mark-downs of wages relative to the expected productivity vary systematically with education.
Such patterns could result from differences across educational groups in the elasticity of labor supply and workers' ability to capture rents. While there is little (if any) evidence on whether the elasticity of labor supply varies systematically with education, it is plausible that labor supply is less elastic for specialized labor, and education generally supports specialization. This inverse relation between education and elasticity suggests that mark-downs increase with education. On the other hand, more educated workers may bargain more effectively and have better abilities to compete for rents. This positive relation between education and bargaining power, in turn, suggests that mark-downs decrease with education. The relative strength of these (and other) factors and how they affect mark-downs across education is at present is not well understood.

\subsection{Information and Assignment}
In our framework, the social returns to education are equal to the direct increase in workers' skills. However, information revealed during schooling might also affect the process of allocating workers to jobs in which they have a comparative advantage. If education improves workers' assignment to employers, this represents an additional channel through which education raises productivity by improving allocative efficiency. A key question then is to whom do these social returns accrue? In a perfectly competitive market, potential employers bid for employees' services based on workers' expected contribution. The highest bidder will employ the workers in their highest productivity job, however, the wages depend on the information revealed during assignment. Therefore it is also not clear to whom the social returns would accrue. More generally, while this mechanism is interesting and could potentially shed light on wage differentials across workers and firms, this mechanism is also not well understood.
	
\subsection{General Equilibrium Effects}

Many instruments rely on large interventions that change educational attainment in the population. These interventions are likely to engender general equilibrium effects, because among other reasons, changes in educational attainment in the population imply movements along production isoquants. Besides, signaling model predicts that the equilibrium wage function changes if interventions change the population wide joint distribution of education and ability. Moreover, because the wage function affects schooling decisions, the signaling aspect of education generates additional feedback from the wage to education decisions.\footnote{See \cite{LangKropp1986} and \cite{Bedard2001} for evidence on such feedback effects.} 

While we abstract from such equilibrium effects, they may affect our analysis. First, our analysis applies best to instruments that affect small population groups (relative to the economy) and are thus unlikely to generate equilibrium effects. Thus, the ideal setting for an empirical application along the lines of our approach would be an experiment that treats only a small subset of the population and sidesteps the concern about equilibrium effects.\footnote{In practice, estimating effects of education interacted with experience using IVs requires large datasets. A complementary approach is to use cross-country variation in educational attainment to identify social return to education, including any general equilibrium effects. For this approach to work, however, we need strong exogeneity assumptions. We refer the interested reader to \cite{LangeTopel2006}.} 

Second, specific to our context, the compulsory schooling reform was gradually implemented throughout Norway between 1960 and 1975. As the reform increasingly applied to the whole country, it is possible that the wage function could have changed due to equilibrium effects. These changes might have interacted with experience so that the variation in the returns to education estimates across different experience levels might confound changes in the wage function over time. Accounting for such effects is beyond the scope of this paper.

\section{Conclusion\label{sec:Conclusion}}
Education policy hinges on the estimates of private and social returns to education, but these returns are difficult to estimate separately. In this paper, we determine conditions under which instruments allow us to separately identify the private and social returns to education within the context of an employer learning model.

We distinguish between two types of IVs: hidden IVs and transparent IVs, where the former are unobserved by employers and thus not directly priced in the wages, while the latter are observed by employers and correctly factored in the wages. We show that a hidden IV identifies the private returns to education. A transparent IV, by contrast, identifies the social returns to education at each level of work experience. If the returns to skill do not vary with experience, than a hidden IV identifies the private and the social returns to education as well as the speed of learning. When returns to skill do vary with experience, access to both a hidden and at transparent IV suffices to identify these parameters. Using data from Norway, we estimate a private internal rate of return to education of 7.9\%, of which 70\% is due to increased productivity and 30\% to signaling. Our estimates also suggest that employers learn workers' abilities quickly.

We believe that IV estimates can provide useful information on the debate between proponents of the human capital and signaling models of the returns to education. To do so requires taking a stand on how informed employers in the market are about the instrument used for identification. We hope that future researchers reporting IV estimates of the returns to education will consider how informed employers are about the IVs and what this means for interpreting their estimates.

\clearpage
\newpage{} 
\bibliographystyle{jole}
\bibliography{Elearning}
\clearpage
\newpage{} 
\begin{center}
\setcounter{page}{1}
\LARGE{\bf Appendix}
\end{center}

\setcounter{footnote}{0}
\setcounter{section}{0}
\setcounter{equation}{0}
\setcounter{table}{0}
\setcounter{figure}{0}
\renewcommand{\thesection}{A}
\renewcommand{\theequation}{A.\arabic{equation}}
\renewcommand{\thetable}{A.\arabic{table}}
\renewcommand{\thefigure}{A.\arabic{figure}}
\section{Additional Results}

\begin{table}[h]
\begin{centering}
\caption{First-Stage Estimates on Years of Schooling -- Controlling for Differential Trends.\label{tab:FS_trends}}
\par\end{centering}
\begin{centering}
{\scriptsize{}{}}%
\begin{tabular*}{1\textwidth}{@{\extracolsep{\fill}}@{\extracolsep{\fill}}lcccccccc}
\toprule 
 & \multicolumn{2}{c}{{\footnotesize{}A. Full Sample}} &  & \multicolumn{2}{c}{{\footnotesize{}B. Hidden IV}}&  & \multicolumn{2}{c}{{\footnotesize{}C. Transparent IV}} \tabularnewline
 & \multicolumn{2}{c}{{\footnotesize{}}} &  & \multicolumn{2}{c}{{\footnotesize{}Sample}}&  & \multicolumn{2}{c}{{\footnotesize{}Sample}} \tabularnewline 
\cmidrule{2-3} \cmidrule{5-6}  \cmidrule{8-9} 
 & {\footnotesize{}Baseline} & {\footnotesize{}Trends} &  & {\footnotesize{}Baseline} & {\footnotesize{}Trends}&  & {\footnotesize{}Baseline} & {\footnotesize{}Trends}\tabularnewline
 & {\footnotesize{}(1)} & {\footnotesize{}(2)} &  & {\footnotesize{}(3)} & {\footnotesize{}(4)}&  & {\footnotesize{}(5)} & {\footnotesize{}(6)}\tabularnewline
\midrule
{\footnotesize{}Instrument:} &  &  &  &  &  &  & \tabularnewline
\emph{\footnotesize{}{}Exposure to Reform} & {\footnotesize{}{}0.237}\textsuperscript{{*}{*}{*}} & {\footnotesize{}{}0.209}\textsuperscript{{*}{*}{*}} &  & {\footnotesize{}{}0.228}\textsuperscript{{*}{*}{*}} & {\footnotesize{}{}0.209}\textsuperscript{{*}{*}{*}} &  & {\footnotesize{}{}0.240}\textsuperscript{{*}{*}{*}} & {\footnotesize{}{}0.199}\textsuperscript{{*}{*}{*}}\tabularnewline
 & {\footnotesize{}{}(0.025)} & {\footnotesize{}{}(0.034)} &  & {\footnotesize{}{}(0.034)} & {\footnotesize{}{}(0.040)} &  & {\footnotesize{}{}(0.032)} & {\footnotesize{}{}(0.038)}\tabularnewline
\midrule
{\footnotesize{}Controls:} &  &  &  &  & \tabularnewline
{\footnotesize{}{}Municipality Fixed Effects} & $\checkmark$ & $\checkmark$ &  & $\checkmark$ & $\checkmark$ &  & $\checkmark$ & $\checkmark$\tabularnewline
{\footnotesize{}{}Cohort Fixed Effects} & $\checkmark$ & $\checkmark$ &  & $\checkmark$ & $\checkmark$ &  & $\checkmark$ & $\checkmark$\tabularnewline
{\footnotesize{}{}Municipality-Specific Trends} &  & $\checkmark$ &  &  & $\checkmark$&  &  & $\checkmark$\tabularnewline 
\midrule
{\footnotesize{}{}Number of Observations} & \multicolumn{2}{c}{\footnotesize{}{}14,746,755} & & \multicolumn{2}{c}{\footnotesize{}{}8,697,979}  & & \multicolumn{2}{c}{\footnotesize{}{}6,048,776} \tabularnewline
\bottomrule
\end{tabular*}
\par\end{centering}
\smallskip{}

\emph{\scriptsize{}{}Note:}{\scriptsize{}{} The full
sample (panel A) consists of Norwegian males born in 1950-1980 observed any time
in earnings data over years 1967-2014 with years of potential experience
between 0 and 30 years and annual earnings above 1 SGA threshold (N=14,746,755).
The hidden IV sample (panel B) further drops individuals who grew
up in the municipality with the largest population size in each of
the 160 labor market regions in Norway (N=8,697,979), while the transparent IV sample (panel C) retains only individuals who grew up in the municipality with the largest population size in each labor market (N=6,048,776). All estimations include fixed effects for birth cohort and childhood municipality. The trends specifications in columns (2), (4) and (6) control
for linear and quadratic municipality-specific trends estimated using data on all pre-reform cohorts born 1930 or later and extrapolated to all post-reform cohorts, separately for
each municipality. Standard errors are clustered at the local labor market region.}{\scriptsize\par}

{\scriptsize{}{}{*} p \textless{} 0.10, {*}{*} \textless{} 0.05,
{*}{*}{*} p \textless{} 0.01.}{\scriptsize\par}
\end{table}

\begin{center}
\begin{table}[t!]
\begin{centering}
\caption{IV Estimates of the Speed of Employer Learning, Initial and
Limit Returns to Education, and the Signaling Value Contribution  --
Controlling for Differential Trends.\label{tab:IV_Speed_InitalLimit_trends}}
\par\end{centering}
\begin{centering}
{\scriptsize{}}%
\begin{tabular*}{1\textwidth}{@{\extracolsep{\fill}}lcc>{\centering}p{1.5cm}cccc}
\toprule 
 {\footnotesize{}Model Specifications:}  & \multicolumn{4}{c}{{\footnotesize{}A. Experience-Varying Returns to Skill}} & & \multicolumn{2}{c}{{\footnotesize{}B. Experience-Invariant}} \tabularnewline
 & \multicolumn{2}{c}{\emph{\footnotesize{}Sequential Estimation}} & \multicolumn{2}{c}{\emph{\footnotesize{}Joint Estimation}} &  & \multicolumn{2}{c}{{\footnotesize{} Returns to Skill}}  \tabularnewline
\cmidrule{2-5} \cmidrule{7-8} 
 & {\footnotesize{}(1)} & {\footnotesize{}(2)} &  {\footnotesize{}(3)} & {\footnotesize{}(4)} & & {\footnotesize{}(5)} & {\footnotesize{}(6)}\tabularnewline
 & {\footnotesize{}Baseline} & {\footnotesize{}Trends} &  {\footnotesize{}Baseline} & {\footnotesize{}Trends} & & {\footnotesize{}Baseline} & {\footnotesize{}Trends}\tabularnewline
\midrule
{\footnotesize{}Parameters of Interest:} &  &  &  &  &  &  & \tabularnewline
\cmidrule{1-1} 
{\footnotesize{}\quad Speed of Learning $\kappa$}  & {\footnotesize{}0.505}\textsuperscript{{*}{*}{*}} & {\footnotesize{}0.506}\textsuperscript{{*}{*}{*}} & {\footnotesize{}0.550}\textsuperscript{{*}{*}{*}} & {\footnotesize{}0.592}\textsuperscript{{*}{*}{*}} & & {\footnotesize{}0.532}\textsuperscript{{*}{*}{*}} & {\footnotesize{}0.565}\textsuperscript{{*}{*}{*}}\tabularnewline
 & {\footnotesize{}(0.107)} & {\footnotesize{}(0.152)} & {\footnotesize{}(0.126)} & {\footnotesize{}(0.205)} &  & {\footnotesize{}(0.058)} & {\footnotesize{}(0.055)} \tabularnewline
{\footnotesize{}\quad Initial Return $b_{0}^{IV^{\mathfrak{h}}}$}  & {\footnotesize{}0.198}\textsuperscript{{*}{*}{*}} & {\footnotesize{}0.209}\textsuperscript{{*}{*}{*}} & {\footnotesize{}0.195}\textsuperscript{{*}{*}{*}} & {\footnotesize{}0.205}\textsuperscript{{*}{*}{*}} & & {\footnotesize{}0.192}\textsuperscript{{*}{*}{*}} & {\footnotesize{}0.204}\textsuperscript{{*}{*}{*}} \tabularnewline
& {\footnotesize{}(0.015)} & {\footnotesize{}(0.023)} & {\footnotesize{}(0.012)} & {\footnotesize{}(0.017)} &  & {\footnotesize{}(0.010)} & {\footnotesize{}(0.010)} \tabularnewline
{\footnotesize{}\quad Limit Return $b_{\infty}^{IV^{\mathfrak{h}}}$}  & {\footnotesize{}0.055}\textsuperscript{{*}{*}{*}} & {\footnotesize{}0.049}\textsuperscript{{*}{*}{*}} & {\footnotesize{}0.055}\textsuperscript{{*}{*}{*}} & {\footnotesize{}0.054}\textsuperscript{{*}{*}{*}} & & {\footnotesize{}0.050}\textsuperscript{{*}{*}{*}} & {\footnotesize{}0.045}\textsuperscript{{*}{*}{*}} \tabularnewline
  & {\footnotesize{}(0.006)} & {\footnotesize{}(0.010)} & {\footnotesize{}(0.008)} & {\footnotesize{}(0.015)} & & {\footnotesize{}(0.003)} & {\footnotesize{}(0.003)} \tabularnewline
{\footnotesize{}\quad Average Return $b^{IV^{\mathfrak{h}}}$} &  {\footnotesize{}0.064}\textsuperscript{{*}{*}{*}} & {\footnotesize{}0.061}\textsuperscript{{*}{*}{*}}&  {\footnotesize{}0.066}\textsuperscript{{*}{*}{*}} & {\footnotesize{}0.060}\textsuperscript{{*}{*}{*}} & &{\footnotesize{}0.067}\textsuperscript{{*}{*}{*}} &  {\footnotesize{}0.062}\textsuperscript{{*}{*}{*}} \tabularnewline
  & {\footnotesize{}(0.003)} & {\footnotesize{}(0.003)} & {\footnotesize{}(0.002)} & {\footnotesize{}(0.004)} & &  {\footnotesize{}(0.007)}&  {\footnotesize{}(0.007)} \tabularnewline
 {\footnotesize{}\quad Average Return $b^{IV^{\mathfrak{t}}}$} &  {\footnotesize{}0.088}\textsuperscript{{*}{*}{*}} & {\footnotesize{}0.058}\textsuperscript{{*}{*}}&  {\footnotesize{}0.089}\textsuperscript{{*}{*}{*}} & {\footnotesize{}0.060}\textsuperscript{{*}{*}{*}} & & -- &  --  \tabularnewline
  & {\footnotesize{}(0.024)} & {\footnotesize{}(0.029)} & {\footnotesize{}(0.019)} & {\footnotesize{}(0.005)} & & &\tabularnewline
{\footnotesize{}Internal Rate of Return:} &  &  &  &  &  &  & \tabularnewline
\cmidrule{1-1} 
{\footnotesize{}\quad Private Return} & {\footnotesize{}0.079} & {\footnotesize{}0.075} & {\footnotesize{}0.076} & {\footnotesize{}0.075} & & {\footnotesize{}0.072} & {\footnotesize{}0.067} \tabularnewline
{\footnotesize{}\quad Social Return}  & {\footnotesize{}0.055} & {\footnotesize{}0.049} & {\footnotesize{}0.055} & {\footnotesize{}0.054} & & {\footnotesize{}0.050} & {\footnotesize{}0.045} \tabularnewline
{\footnotesize{}\quad Signaling Value}  & {\footnotesize{}30.4\%} & {\footnotesize{}34.6\%} & {\footnotesize{}27.6\%} & {\footnotesize{}28.0\%} & & {\footnotesize{}30.6\%} & {\footnotesize{}32.8\%} \tabularnewline
\midrule
{\footnotesize{}Controls:} &  &  &  &  &  &  & \tabularnewline
\cmidrule{1-1} 
{\footnotesize{}\quad Municipality Fixed Effects} & $\checkmark$ & $\checkmark$ &  $\checkmark$ & $\checkmark$ & & $\checkmark$ & $\checkmark$\tabularnewline
{\footnotesize{}\quad Cohort Fixed Effects} & $\checkmark$ & $\checkmark$ &  $\checkmark$ & $\checkmark$ & & $\checkmark$ & $\checkmark$\tabularnewline
{\footnotesize{}\quad Differential Trends} &  & $\checkmark$ &  & $\checkmark$ & & & $\checkmark$\tabularnewline
\bottomrule
\end{tabular*}{\scriptsize\par}
\par\end{centering}
\begin{singlespace}
{\scriptsize{}}{\scriptsize\par}

\smallskip{}

\emph{\scriptsize{}Note:}{\scriptsize{}The estimation sample consists of Norwegian males born 1950-1980 observed in earnings data over years 1967-2014 with years of potential experience between 0 and 30 years and annual earnings above 1 SGA threshold (N=14,746,755), partitioned in a hidden IV sample (N=8,697,979) and a transparent IV sample (N=6,048,776) as discussed in Section \ref{sec:Setting}. Parameter estimates for the model specification with experience-varying returns to skill (panel A) are based on a combination of the hidden and the transparent IV estimates plotted in Figure \ref{fig:IVestimates}(a)-(b). In the sequential estimation approach (columns (1)-(2)), we first estimate the $\hat{\lambda}_{t}$ profile based on the transparent IV estimates using  (\ref{eq:IV_transparent_time}) under location normalization $\lambda_{0}=1$, and then insert $\hat{\lambda}_{t}$ in  (\ref{eq:IV_hidden_time}) and solve for the model parameters using the non-linear least squares estimation. In the joint estimation approach (columns (3)-(4)), we jointly solve for the model parameters and $\lambda_{t}$ using (\ref{eq:IV_hidden_time})-(\ref{eq:IV_transparent_time}) by non-linear least squares estimation. Parameters estimates for the model specification with experience-invariant returns to skill (panel B) are based on the hidden IV estimates plotted in Figure \ref{fig:IVestimates}(a) and constructed using non-linear least squares estimation of (\ref{eq:IV_hidden}). Columns (2), (4) and (6) rely on IV estimates that control for linear and quadratic municipality-specific trends estimated using data on all pre-reform cohorts born 1930 or later and extrapolated to all post-reform cohorts.}{\scriptsize\par}

{\scriptsize{}{*} p \textless{} 0.10, {*}{*} \textless{} 0.05, {*}{*}{*}
p \textless{} 0.01.}{\scriptsize\par}
\end{singlespace}
\end{table}
\par\end{center}

\begin{center}
\begin{sidewaystable}[t!]
\begin{centering}
\caption{Alternative Hidden IV Estimates of the Speed of Learning, Initial and
Limit Return.\label{tab:IV_Speed_InitalLimit_robust}}
\par\end{centering}
\begin{centering}
{\scriptsize{}}%
\begin{tabular*}{1\textwidth}{@{\extracolsep{\fill}}lcccccccccccc}
\toprule 
 & {\footnotesize{}A. Baseline} & \multicolumn{3}{c}{{\footnotesize{}B. Sensitivity to Unbalanced}} &  & \multicolumn{3}{c}{{\footnotesize{}C. Sensitivity to Alternative}} &  & \multicolumn{3}{c}{{\footnotesize{}D. Sensitivity to Alternative}}  \tabularnewline
 & {\footnotesize{}} & \multicolumn{3}{c}{{\footnotesize{}Panel of Worker Careers}} &  & \multicolumn{3}{c}{{\footnotesize{}Earnings Thresholds}} &  & \multicolumn{3}{c}{{\footnotesize{}Definitions of Hidden IV Sample}}  \tabularnewline
\cmidrule{3-5} \cmidrule{7-9}  \cmidrule{11-13} 
 & {\footnotesize{}} & {\footnotesize{}Years of} & {\footnotesize{}Birth} & {\footnotesize{}Birth}  & & {\footnotesize{}Earnings} & {\footnotesize{}Earnings} & {\footnotesize{}Earnings} & & {\footnotesize{}Population} & {\footnotesize{}Population} & {\footnotesize{}Population}\tabularnewline
  & {\footnotesize{}} & {\footnotesize{}Experience} & {\footnotesize{}Cohorts} & {\footnotesize{}Cohorts}  & & {\footnotesize{}$>.5$} & {\footnotesize{}$>.75$} & {\footnotesize{}$>1.5$}  & & {\footnotesize{}$< 5,000$} & {\footnotesize{}$< 10,000$} &  {\footnotesize{}$< 25,000$} \tabularnewline
  & {\footnotesize{}} & {\footnotesize{} $t \le 10$} & {\footnotesize{}1950--65} & {\footnotesize{}1940--80}  & & {\footnotesize{}SGA} & {\footnotesize{}SGA} & {\footnotesize{}SGA} \tabularnewline
 & {\footnotesize{}(1)} & {\footnotesize{}(2)} & {\footnotesize{}(3)} & {\footnotesize{}(4)}  & & {\footnotesize{}(5)} & {\footnotesize{}(6)} & {\footnotesize{}(7)} & & {\footnotesize{}(8)} & {\footnotesize{}(9)} & {\footnotesize{}(10)} \tabularnewline
\midrule
{\footnotesize{}Parameters:} &  &  &  &  &  &  & \tabularnewline
\cmidrule{1-1} 
{\footnotesize{}Speed of Learning $\kappa$} & {\footnotesize{}0.532}\textsuperscript{{*}{*}{*}} & {\footnotesize{}0.605}\textsuperscript{{*}{*}{*}} & {\footnotesize{}0.753}\textsuperscript{{*}{*}{*}} & {\footnotesize{}0.485}\textsuperscript{{*}{*}{*}} & & {\footnotesize{}0.594}\textsuperscript{{*}{*}{*}} & {\footnotesize{}0.498}\textsuperscript{{*}{*}{*}} & {\footnotesize{}0.656}\textsuperscript{{*}{*}{*}}& & {\footnotesize{}0.690}\textsuperscript{{*}{*}{*}}& {\footnotesize{}0.621}\textsuperscript{{*}{*}{*}}& {\footnotesize{}0.562}\textsuperscript{{*}{*}{*}}\tabularnewline
 & {\footnotesize{}(0.058)} & {\footnotesize{}(0.065)} &  {\footnotesize{}(0.079)}  & {\footnotesize{}(0.049)} & & {\footnotesize{}(0.070)} & {\footnotesize{}(0.062)} & {\footnotesize{}(0.041)} &  & {\footnotesize{}(0.099)}  & {\footnotesize{}(0.063)}  & {\footnotesize{}(0.070)}\tabularnewline
{\footnotesize{}Initial Return $b_{0}^{IV^{\mathfrak{h}}}$} & {\footnotesize{}0.192}\textsuperscript{{*}{*}{*}} & {\footnotesize{}0.193}\textsuperscript{{*}{*}{*}} & {\footnotesize{}0.159}\textsuperscript{{*}{*}{*}}  & {\footnotesize{}0.161}\textsuperscript{{*}{*}{*}} & & {\footnotesize{}0.209}\textsuperscript{{*}{*}{*}} & {\footnotesize{}0.231}\textsuperscript{{*}{*}{*}} & {\footnotesize{}0.189}\textsuperscript{{*}{*}{*}}& & {\footnotesize{}0.174}\textsuperscript{{*}{*}{*}} & {\footnotesize{}0.214}\textsuperscript{{*}{*}{*}} & {\footnotesize{}0.185}\textsuperscript{{*}{*}{*}}\tabularnewline
 & {\footnotesize{}(0.010)} & {\footnotesize{}(0.008)} &  {\footnotesize{}(0.010)} & {\footnotesize{}(0.007)} & & {\footnotesize{}(0.015)} & {\footnotesize{}(0.011)} & {\footnotesize{}(0.007)}& & {\footnotesize{}(0.015)} & {\footnotesize{}(0.012)} & {\footnotesize{}(0.010)}\tabularnewline
{\footnotesize{}Limit Return $b_{\infty}^{IV^{\mathfrak{h}}}$} & {\footnotesize{}0.050}\textsuperscript{{*}{*}{*}} & {\footnotesize{}0.056}\textsuperscript{{*}{*}{*}} & {\footnotesize{}0.041}\textsuperscript{{*}{*}{*}}  & {\footnotesize{}0.042}\textsuperscript{{*}{*}{*}} & & {\footnotesize{}0.054}\textsuperscript{{*}{*}{*}} & {\footnotesize{}0.049}\textsuperscript{{*}{*}{*}} & {\footnotesize{}0.050}\textsuperscript{{*}{*}{*}} & & {\footnotesize{}0.042}\textsuperscript{{*}{*}{*}} & {\footnotesize{}0.046}\textsuperscript{{*}{*}{*}} & {\footnotesize{}0.059}\textsuperscript{{*}{*}{*}}\tabularnewline
 & {\footnotesize{}(0.003)} & {\footnotesize{}(0.005)} & {\footnotesize{}(0.002)} & {\footnotesize{}(0.002)} & & {\footnotesize{}(0.004)} & {\footnotesize{}(0.004)} & {\footnotesize{}(0.002)}& & {\footnotesize{}(0.004)} & {\footnotesize{}(0.003)} & {\footnotesize{}(0.003)}\tabularnewline
\midrule
{\footnotesize{}Controls:} &  &  &  &  & &  &  & \tabularnewline
\cmidrule{1-1} 
{\footnotesize{}Municipality FEs} & $\checkmark$ & $\checkmark$ & $\checkmark$  & $\checkmark$& & $\checkmark$ & $\checkmark$ & $\checkmark$& & $\checkmark$ & $\checkmark$ & $\checkmark$\tabularnewline
{\footnotesize{}Cohort FEs} & $\checkmark$ & $\checkmark$ & $\checkmark$  & $\checkmark$ & & $\checkmark$ & $\checkmark$ & $\checkmark$& & $\checkmark$ & $\checkmark$ & $\checkmark$\tabularnewline
\bottomrule
\end{tabular*}{\scriptsize\par}
\par\end{centering}
\begin{singlespace}
{\scriptsize{}}{\scriptsize\par}

\smallskip{}

\emph{\scriptsize{}Note:}{\scriptsize{} 
The baseline hidden IV sample in column (1) consists of Norwegian males born 1950-1980 observed in earnings data over years 1967-2014 with years of potential experience between 0 and 30 years and annual earnings above 1 SGA threshold, excluding individuals who grew up in the municipality with the largest population size in each of the 160 labor market regions in Norway (N=8,697,979). In column (2), we restrict estimations to years of potential experience between 0 and 10 years for the same sample as in column (1). In column (3), we restrict the sample in column (1) to males born 1950-1965 (N=5,326,336), while in column (4), we instead expand this sample to males born 1950-1980 and also include observations with missing IQ data (N=13,172,738). In column (5), we expand the sample in column (1) to include individuals with annual earnings above 50\% of the SGA threshold (N=8,918,180), while in column (6), we similarly include individuals with annual earnings above 75\% of the SGA threshold (N=8,803,482). In column (7), we instead restrict the sample in column (1) to include individuals with annual earnings above 150\% of the SGA threshold (N=8,486,015). In columns (8)-(10), we change the definition of hidden IV sample to individuals who grew up in municipalities with, respectively, less than 5,000 (N=4,105,767), 10,000  (N=6,249,821) or 25,000  (N=7,617,157) residents in 1960, besides excluding individuals who grew up in the municipality with the largest population size in each labor market region as in our baseline.}{\scriptsize\par}

{\scriptsize{}{*} p \textless{} 0.10, {*}{*} \textless{} 0.05, {*}{*}{*}
p \textless{} 0.01.}{\scriptsize\par}
\end{singlespace}
\end{sidewaystable}
\par\end{center}

\begin{table}[t!]
\begin{centering}
\caption{IV Estimates of Years of Schooling on Standardized IQ Test Scores.\label{tab:IQ-IV}}
\par\end{centering}
\begin{centering}
{\scriptsize{}{}}%
\begin{tabular*}{1\textwidth}{@{\extracolsep{\fill}}@{\extracolsep{\fill}}lcccccccc}
\toprule 
 & \multicolumn{2}{c}{{\footnotesize{}A. Full Sample}} &  & \multicolumn{2}{c}{{\footnotesize{}B. Hidden IV}} &  & \multicolumn{2}{c}{{\footnotesize{}C. Transparent IV}}\tabularnewline
 & \multicolumn{2}{c}{{\footnotesize{}}} &  & \multicolumn{2}{c}{{\footnotesize{}Sample}} &  & \multicolumn{2}{c}{{\footnotesize{}Sample}}\tabularnewline
\cmidrule{2-3} \cmidrule{5-6}  \cmidrule{8-9} 
 & {\footnotesize{}Baseline} & {\footnotesize{}Trends} &  & {\footnotesize{}Baseline} & {\footnotesize{}Trends} &  & {\footnotesize{}Baseline} & {\footnotesize{}Trends}\tabularnewline
 & {\footnotesize{}(1)} & {\footnotesize{}(2)} &  & {\footnotesize{}(3)} & {\footnotesize{}(4)} &  & {\footnotesize{}(5)} & {\footnotesize{}(6)}\tabularnewline
\midrule
{\footnotesize{}Reduced Form:} &  &  &  &  & \tabularnewline
\emph{\footnotesize{}{}Exposure to Reform} & {\footnotesize{}{}0.041}\textsuperscript{{*}{*}{*}} & {\footnotesize{}{}0.036}\textsuperscript{{*}{*}{*}} &  & {\footnotesize{}{}0.047}\textsuperscript{{*}{*}{*}} & {\footnotesize{}{}0.037}\textsuperscript{{*}{*}{*}} &  & {\footnotesize{}{}0.037}\textsuperscript{{*}{*}{*}} & {\footnotesize{}{}0.035}\textsuperscript{{*}{*}{*}}\tabularnewline
 & {\footnotesize{}{}(0.009)} & {\footnotesize{}{}(0.013)} &  & {\footnotesize{}{}(0.014)} & {\footnotesize{}{}(0.016)} &  & {\footnotesize{}{}(0.010)} & {\footnotesize{}{}(0.011)}\tabularnewline
{\footnotesize{}IV Estimates:} &  &  &  &  & \tabularnewline
\emph{\footnotesize{}{}Years of Schooling at Age 18} & {\footnotesize{}{}0.265}\textsuperscript{{*}{*}{*}} & {\footnotesize{}{}0.235}\textsuperscript{{*}{*}{*}} &  & {\footnotesize{}{}0.318}\textsuperscript{{*}{*}{*}} & {\footnotesize{}{}0.258}\textsuperscript{{*}{*}} &  & {\footnotesize{}{}0.229}\textsuperscript{{*}{*}{*}} & {\footnotesize{}{}0.219}\textsuperscript{{*}{*}}\tabularnewline
 & {\footnotesize{}{}(0.055)} & {\footnotesize{}{}(0.074)} &  & {\footnotesize{}{}(0.075)} & {\footnotesize{}{}(0.099)}&  & {\footnotesize{}{}(0.060)} & {\footnotesize{}{}(0.068)}\tabularnewline
\midrule
{\footnotesize{}{}Municipality Fixed Effects} & $\checkmark$ & $\checkmark$ &  & $\checkmark$ & $\checkmark$ &  & $\checkmark$ & $\checkmark$\tabularnewline
{\footnotesize{}{}Cohort Fixed Effects} & $\checkmark$ & $\checkmark$ &  & $\checkmark$ & $\checkmark$ &  & $\checkmark$ & $\checkmark$\tabularnewline
{\footnotesize{}{}Municipality-Specific Trends} &  & $\checkmark$ &  &  & $\checkmark$ &  &  & $\checkmark$\tabularnewline
\bottomrule
\end{tabular*}
\par\end{centering}
\smallskip{}

\emph{\scriptsize{}{}Note:}{\scriptsize{}{}The full
sample (panel A) consists of Norwegian males born in 1950-1980 observed any time
in earnings data over years 1967-2014 with years of potential experience
between 0 and 30 years and annual earnings above 1 SGA threshold (N=14,746,755). The hidden IV sample (panel B) further drops individuals who grew up in the municipality with the largest population size in each of
the 160 labor market regions in Norway (N=8,697,979), while the transparent IV sample (panel C) retains only individuals who grew up in the municipality with the largest population size in each labor market (N=6,048,776). All estimations include fixed effects for birth cohort and childhood municipality. The trends specifications in columns (2), (4) and (6) further also controls
for municipality-specific trends estimated using data on all pre-reform cohorts born 1930
or later and extrapolated to all post-reform cohorts, separately for
each municipality. Standard errors are clustered at the local labor
market region.}{\scriptsize\par}

{\scriptsize{}{}{*} p \textless{} 0.10, {*}{*} \textless{} 0.05,
{*}{*}{*} p \textless{} 0.01.}{\scriptsize\par}
\end{table}
\clearpage

\begin{figure}[h!]
\begin{centering}
\includegraphics[width=0.8\columnwidth]{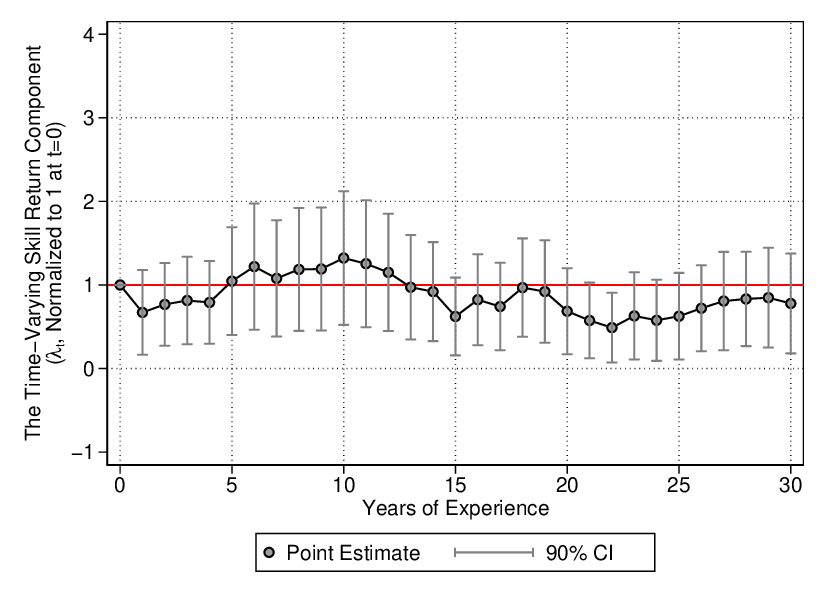}
\par\end{centering}
\caption{The Experience-Varying Component of Returns to Skill -- $\lambda_t$.\label{fig:IV_lambda}}
\emph{\scriptsize{}Note:}{\scriptsize{} The $\hat{\lambda}_t$ estimates are constructed using the IV estimates displayed in Figure \ref{fig:IVestimates}-(b) for the transparent IV sample. We use the formula $\hat{\lambda}_{t}=({\hat{b}_{t}^{IV^{\mathfrak t}}} \slash {\hat{b}_{0}^{IV^{\mathfrak t}}})$ for $t>0$ and $\lambda_0=1$ (location normalization), and employ the delta-method to construct standard errors. The 90\% confidence intervals corresponding to each point estimate are displayed as vertical bars. The joint test for the hypothesis that for all $t\in\mathbb{T}_{>0}$ the ratio $\hat{\lambda}_{t}=1$ provides an F-statistic of 0.9, which means that we cannot statistically reject the hypothesis of constant social returns. The transparent IV sample consists of Norwegian males born 1950-1980 observed in earnings data over years 1967-2014 with years of potential experience between 0 and 30 years with annual earnings above 1 SGA threshold who who grew up in the municipality with the largest population size in each labor market (N=6,048,776). }{\scriptsize\par}
\end{figure}

\begin{figure}[h]
\begin{centering}
\subfloat[Full Sample]{\centering{}\includegraphics[width=0.5\columnwidth]{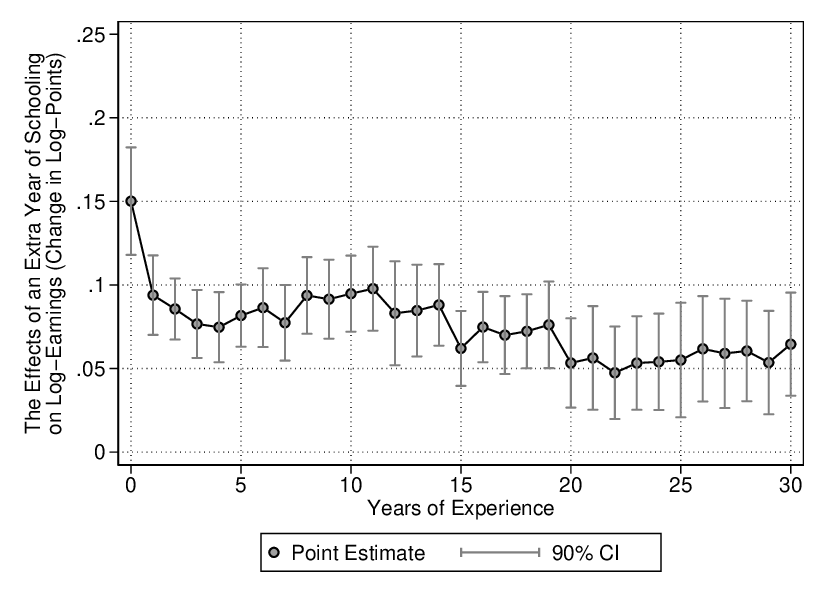}}
\subfloat[Full Sample w/ Differential Trends]{\centering{}\includegraphics[width=0.5\columnwidth]{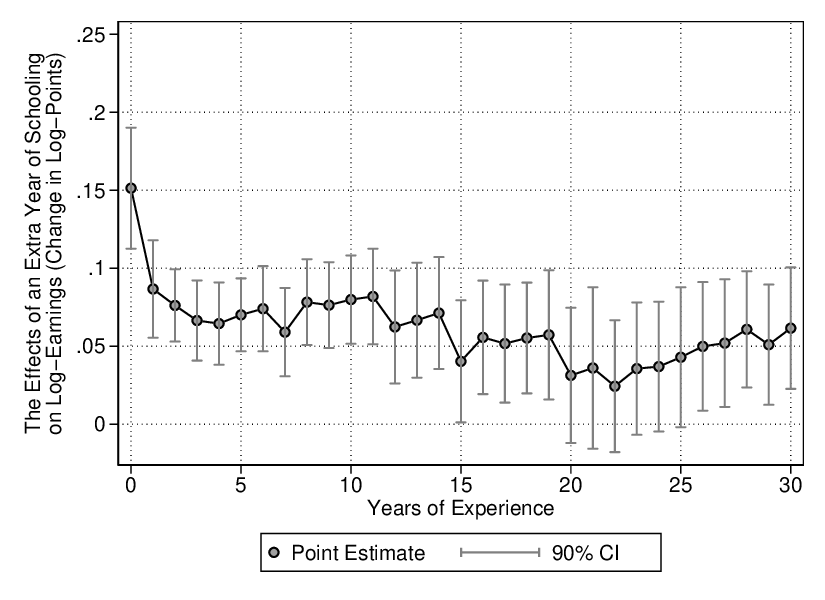}}
\par\end{centering}
\begin{centering}
\subfloat[Hidden IV Sample]{\centering{}\includegraphics[width=0.5\columnwidth]{fig.eps/fig_logL_IV_noiq_dropm}}
\subfloat[Hidden IV Sample w/ Differential Trends]{\centering{}\includegraphics[width=0.5\columnwidth]{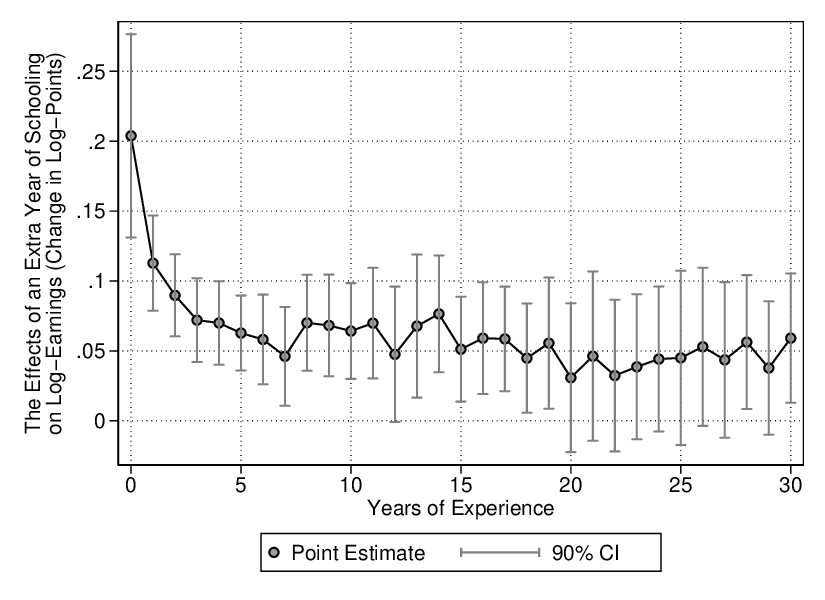}}
\par\end{centering}
\begin{centering}
\subfloat[Transparent IV Sample]{\centering{}\includegraphics[width=0.5\columnwidth]{fig.eps/fig_logL_IV_noiq_keepm}}
\subfloat[Transparent IV Sample w/ Differential Trends]{\centering{}\includegraphics[width=0.5\columnwidth]{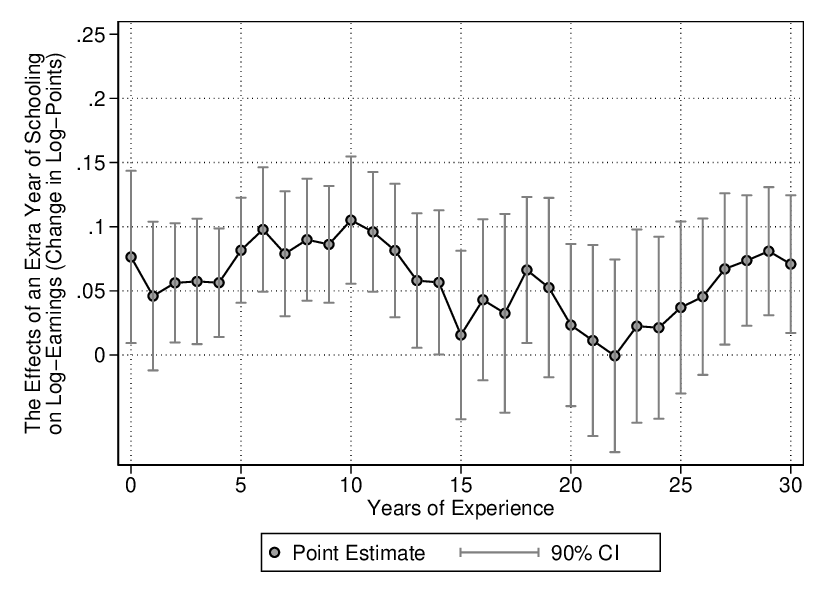}}
\par\end{centering}
\caption{IV Estimates of the Returns to Schooling -- Controlling for Differential Trends.\label{fig:IVestimates_trends}}

\emph{\scriptsize{}Note:}{\scriptsize{} 
See notes below Figure \ref{fig:IVestimates} for details on each estimation. Plots (b), (d) and (f) control for linear and quadratic municipality-specific trends estimated using data on all pre-reform cohorts born 1930 or later and extrapolated to post-reform cohorts.}{\scriptsize\par}
\end{figure}

\begin{sidewaysfigure}[h]
\begin{centering}
\subfloat[Baseline Hidden IV]{\centering{}\includegraphics[width=0.33\columnwidth]{fig.eps/fig_logL_IV_noiq_dropm}}
\subfloat[Birth Cohorts 1950--65]{\centering{}\includegraphics[width=0.33\columnwidth]{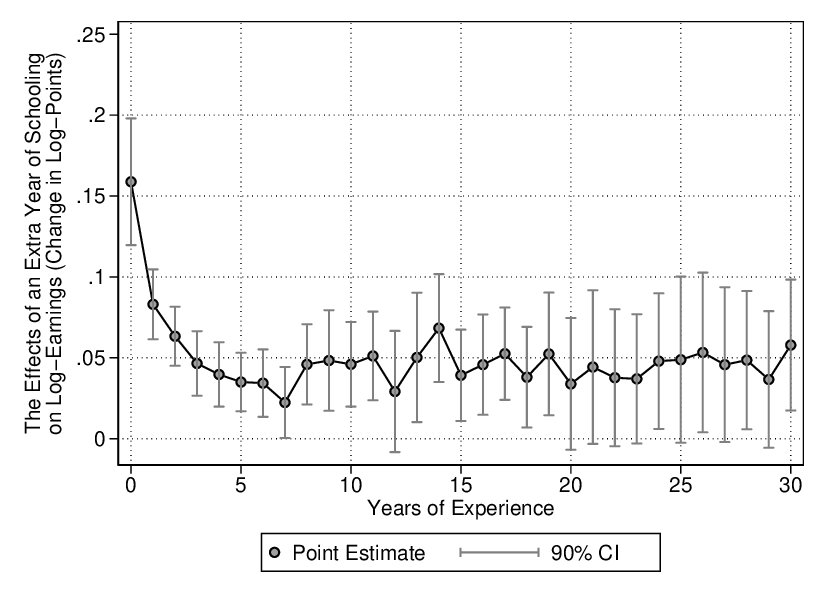}}
\subfloat[Birth Cohorts 1940--80]{\centering{}\includegraphics[width=0.33\columnwidth]{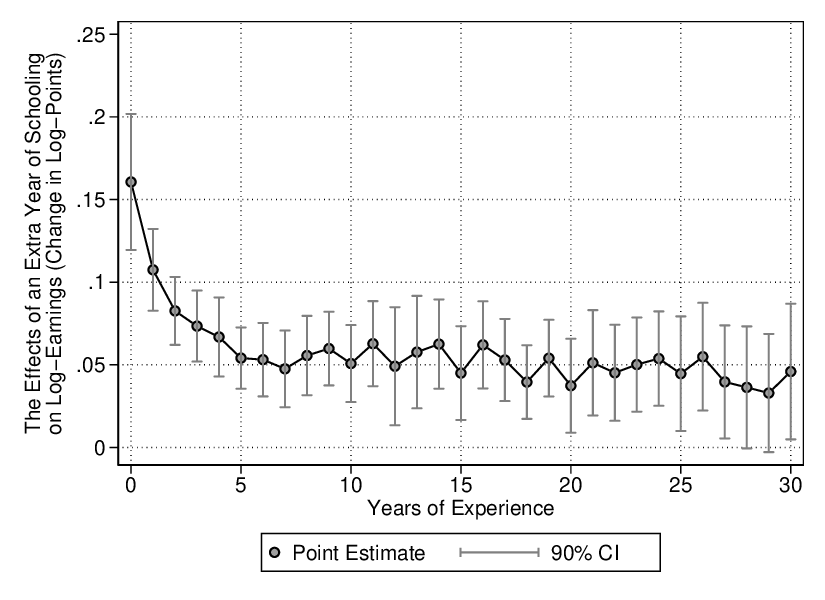}}
\par\end{centering}
\begin{centering}
\subfloat[Annual Earnings $>.5$ SGA]{\centering{}\includegraphics[width=0.33\columnwidth]{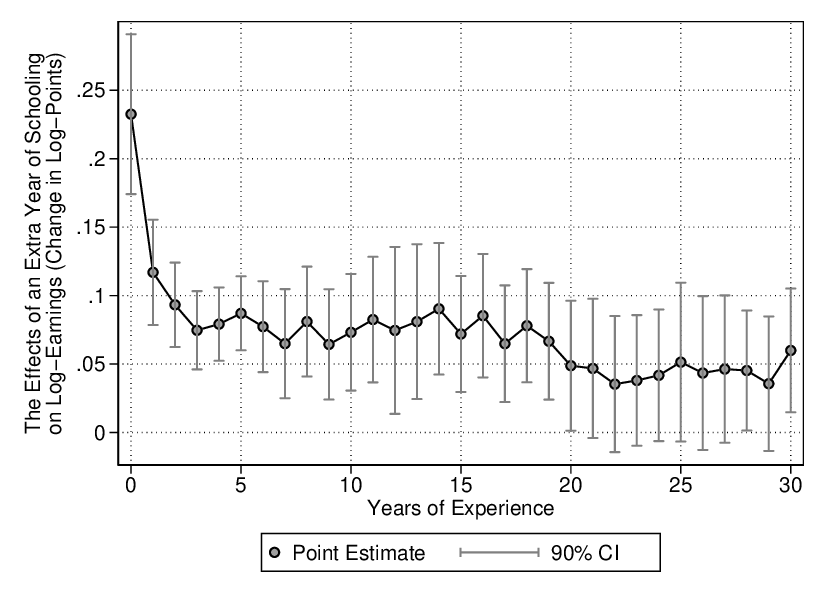}}
\subfloat[Annual Earnings $>.75$ SGA]{\centering{}\includegraphics[width=0.33\columnwidth]{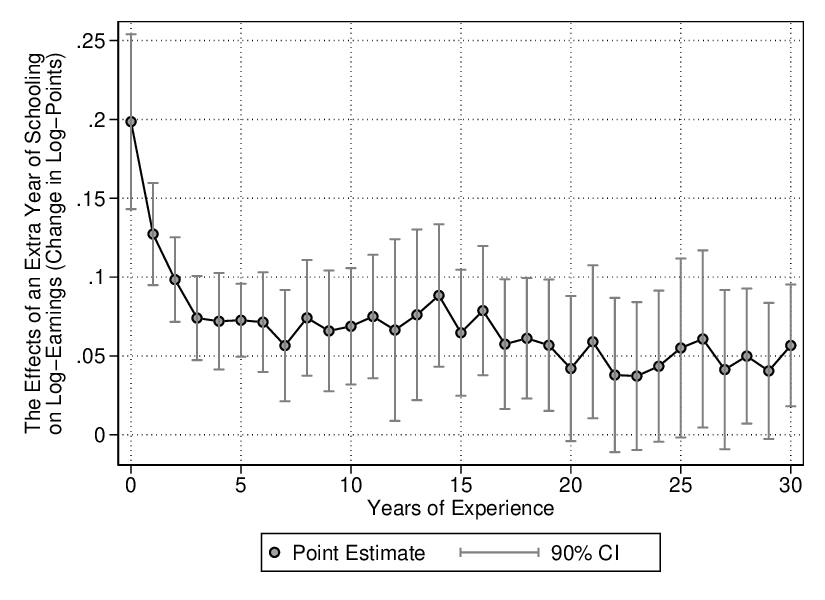}}
\subfloat[Annual Earnings $>1.5$ SGA]{\centering{}\includegraphics[width=0.33\columnwidth]{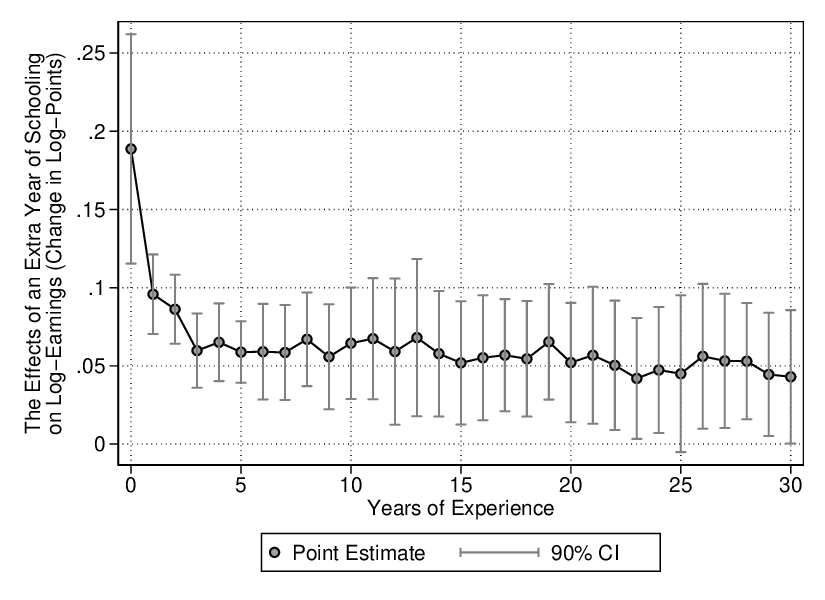}}
\par\end{centering}
\begin{centering}
\subfloat[Population $<$ 5,000]{\centering{}\includegraphics[width=0.33\columnwidth]{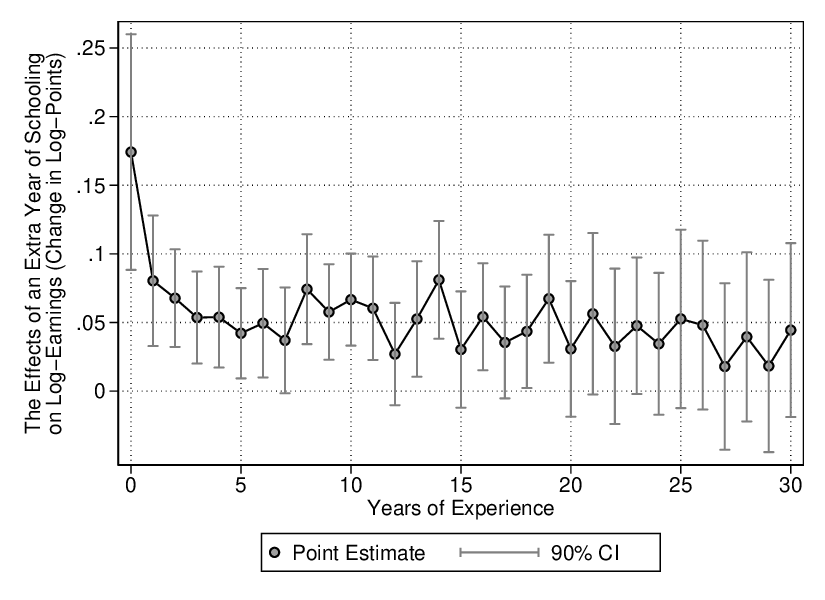}}
\subfloat[Population $<$ 10,000]{\centering{}\includegraphics[width=0.33\columnwidth]{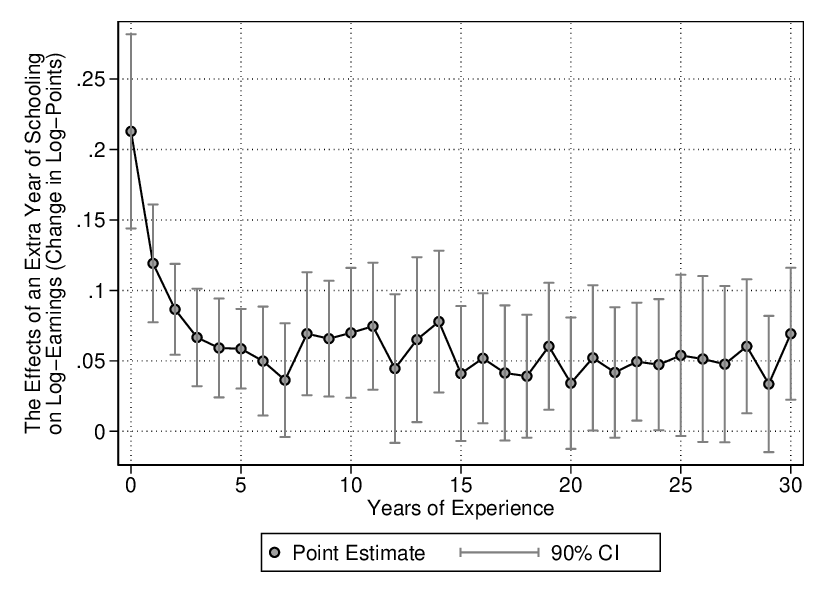}}
\subfloat[Population $<$ 25,000]{\centering{}\includegraphics[width=0.33\columnwidth]{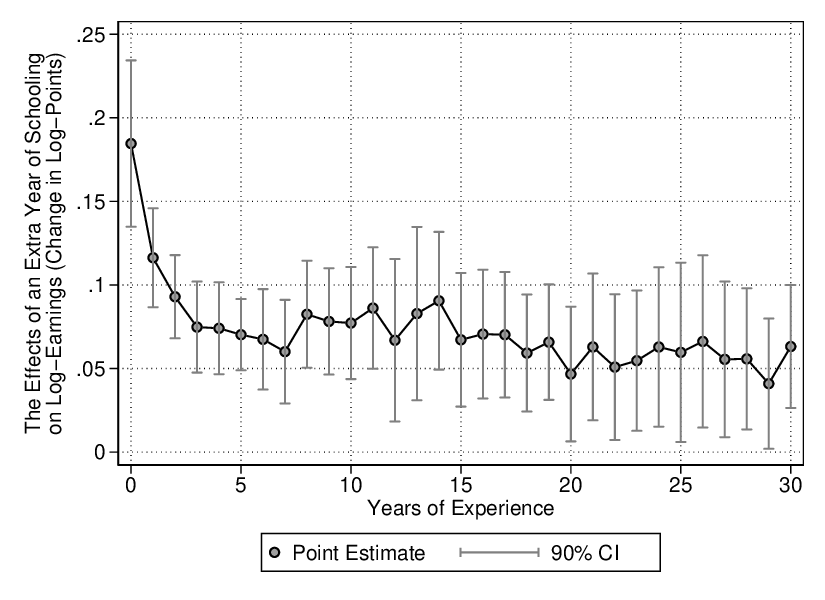}}
\par\end{centering}
\caption{Alternative Hidden IV Estimates of the Returns to Schooling.\label{fig:IVestimates_robust}}

\emph{\scriptsize{}Note:}{\scriptsize{} 
See notes below Table \ref{tab:IV_Speed_InitalLimit_robust} for details on each estimation.}{\scriptsize\par}
\end{sidewaysfigure}

\setcounter{section}{0}
\setcounter{equation}{0}
\setcounter{table}{0}
\setcounter{figure}{0}
\renewcommand{\thesection}{B}
\renewcommand{\theequation}{B.\arabic{equation}}
\renewcommand{\thetable}{B.\arabic{table}}
\renewcommand{\thefigure}{B.\arabic{figure}}

\section{Identification with Partially-Transparent Instrument\label{sec:Extension_PartialIVappendix}}
In this section, we extend the identification with partially-transparent instrument variables. 
Let $D^{{\mathfrak p}}\in\{0,1\}$ denote the partially-transparent IV such that a $\rho\in[0,1]$ fraction of workers be ``exposed" to a transparent IV, and $(1-\rho)$ fraction to a hidden IV. 
Here, the fraction $\rho$ is unobserved to the researchers. 

\paragraph{Experience-Invariant Returns to Skill.} 
$D^{{\mathfrak {p}}}$ satisfies Assumption \ref{ass:1}, and with experience-invariant returns to skill, the conditional mean of log wages at $t$ given $D^{{\mathfrak {p}}}$ is 
{ \begin{eqnarray}
\E\left[\ln W_{it}|D^{{\mathfrak {p}}}_{i},t\right]&=&\rho\times \E\left[\ln W_{it}|D^{{\mathfrak t}}_{i},t\right]+(1-\rho)\times\E\left[\ln W_{it}|D^{{\mathfrak h}}_{i},t\right]\notag\\ 
&=&\rho\times \delta^{\psi|S}\times\E\left[S_{i}|D^{{\mathfrak t}}_{i}\right]+(1-\rho)\times \left(\delta^{\psi|S}+\theta_{t}\phi_{A|S}\right)\E\left[S_{i}|D^{{\mathfrak h}}_{i}\right],\quad\qquad\label{eq:partial1}
\end{eqnarray} }
which is the weighted average of the conditional log wage under transparent and hidden IV.
For notational simplicity, and without loss of generality, we assume that the first-stage effect of $D^{{\mathfrak {p}}}$ on schooling does not depend on whether it is hidden or transparent, i.e., $\E[S_{i}|D^{{\mathfrak h}}_{i}=1]-\E[S_{i}|D^{{\mathfrak h}}_{i}=0]=\E\left[S_{i}|D^{{\mathfrak t}}_{i}=1\right]-\E\left[S_{i}|D^{{\mathfrak t}}_{i}=0\right]$. Then from (\ref{eq:Binary Instrument}), with $\lambda_t=1$  
\begin{eqnarray}
\texttt{plim}\!\!\quad\!\!\hat{b}^{IV^{{\mathfrak {p}}}}_{t} & = & \frac{\E\left[\ln W_{it}|D_{i}^{{\mathfrak {p}}}=1,t\right]-\E\left[\ln W_{it}|D_{i}^{{\mathfrak {p}}}=0,t\right]}{\E\left[S_{i}|D_{i}^{{\mathfrak {p}}}=1,t\right]-\E\left[S_{i}|D_{i}^{{\mathfrak {p}}}=0,t\right]}\notag\\
& = & \delta^{\psi|S} \rho+\left(\delta^{\psi|S}+\theta_{t}\phi_{A|S}\right) (1-\rho)\notag\\
&=& \delta^{\psi|S}+\theta_{t}\phi_{A|S}(1-\rho),\quad\qquad \label{eq:IV_partial}
\end{eqnarray}
where the second equality follows from (\ref{eq:partial1}). 
Thus, a partially-transparent IV identifies a \emph{lower bound} on the private return to education. 
To identify the social return we have to rely on the information at $t=\infty$, because $\lim_{t\rightarrow\infty}\theta_t=0$, and from (\ref{eq:IV_partial}) we get 
$
\texttt{plim}\!\!\quad\!\!\left(\lim_{t\rightarrow\infty}\hat{b}^{IV^{{\mathfrak {p}}}}_{t}\right)=\delta^{\psi|S}.
$
Then, subtracting $\texttt{plim}\!\!\quad\!\!\left(\lim_{t\rightarrow\infty}\hat{b}^{IV^{{\mathfrak {p}}}}_{t}\right)$ from (\ref{eq:IV_partial}) evaluated at two finite experience levels, $t\neq t'$, and taking their ratios identify $\theta_t/\theta_{t'}$, identifies the speed of learning parameter $\kappa$ and with it the lower bound of the signaling value. 

\paragraph{Experience-Varying Returns to Skill.} 
When returns to skill vary with experience, access to a partially-transparent IV is insufficient to bound the returns to signaling because both $\lambda_t$ and $\theta_t$ vary with $t$. 
However, if we have access to a transparent IV and a partially-transparent IV, then under Assumptions \ref{ass:1}-\ref{ass:4}, we can identify the lower bound of the returns to signaling. Then we can use (\ref{eq:IV_transparent_time}) to identify $\{\lambda_t: t\in\mathbb{T}\}$, and the rest of the identification strategy follows the same steps as with the experience-invariant returns to skill. 

Furthermore, if we have a hidden IV and a partially-transparent IV, we can point-identify private returns to education and the returns to signaling. As we show next, for this identification result, we rely on the homogeneity of social returns across the two IV samples. 

We begin with observation that as with (\ref{eq:IV_partial}), $D^{{\mathfrak {p}}}$ identifies a mixture of social and private returns, i.e., from (\ref{eq:Binary Instrument}) we get $
\texttt{plim}\!\!\quad\!\!\hat{b}^{IV^{{\mathfrak {p}}}}_{t}= \lambda_t\times \left( \delta^{\psi|S}+\theta_{t}\phi_{A|S} (1-\rho)\right). 
$
Simplifying further and using $b^{IV^{\mathfrak h}}_{0}= \delta^{\psi|S}+\phi_{A|S}$ and $b^{IV^{\mathfrak h}}_{\infty}=\delta^{\psi|S}$ gives 
\begin{eqnarray}
\texttt{plim}\!\!\quad\!\!\hat{b}^{IV^{{\mathfrak {p}}}}_{t} 
&=&\lambda_t\times \left(\theta_{t}\times b^{IV^{\mathfrak h}}_{0}+\left(1-\theta_{t}\right)\times b^{IV^{\mathfrak h}}_{\infty}\right)-\lambda_t\times\theta_t \times \left(\rho \times\phi_{A|S}\right)\notag\\
&=&\texttt{plim}\!\!\quad\!\! \hat{b}^{IV^{\mathfrak h}}_{t}-\lambda_t\times\theta_t \times \rho \times (b^{IV^{\mathfrak h}}_{0}-b^{IV^{\mathfrak t}}_{0}),\qquad\label{eq:partial2}
\end{eqnarray}
where the last equality follows from (\ref{eq:nlls}).
 Next, we make the following assumption. 
 
\begin{thmbis}{ass:4} (Homogeneity of Social Returns) Let the social returns $\lambda_t\times \delta^{\psi|S}$ be homogenous across the hidden IV sample and partially-transparent IV sample at each $t\in\mathbb{T}$.
\end{thmbis} 
Although Assumption \ref{ass:4}$'$ is stronger than Assumption \ref{ass:4} it has a testable implication. 
In particular, it implies that the hidden IV estimate is always larger than the partially-transparent IV estimate at every $t>0$.\footnote{For instance, this assumption is rejected in our sample. As we can see from the estimates in Figure \ref{fig:IVestimates}, for some intermediate $t$, the social returns estimated from transparent IV sample (Figure \ref{fig:IVestimates}-(b)) is larger than the private returns estimated from hidden IV sample (Figure \ref{fig:IVestimates}-(a)).} 
Suppose $D^{\mathfrak h}$ and $D^{\mathfrak p}$ satisfy Assumptions \ref{ass:1}-\ref{ass:3} and \ref{ass:4}$'$.  
Evaluating (\ref{eq:partial2}) at $t=0$ and using $\theta_0=1$ and $\lambda_0=1$ give 
\begin{eqnarray}
\texttt{plim}\!\!\quad\!\! \hat{b}^{IV^{\mathfrak h}}_{0}-\texttt{plim}\!\!\quad\!\!\hat{b}^{IV^{{\mathfrak {p}}}}_{0}=\rho\times (b^{IV^{\mathfrak h}}_{0}-b^{IV^{\mathfrak t}}_{0}), \label{eq:p}
\end{eqnarray}
which identifies $\rho$ up to $(b^{IV^{\mathfrak h}}_{0}-b^{IV^{\mathfrak t}}_{0})$.
Substituting (\ref{eq:p}) in (\ref{eq:partial2}) gives $\lambda_t\times\theta_t = \frac{\texttt{plim}\!\!\quad\!\! \hat{b}^{IV^{\mathfrak h}}_{t}-\texttt{plim}\!\!\quad\!\!\hat{b}^{IV^{{\mathfrak {p}}}}_{t}}{\texttt{plim}\!\!\quad\!\! \hat{b}^{IV^{\mathfrak h}}_{0}-\texttt{plim}\!\!\quad\!\!\hat{b}^{IV^{{\mathfrak {p}}}}_{0}},
$
and substituting this in $\texttt{plim}\!\!\quad\!\! \hat{b}^{IV^{\mathfrak h}}_{t}$ for $1\leq t\leq\infty$, and simplifying gives 
\begin{eqnarray}
\texttt{plim}\!\!\quad\!\! \hat{b}^{IV^{\mathfrak h}}_{t}\times\left( \frac{\texttt{plim}\!\!\quad\!\! \hat{b}^{IV^{\mathfrak h}}_{t}-\texttt{plim}\!\!\quad\!\!\hat{b}^{IV^{{\mathfrak {p}}}}_{t}}{\texttt{plim}\!\!\quad\!\! \hat{b}^{IV^{\mathfrak h}}_{0}-\texttt{plim}\!\!\quad\!\!\hat{b}^{IV^{{\mathfrak {p}}}}_{0}}\right)^{-1} &=& b^{IV^{\mathfrak h}}_{0}+b^{IV^{\mathfrak h}}_{\infty}\times\left(\frac{1-\theta_t}{\theta_t}\right).\label{eq:partial4}
\end{eqnarray}
Assumption \ref{ass:4}$'$ implies that $\texttt{plim}\!\!\quad\!\! \hat{b}^{IV^{\mathfrak h}}_{t}>\texttt{plim}\!\!\quad\!\!\hat{b}^{IV^{{\mathfrak {p}}}}_{t}$ and $\texttt{plim}\!\!\quad\!\! \hat{b}^{IV^{\mathfrak h}}_{0}\geq \texttt{plim}\!\!\quad\!\!\hat{b}^{IV^{{\mathfrak {p}}}}_{0}$ for $t>0$. 
Thus, with a sufficiently large panel, we can use the NLLS method to estimate the right-hand side parameters of (\ref{eq:partial4}) and from that the speed of learning parameter $\kappa$. 
 
\setcounter{section}{0}
\setcounter{equation}{0}
\setcounter{table}{0}
\setcounter{figure}{0}
\renewcommand{\thesection}{C}
\renewcommand{\theequation}{C.\arabic{equation}}
\renewcommand{\thetable}{C.\arabic{table}}
\renewcommand{\thefigure}{C.\arabic{figure}}
\section{Identification with Heterogeneous Returns \label{sec:heterogeneous}}
In this section, we extend the employer learning model to allow heterogeneous returns to education and determine conditions under which we can use IV to identify key model parameters. We use the binary potential outcomes framework of Neyman-Rubin.
Let schooling takes two values, $S_i\in \{0,1\}$, where $S_i=1$ (respectively, $0$) denotes a higher (respectively, lower) level of schooling. 
Worker $i$ is characterized by a vector of potential outcomes
$\left\{ \psi_{0,i},\psi_{1,i}\right\} $, where $\psi_{S,i}$ is the experience-invariant component of $i$'s productivity, which subsumes $A_{i}$ in (\ref{eq:chi}).

As in (\ref{eq:chi}), let $\varepsilon_{i,t}\stackrel{i.i.d}{\sim}{\mathcal N}\left(0,\sigma_{\varepsilon}^{2}\right)$ are mean-zero ``noise" in the production process that are independent of the model primitives.
Then the realized productivity $\psi_{i,t}$ at time $t$ is:
\begin{eqnarray}
\psi_{i,t}&=&S_{i} \times \left[{\psi}_{1,i} + \varepsilon_{1,i,t}\right]+(1-S_{i}) \times \left[{\psi}_{0,i} + \varepsilon_{0,i,t}\right]+H(t). \label{eq:Def psi_time}
\end{eqnarray}
For notational ease, we suppress $H(t)$ in the following.\footnote{Note that productivity $\psi_{i,t}$ is expressed in levels and not in logs, because with this level of generality it is easier to work in levels. We continue to maintain the assumption that $S$ and $H(t)$ are additively separable.}
Worker $i$ knows his potential outcomes $\left\{ \psi_{0,i},\psi_{1,i}\right\} $,
but employers only observe $(S_{i},\psi_{i}^{t})$, where $\psi_{i}^{t}:=\left\{ \psi_{i,\tau}\right\} _{\tau<t}$ are observed only for $S_{i}=S$. Note that observing $\psi_{i,t}$, conditional on $S_{i}$ is informationally equivalent to observing $\xi_{S_i,i,t}={\psi}_{S_i,i}+\varepsilon_{S_i,i,t}$. We can thus denote the employers' information set by $\mathcal{E}_{i,t}^{S_i}=(S_{i},\xi_{S_i,i}^{t})$, where $\xi_{S_i,i}^{t}=\left\{ \xi_{S_i,i,\tau}\right\} _{\tau<t}$. Wages are set equal to the expected productivity, conditional on information $\mathcal{E}_{i,t}^{S_i}$. Let $W_{S_i,i,t}:= \mathbb{E}\left[{\psi}_{1,i}\Big{|}\mathcal{E}_{i,t}^{S_i}\right] $ denotes potential wage outcomes for different $S_{i}$. Then we can write the wage equation as 
\begin{eqnarray}
W_{i,t}&=&\mathbb{E}\left[\psi_{i,t}\Big{|}\mathcal{E}_{i,t}^{S_i}\right]=S_i\times W_{1,i,t}+(1-S_i)\times W_{0,i,t} ,\label{eq:wages}
\end{eqnarray}
where the second equality follows from (\ref{eq:Def psi_time}) and the independence and zero-mean properties of $\varepsilon_{S_i,i,t}$. 

Then, the social returns and the private returns to schooling for $i$ are, respectively 
\begin{eqnarray}
\delta^{\psi|S_i}_{i} &:=& \psi_{1,i}-\psi_{0,i};\label{eq:Individual Social Returns}\\
\delta^{W|S_i}_{i,t} &:=& W_{1,i,t}-W_{0,i,t} = \mathbb{E}\left[{\psi}_{1,i}|\mathcal{E}_{i,t}^{1}\right]-\mathbb{E}\left[{\psi}_{0,i}|\mathcal{E}_{i,t}^{0}\right].\label{eq:Individual Private Returns}
\end{eqnarray}
Note that both the social returns $\delta^{\psi|S}_{i}$ and the private returns $\delta^{W|S}_{i,t}$ are individual-specific. The \emph{average social returns} and \emph{average private returns} are then the population averages of (\ref{eq:Individual Social Returns}) and (\ref{eq:Individual Private Returns}), respectively, while measures such as \emph{social returns for the treated} and \emph{private returns for the treated} are averages across the corresponding populations.

\subsection*{Identification Using Instrumental Variables}
To understand what a binary IV identifies, we proceed analogously
to \cite{ImbensAngrist1994}.
Let $S_{i}\left(D_{i}\right)$ denote potential schooling, which is a function of $D_{i}\in\{0,1\}$, and define compliers as $\mathbb{C\equiv}\left\{ i|S_{i}\left(1\right)=1, \text{and}\quad\!\!\! S_{i}\left(0\right)=0\right\}$ and defiers as $\mathbb{D\equiv}\left\{ i|S_{i}\left(1\right)=0, \text{and}\quad\!\!\! S_{i}\left(0\right)=1\right\}$.
Similarly, we can define always-takers to be $\mathbb{A\equiv}\left\{ i|S_{i}\left(1\right)=1, \text{and}\quad\!\!\! S_{i}\left(1\right)=1\right\} $
and never-takers to be $\mathbb{N\equiv}\left\{ i|S_{i}\left(1\right)=0, \text{and}\quad\!\!\!S_{i}\left(0\right)=0\right\}$.
Then, as before, the Wald estimator gives 
\begin{eqnarray}
\texttt{plim}\quad\!\! \hat{b}_{t}^{IV} & := & \frac{\mathbb{E}\left[W_{i,t}|D_{i}=1\right]-\mathbb{E}\left[W_{i,t}|D_{i}=0\right]}{\mathbb{E}\left[S_{i}|D_{i}=1\right]-\mathbb{E}\left[S_{i}|D_{i}=0\right]}.\label{eq:Wald estimator}
\end{eqnarray}
As $D$ satisfies the monotonicity condition, $\Pr(\mathbb{D})=0$, (\ref{eq:Wald estimator})'s denominator simplifies to  
\begin{eqnarray}
\mathbb{E}\left[S_{i}|D_{i}=1\right]-\mathbb{E}\left[S_{i}|D_{i}=0\right]&=&(\mathbb{E}\left[S_{i}|D_{i}=1, \mathbb{A}\right]-\mathbb{E}\left[S_{i}|D_{i}=0, \mathbb{A}\right])\times\Pr(\mathbb{A})\notag \\&+& (\mathbb{E}\left[S_{i}|D_{i}=1, \mathbb{N}\right]-\mathbb{E}\left[S_{i}|D_{i}=0, \mathbb{N}\right])\times\Pr(\mathbb{N})\notag\\&+& (\mathbb{E}\left[S_{i}|D_{i}=1, \mathbb{C}\right]-\mathbb{E}\left[S_{i}|D_{i}=0, \mathbb{C}\right])\times\Pr(\mathbb{C})\notag\\&+&(\mathbb{E}\left[S_{i}|D_{i}=1, \mathbb{D}\right]-\mathbb{E}\left[S_{i}|D_{i}=0, \mathbb{D}\right])\times\Pr(\mathbb{D})=\Pr(\mathbb{C}).\qquad\quad\label{eq:denominator}
\end{eqnarray}
As before, $D_i^{\mathfrak h}\in\{0,1\}$ denotes a hidden IV and $D_i^{\mathfrak t}\in\{0,1\}$ a transparent IV. With a hidden IV, we also know that $i$'s wage conditional on employer information $\mathcal{E}_{i,t}^{S}$ does not depend on the IV itself. Thus, for $D_i^{\mathfrak h}$, (\ref{eq:Wald estimator})'s numerator can be expressed as  
\begin{eqnarray}
\mathbb{E}[W_{i,t}|D_{i}^{\mathfrak h}=1]-\mathbb{E}[W_{i,t}|D_{i}^{\mathfrak h}=0] & = & \left(\mathbb{E}[W_{i,t}|D_{i}^{\mathfrak h}=1, \mathbb{C}]-\mathbb{E}[W_{i,t}|D_{i}^{\mathfrak h}=0, \mathbb{C}]\right)\times \Pr(\mathbb{C})\nonumber \\
 & = & \left(\mathbb{E}[W_{1,i,t}|\mathbb{C}]-\mathbb{E}[W_{0,i,t}|\mathbb{C}]\right)\times \Pr(\mathbb{C})=\mathbb{E}[\delta^{W|S}_{i,t}|\mathbb{C}]\times \Pr(\mathbb{C}).\qquad\quad\label{eq:Hidden IV estimate}
\end{eqnarray}
The first equality follows from the law of total expectation, the second from the definition of a complier and substituting for the potential outcomes from (\ref{eq:wages}) and using the properties of a hidden IV, and the last from the definition of private returns in (\ref{eq:Individual Private Returns}). Using (\ref{eq:denominator}) and (\ref{eq:Hidden IV estimate}) in (\ref{eq:Wald estimator}) with a hidden IV, gives $\texttt{plim}\quad\!\! \hat{b}_{t}^{IV^{\mathfrak h}} =\mathbb{E}\left[\delta^{W|S}_{i,t}\big{|}\mathbb{C}\right]$. Thus, a binary hidden IV identifies the \emph{average private returns} to education among compliers.

Next, we consider the identification with transparent IV, $D_i^{\mathfrak t}\in\{0,1\}$. Wages equal expected productivity given employer information $(\mathcal{E}_{i,t}^{S},D_i^{\mathfrak t})$, i.e., $
W_{i,t}=\mathbb{E}[\psi_{i,t}|\mathcal{E}_{i,t}^{S},D_i^{\mathfrak t}]$ and from the law of total expectation we get 
$
\mathbb{E}\left[W_{i,t}|D_{i}^{\mathfrak t}\right]=\mathbb{E}[\ensuremath{\mathbb{E}}[\psi_{i,t}|\mathcal{E}_{i,t}^{S_i},D_i^{\mathfrak t}]|D_{i}^{\mathfrak t}]=\mathbb{E}[\psi_{i,t}|D_{i}^{\mathfrak t}]$. 
Conditional on $D_i^{\mathfrak t}$, the average wages equals
the average product, and hence 
\begin{eqnarray}
\mathbb{E}\left[W_{i,t}|D_{i}^{\mathfrak t}=1\right]-\mathbb{E}[W_{i,t}|D_{i}^{\mathfrak t}=0] & = & \mathbb{E}[\psi_{i,t}|D_{i}^{\mathfrak t}=1]-\mathbb{E}[\psi_{i,t}|D_{i}^{\mathfrak t}=0]\notag\\&=& \mathbb{E}[\psi_{1,i}-\psi_{0,i}|\mathbb{C}]\times\Pr\left(\mathbb{C}\right)=\mathbb{E}[\delta^{\psi|S}_{i}|\mathbb{C}]\times\Pr\left(\mathbb{C}\right).\qquad\qquad
\label{eq:Transparent IV estimate}
\end{eqnarray}

Using (\ref{eq:denominator}) and (\ref{eq:Transparent IV estimate}) in (\ref{eq:Wald estimator}) gives $\texttt{plim}\quad\!\! \hat{b}^{IV^{\mathfrak t}} =\mathbb{E}[\delta^{\psi|S}_{i}|\mathbb{C}]$. Thus, a binary transparent IV identifies the \emph{average social returns} to education among compliers. 
Therefore, with heterogeneous returns, a transparent IV identifies the average social returns and a hidden IV identifies the average private returns to education for the compliers. 

\paragraph{Speed of Learning.}
Next, we consider identifying the speed of learning by determining how quickly the market learns workers' ability. 
The speed of learning will depend on schooling and the selection between schooling and unobserved ability. 
Let the unobserved components of productivity, conditional on schooling $S$, follow a Normal distribution, i.e., ${\psi}_{S,i}\big{|}S\sim {\mathcal N}\left(\mu_{\psi,S}, \sigma_{\psi}^{2}\right)$. 
Using the potential wage outcomes defined in (\ref{eq:wages}), and the Kalman property as in (\ref{eq:Kalman}), we can express wages as
\begin{eqnarray}
W_{S_i,i,t}=\mathbb{E}\left[{\psi}_{S_i,i}|\mathcal{E}_{i,t}^{S_i}\right]=\theta_{t}\times\mu_{\psi,S}+\left(1-\theta_{t}\right)\times\bar{\xi}_{S_i,i}^{t},\label{eq:Wagenew}
\end{eqnarray}
where, $\theta_{t}=\frac{1-\kappa}{1+\left(t-1\right)\kappa}$,  $\kappa=\frac{\sigma_{\psi}^{2}}{\sigma_{\psi}^{2}+\sigma_{\varepsilon}^{2}}$ is the speed of learning, and $\bar{\xi}_{S_i,i}^{t}=\frac{1}{t}\sum_{\tau<t}\xi_{S_i,i,\tau}$ is the average of signals up to period $t$. 
Conditional expectation of (\ref{eq:Wagenew}) for $(S_{i}, {\psi}_{S_i,i})$ gives 
\begin{eqnarray}
\mathbb{E}[W_{S_i,i,t}|S_{i},{\psi}_{S_i,i}] & = & \theta_{t}\times\mu_{\psi,S}+\left(1-\theta_{t}\right)\times\mathbb{E}\left[\bar{\xi}_{S_i,i}^t|S_{i},\tilde{\psi}_{S_i,i}\right]=\theta_{t}\times\mu_{\psi,S}+\left(1-\theta_{t}\right)\times{\psi}_{S,i}\nonumber \\
 & = & \mu_{\psi,S}+\left(1-\theta_{t}\right)\times\left({\psi}_{S_i,i}-\mu_{\psi,S}\right).\label{eq:ExpWage given Inf}
\end{eqnarray}
Thus, the wage of a worker with schooling $S_i$ at with $t$ years of work experience is the sum of the average productivity $\mu_{\psi,S}$ and the deviation of $i$'s productivity from its mean $\left({\psi}_{S_i,i}-\mu_{\psi,S}\right)$ augmented by employers' learning $(1-\theta_{t})$. So, at the start, i.e., $t=0$, $i$'s wage is the average productivity because $\theta_{0}=1$, and as information about $i$'s ability is accumulated in the market, $i$'s wage becomes more responsive to $i$'s true productivity as $\lim_{t\rightarrow\infty}\theta_{t}=0$. 
Using (\ref{eq:Individual Private Returns}), (\ref{eq:Hidden IV estimate}) and (\ref{eq:ExpWage given Inf}) in (\ref{eq:Wald estimator}) the hidden IV estimate becomes  
{\footnotesize\begin{eqnarray}
\texttt{plim}\quad\!\!\hat{b}^{IV^{\mathfrak h}}_{t} & = & \mathbb{E}\left[\left(\mu_{\psi,1}-\mu_{\psi,0}\right)+\left(1-\theta_{t}\right)\left({\psi}_{1,i}-\mu_{\psi,1}\right)-\left(1-\theta_{t}\right)\left({\psi}_{0,i}-\mu_{\psi,0}\right)\Big{|}\mathbb{C}\right]\nonumber \\
 & = & \underbrace{\mathbb{E}\left[\left(\mu_{\psi,1}-\mu_{\psi,0}\right)|\mathbb{C}\right]}_{:=\Upsilon}+\left(1-\theta_{t}\right)\Big(\underbrace{\mathbb{E}\left[\left({\psi}_{1,i}-\mu_{\psi,1}\right)|\mathbb{C}\right]}_{:=\Upsilon_1}-\underbrace{\mathbb{E}\left[\left({\psi}_{0,i}-\mu_{\psi,0}\right)|\mathbb{C}\right]}_{:=\Upsilon_0}\Big)\notag\\&:=& \Upsilon+\left(1-\theta_{t}\right)(\Upsilon_{1}-\Upsilon_{0}).\label{eq:IV as mixture of learning}
\end{eqnarray}}
Furthermore, because private returns are greater than social returns, it follows from (\ref{eq:Individual Social Returns}) and (\ref{eq:Individual Private Returns}) that $ \Upsilon_{1}< \Upsilon_{0}$, and $T$ is sufficiently large.
Then we can use (\ref{eq:IV as mixture of learning}) to identify $\theta_t$ and the speed of learning parameter $\kappa$. For instance, at $t=0$ we can get  $\texttt{plim}\quad\!\!\hat{b}^{IV^{\mathfrak h}}_{0}=\Upsilon$ and at $t\rightarrow \infty$ we get $\texttt{plim}\quad\!\!\hat{b}^{IV^{\mathfrak h}}_{\infty}=\Upsilon+(\Upsilon_{1}-\Upsilon_{0})$. So, for $0<t<\infty$ we identify $\theta_t = \frac{\texttt{plim}\quad\!\!\hat{b}^{IV^{\mathfrak h}}_{t}-\texttt{plim}\quad\!\!\hat{b}^{IV^{\mathfrak h}}_{0}}{\texttt{plim}\quad\!\!\hat{b}^{IV^{\mathfrak h}}_{\infty}-\texttt{plim}\quad\!\!\hat{b}^{IV^{\mathfrak h}}_{0}}$. 

\paragraph{Experience-Varying Returns to Skills.}
We end this section by briefly considering the identification with experience-varying returns to skill. 
First note that with experience-varying returns to skill, realized productivity (\ref{eq:Def psi_time}) becomes 
\begin{eqnarray}
\psi_{i,t}=\lambda_t\times\left(S_{i} \times \left[ {\psi}_{1,i} + \varepsilon_{1,i,t}\right]+(1-S_{i}) \times \left[{\psi}_{0,i} + \varepsilon_{0,i,t}\right]\right)+H(t). \label{eq:Def psi_time_new}
\end{eqnarray}
Once we re-define social and private returns to include the effect of $\lambda_t$, following the same identification strategy as above, it follows that the hidden IV identifies the average private returns and transparent IV identifies the average social returns, among the compliers, at each work experience $t\in\mathbb{T}$. 
Suppose, as before, we have access to a hidden IV and a transparent IV and suppose $\{\lambda_t: t\in \mathbb{T}\}$ satisfy Assumption \ref{ass:4}. 

The hidden IV and the transparent IV estimates the (average) private and (average) social returns for different complier groups. Under Assumption \ref{ass:4}, however, we can combine the two sets of estimates to identify the model with experience-varying returns for hidden IV compliers, even though the transparent IV estimates relate to a different set of compliers. 

In particular, following the same step as in  (\ref{eq:IV as mixture of learning}) we get 
{\footnotesize\begin{eqnarray}
\texttt{plim}\quad\!\!\hat{b}^{IV^{\mathfrak h}}_{t} &=& \lambda_t\Upsilon+\lambda_t\left(1-\theta_{t}\right)(\Upsilon_{1}-\Upsilon_{0});\label{eq:secondlast}\\
\texttt{plim}\quad\!\!\hat{b}^{IV^{\mathfrak t}}_{t} & = &\lambda_t\Upsilon+\lambda_t(\Upsilon_{1}-\Upsilon_{0})\label{eq:last}.
\end{eqnarray}}
Evaluating (\ref{eq:secondlast}) at $t=0$, and using $\lambda_0=1$, identifies $\texttt{plim}\quad\!\!\hat{b}^{IV^{\mathfrak h}}_{0}=\Upsilon$. Substituting it in (\ref{eq:last}) at $t=0$ identifies $\texttt{plim}\quad\!\!\hat{b}^{IV^{\mathfrak t}}_{0}-\texttt{plim}\quad\!\!\hat{b}^{IV^{\mathfrak h}}_{0}=(\Upsilon_{1}-\Upsilon_{0})$. Then from the transparent IV (\ref{eq:last}) we can identify $\{\lambda_t: t>0\}$. Then using these variables in (\ref{eq:secondlast}) we identify $\kappa$. 

\setcounter{section}{0}
\setcounter{equation}{0}
\setcounter{table}{0}
\setcounter{figure}{0}
\renewcommand{\thesection}{D}
\renewcommand{\theequation}{D.\arabic{equation}}
\renewcommand{\thetable}{D.\arabic{table}}
\renewcommand{\thefigure}{D.\arabic{figure}}
\section{Identification with Employer-Observed Correlate\label{section:IQ_new}}
In this section, we extend our primary model to allow employers to observe a correlate of ability $Q$ that the researcher does not observe. Throughout this section, we maintain all other assumptions from our model.  
Worker $i$'s log-productivity for $t\in\mathbb{T}$ is given by 
\begin{eqnarray}
\psi_{it} := \ln \chi_{it} =\lambda_{t}\times\left(\beta_{ws}S_{i}+\beta_{wq}Q_{i}+A_{i}+\varepsilon_{it}\right)+H(t),\label{eq:chi_new}
\end{eqnarray}
where $Q$ is a correlate of ability observed by employers and is possibly correlated with $A$. An example of a $Q$ could be knowledge of a foreign language, which is typically mentioned in job applicants' r\'esum\'es, and can be verified by the employers.

To model employer learning in addition to $\varepsilon_{it}\stackrel{i.i.d}{\sim}\mathcal{N}(0,\sigma_{\varepsilon}^{2})$ let $(S_i,Q_i,A_i)\stackrel{i.i.d}{\sim}\mathcal{N}(\bm{\mu},\Sigma)$, across workers and
across time. The joint normality assumption allows us to express $A$ as a linear function of $(S,Q)$
\begin{equation}
A_{i}=\phi_{A|S}S_{i}+\phi_{A|Q}Q_{i}+\varepsilon_{A_i|S_i,Q_i},\label{eq:corr A on S,Q_new}
\end{equation}
where $\varepsilon_{A_i|S_i,Q_i}:= A_{i}-\E\left[A_{i}|S_i,Q_i\right]$. 
Under perfect competition, log wages is   
\begin{eqnarray}
\ln W_{it} & = & \lambda_t\times (\beta_{ws}S_{i}+\beta_{wq}Q_{i}+\E\left[A_{i}|\mathcal{E}_{it}\right])+\tilde{H}\left(t\right),\label{eq:Wage_new}
\end{eqnarray}
where, as before, $\tilde{H}(t)\equiv H\left(t\right)+\frac{1}{2}v_{t}$ collects the terms that vary only with $t$ but not across the realizations
of $\xi_{i}^{t}$. 
For notational simplicity, we suppress $\tilde{H}\left(t\right)$ until the empirical analysis.

The normality assumptions also allow us to use the Kalman filter to write the conditional expectation of ability $\E\left[A_{i}|\mathcal{E}_{it}\right]$ in linear form as 
\begin{eqnarray}
\E\left[A_{i}|\mathcal{E}_{it}\right]=\theta_{t}\E\left[A_{i}|S_i,Q_i\right]+\left(1-\theta_{t}\right)\bar{\xi_{i}^{t}},\label{eq:Kalman_new}
\end{eqnarray}
where $\bar{\xi_{i}^{t}}=\frac{1}{t}\sum_{\tau<t}\xi_{i\tau}$ is the
average of signals up to period $t$ and $\theta_{t}=\frac{1-\kappa}{1+(t-1)\kappa}\in[0,1]$ is the weight on the initial signal $(S_i,Q_i)$ with $\kappa=\frac{\sigma_{0}^{2}}{\sigma_{0}^{2}+\sigma_{\varepsilon}^{2}}\in[0,1]$. 
Next, we define the social and private returns to education. 
Recall the notation that for $Y$, $\delta^{Y|S}$ denotes the causal effect of $S$ on $Y$ and $\tilde{Y}$ denote the part of $Y$ that is not caused by schooling $S$ but may correlate with $S$. Using these notations and assumptions for $Y=Q$ and $Y=A$, we get, respectively,
\begin{eqnarray}
Q_{i} = \delta^{Q|S}S_{i}+\tilde{Q_{i}};\quad \text{and}\quad 
A_{i} =\delta^{A|S}S_{i}+\tilde{A}_{i}.\label{eq:causal A on SQ_new}
\end{eqnarray}
Then, substituting (\ref{eq:causal A on SQ_new}) into (\ref{eq:chi_new}),
we obtain 
\begin{eqnarray}
\psi_{it} = \lambda_t\times\left(\beta_{ws}+\beta_{wq}\delta^{Q|S}+\delta^{A|S}\right)S_{i}+\lambda_t\times\left(\beta_{wq}\tilde{Q}_{i}+\tilde{A}_{i}+\varepsilon_{it}\right):=\delta_{t}^{\psi|S}\times S_{i}+u_{it}.\label{eq: causal schooling 2_new}
\end{eqnarray}

The coefficient ($\delta_{t}^{\psi|S}$) in (\ref{eq: causal schooling 2_new}) is the total causal effect of schooling on productivity --the social return to education-- and it captures the direct and indirect effect on other ability components, i.e., $(Q,A)$.
Thus (\ref{eq: causal schooling 2_new}) shows that an extra year of $S$ increases $Q$ by $\delta^{Q|S}$ and $A$ by $\delta^{A|S}$ units, and in turn, they increase the productivity by $\lambda_t\beta_{wq}$ and $\lambda_t$, respectively. 

Consider now the private returns to education. 
Substituting (\ref{eq:corr A on S,Q_new}) and (\ref{eq:Kalman_new})
in (\ref{eq:Wage_new}), and using $\bar{\xi_{i}^{t}}:=\frac{1}{t}\Sigma_{\tau<t}\left(A_{i}+\varepsilon_{i\tau}\right)=A_{i}+\overline{\varepsilon}_{i}^{t}$, and the fact that $\E(A_i|S_i,Q_i)$ is linear and separable in $S_i$ and $Q_i$, we can write the log-earnings as 
$
\ln W_{it}=\lambda_t\times\left(\beta_{ws}+\theta_{t}\phi_{A|S}\right) S_{i}+\lambda_t\times\left(\beta_{wq}+\theta_{t}\phi_{A|Q}\right) Q_{i}+\lambda_t\times\left(1-\theta_{t}\right)\left(A_{i}+\overline{\varepsilon}_{i}^{t}\right).
$
Then, using (\ref{eq:causal A on SQ_new}) to replace $Q$ and $A$ gives  
{\begin{eqnarray}
\ln W_{it}
&=&\lambda_t\left(\beta_{ws}+\beta_{wq}\delta^{Q|S}+ \delta^{A|S}+ \theta_{t}(\phi_{A|S}+\phi_{A|Q}\delta^{Q|S}-\delta^{A|S})\right)S_{i}\notag\\&&+\lambda_t\left(\beta_{wq}+\theta_{t}\phi_{A|Q}\right)\tilde{Q}_{i}+\lambda_t\left(1-\theta_{t}\right)\left(\tilde{A}_{i}+\overline{\varepsilon}_{i}^{t}\right):=\delta_{t}^{W|S}\times S_{i}+\tilde{u}_{it}. \label{eq:final_wage_eq_new} 
\end{eqnarray}}
The coefficient of schooling ($\delta_{t}^{W|S}$) in (\ref{eq:final_wage_eq_new}) is the private return to education. 
Comparing this coefficient with the coefficient of schooling $\delta_{t}^{\psi|S}$ in (\ref{eq: causal schooling 2}), gives: 
\begin{equation}
\delta_{t}^{W|S}=\delta_{t}^{\psi|S}+\theta_{t}\times\lambda_t \times\left(\phi_{A|S}+\phi_{A|Q}\delta^{Q|S}-\delta^{A|S}\right).\label{eq:private return_new}
\end{equation}
Once we have augmented the definition of the social returns and the adjustment term to capture the effect of $Q$ in (\ref{eq:private return_new}), the rest of the identification results applies verbatim. In particular, when the returns to skill is experience-invariant and the hidden IV, $D^{{\mathfrak h}}$, satisfies the assumption $\ln W_{it}\text{\ensuremath{\perp}}D^{{\mathfrak h}}_{i}|(S_{i},Q_{i},\xi^{t}_{i})$ then it identifies the private returns to education, i.e., 
\begin{eqnarray}
\texttt{plim}\!\!\quad\!\! \hat{b}^{IV^{\mathfrak h}}_{t} = \frac{\E\left[\ln W_{it}|D^{{\mathfrak h}}_{i}=1,t\right]-\E\left[\ln W_{it}|D^{{\mathfrak h}}_{i}=0,t\right]}{\E\left[S_{i}|D^{{\mathfrak h}}_{i}=1\right]-\E\left[S_{i}|D^{{\mathfrak h}}_{i}=0\right]}=\delta^{\psi|S}+\theta_{t}\left(\phi_{A|S}+\phi_{A|Q}\delta^{Q|S}-\delta^{A|S}\right).\quad\label{eq:IV_hidden_new}
\end{eqnarray}
Comparing (\ref{eq:IV_hidden_new}) with the private returns defined in (\ref{eq:private return_new}), we can conclude that, for every work experience level $t$, the hidden IV identifies the private returns to education, i.e., $\texttt{plim}\!\!\quad\!\!\hat{b}^{IV^{\mathfrak h}}_{t}=\delta^{W|S}_{t}$. Hidden IV also identifies the speed of learning.  
Likewise, the Wald estimator for a transparent IV, $D^{{\mathfrak t}}$, identifies the social returns to education at all $t$, i.e.,
\begin{eqnarray}
\texttt{plim}\!\!\quad\!\! \hat{b}^{IV^{\mathfrak t}}_{t} &=& \frac{\E\left[\ln W_{it}|D^{{\mathfrak t}}_{i}=1,t\right]-\E\left[\ln W_{it}|D^{{\mathfrak t}}_{i}=0,t\right]}{\E\left[S_{i}|D^{{\mathfrak t}}_{i}=1\right]-\E\left[S_{i}|D^{{\mathfrak t}}_{i}=0\right]}=\delta^{\psi|S}.\quad\label{eq:IV_transparent_new}
\end{eqnarray}

Next, we consider the case when the returns to skill vary with experience. 
As before, a hidden IV identifies the private returns to education, and a transparent IV identifies the experience-varying social returns to education. 
Formally, following the same steps as in (\ref{eq:IV_hidden_new}), $D^{\mathfrak h}$ and $D^{\mathfrak t}$ at $t$, respectively, identify the private and social returns as 
\begin{eqnarray}
\texttt{plim}\!\!\quad\!\!\hat{b}^{IV^{\mathfrak h}}_{t} \!\!\!\!\!\!& = &\!\!\!\!\! \frac{\E\left[\ln W_{it}|D_{i}^{{\mathfrak h}}=1,t\right]-\E\left[\ln W_{it}|D_{i}^{{\mathfrak h}}=0,t\right]}{\E\left[S_{i}|D_{i}^{{\mathfrak h}}=1,t\right]-\E\left[S_{i}|D_{i}^{{\mathfrak h}}=0,t\right]}\!=\!\lambda_t\!\left(\delta^{\psi|S}+ \theta_{t}(\phi_{A|S}+\phi_{A|Q}\delta^{Q|S}-\delta^{A|S})\right);\quad\qquad\label{eq:IV_hidden_time_new} \\ 
\texttt{plim}\!\!\quad\!\!\hat{b}_{t}^{IV^{\mathfrak t}} \!\!\!& = &\!\!\! \frac{\E\left[\ln W_{it}|D_{i}^{{\mathfrak t}}=1,t\right]-\E\left[\ln W_{it}|D_{i}^{{\mathfrak t}}=0,t\right]}{\E\left[S_{i}|D_{i}^{{\mathfrak t}}=1,t\right]-\E\left[S_{i}|D_{i}^{{\mathfrak t}}=0,t\right]} =\lambda_{t}\times\delta^{\psi|S}:= \delta_{t}^{\psi|S}.\label{eq:IV_transparent_time_new}
\end{eqnarray}
Note that with access to both IVs that satisfy Assumption \ref{ass:4}, we can identify $\{\lambda_t: t\in\mathbb{T}\}$.

\end{document}